\documentclass[onecolumn,tighten,times]{aastex62}
\usepackage{mathrsfs}
\usepackage{amsmath}
\usepackage{url}
\usepackage[normalem]{ulem}
\usepackage{graphicx}

\usepackage{natbib}
\usepackage{color}
\bibpunct{(}{)}{;}{a}{}{,}

\newcommand\aproxgt{\mathrel{%
      \rlap{\raise 0.511ex \hbox{$>$}}{\lower 0.511ex \hbox{$\sim$}}}}
\newcommand\aproxlt{\mathrel{%
      \rlap{\raise 0.511ex \hbox{$<$}}{\lower 0.511ex \hbox{$\sim$}}}}

\newcommand{\ignore}[1]{}

\newcommand{\hst}{{\it HST\,}}

\newcommand{\mbh}{\mbox{$M_{\rm BH}$}}
\newcommand{\msun}{\mbox{$M_\odot$}}
\newcommand{\kms}{\mbox{\rm km~s$^{-1}$}}
\newcommand{\ergsec}{\mbox{\rm erg\,s$^{-1}$}}
\newcommand{\sigl}{\mbox{$\sigma_{\rm line}$}}
\newcommand{\sigr}{\mbox{$\sigma_{\rm R}$}}
\newcommand{\sigm}{\mbox{$\sigma_{\rm M}$}}
\newcommand{\fwr}{\mbox{${\rm FWHM}_{\rm R}$}}
\newcommand{\fwm}{\mbox{${\rm FWHM}_{\rm M}$}}
\newcommand{\murm}{\mbox{$\mu_{\rm RM}$}}
\newcommand{\muse}{\mbox{$\mu_{\rm SE}$}}
\newcommand{\mdot}{\mbox{$\dot{m}$}}
\newcommand{\cc}{\mbox{${\rm cm}^{-3}$}}
\newcommand{\EVone}{\mbox{Eigenvector 1}}
\newcommand{\RFe}{\mbox{${\cal R}$(Fe\,{\sc ii})}}

\newcommand{\hb}{\mbox{\rm H$\beta$}}
\newcommand{\civ}{\mbox{\rm C\,{\sc iv}}}
\newcommand{\feii}{\mbox{\rm Fe\,{\sc ii}}}
\newcommand{\mgii}{\mbox{\rm Mg\,{\sc ii}}}
\newcommand{\siiv}{\mbox{\rm Si\,{\sc iv}}}
\newcommand{\heii}{\mbox{\rm He\,{\sc ii}}}

\newcommand{\oiii}{\mbox{\rm [O\,{\sc iii}]}}
\newcommand{\oiv}{\mbox{\rm O\,{\sc iv}]}}

\shorttitle{QUASAR BLACK HOLE MASSES}
\shortauthors{Dalla Bont\`{a} et al.}
 
\begin{document}

\title{THE SLOAN DIGITAL SKY SURVEY REVERBERATION MAPPING PROJECT:
ESTIMATING MASSES OF BLACK HOLES IN QUASARS WITH SINGLE-EPOCH SPECTROSCOPY}

\author[0000-0001-9931-8681]{Elena~Dalla~Bont\`{a}}
\affiliation{Dipartimento di Fisica e Astronomia ``G. Galilei,'' Universit\`{a} di Padova, Vicolo dell'Osservatorio 3, I-35122 Padova, Italy}
\affiliation{INAF-Osservatorio Astronomico di Padova, Vicolo dell'Osservatorio 5, I-35122, Padova, Italy}

\author[0000-0001-6481-5397]{Bradley~M.~Peterson}
\affiliation{Department of Astronomy, The Ohio State University, 140 W 18th Ave, Columbus, OH 43210, USA}
\affiliation{Center for Cosmology and AstroParticle Physics, 191 Woodruff Ave., Columbus, OH 43210, USA}
\affiliation{Space Telescope Science Institute, 3700 San Martin Drive, Baltimore, MD 21218, USA}

\author[0000-0002-2816-5398]{Misty~C.~Bentz}
\affiliation{Department of Physics and Astronomy, Georgia State University, 25 Park Place, Suite 605, Atlanta, GA 30303, USA}

\author[0000-0002-0167-2453]{W.~N.~Brandt}
\affiliation{Department of Astronomy and Astorphysics, Eberly College of Science, The Pennsylvania State
University, 525 Davey Laboratory, University Park, PA 16802, USA}
\affiliation{Institute for Gravitation and the Cosmos, The Pennsylvania State
University, University Park, PA 16802, USA}
\affiliation{Department of Physics, The Pennsylvania State
University, 525 Davey Laboratory, University Park, PA 16802, USA}

\author[0000-0001-9539-3940]{S.~Ciroi}
\affiliation{Dipartimento di Fisica e Astronomia ``G. Galilei,'' Universit\`{a} di Padova, Vicolo dell'Osservatorio 3, I-35122 Padova, Italy}

\author[0000-0003-3242-7052]{Gisella~De~Rosa}
\affiliation{Space Telescope Science Institute, 3700 San Martin Drive, Baltimore, MD 21218, USA}

\author[0000-0003-0042-6936]{Gloria~Fonseca~Alvarez}
\affiliation{Department of Physics, University of Connecticut, 2152 Hillside Rd., Unit 3046,
Storrs, CT 06269, USA}

\author[0000-0001-9920-6057]{Catherine~J.~Grier}
\affiliation{Steward Observatory, University of Arizona, 933 North Cherry Avenue, Tucson, AZ 85721, USA}

\author[0000-0002-1763-5825]{P.~B.~Hall}
\affiliation{Department of Physics and Astronomy, York University, Toronto, ON M3J 1P3, Canada}

\author[0000-0002-6733-5556]{Juan~V.~Hern\'{a}ndez~Santisteban}
\affiliation{SUPA Physics and Astronomy, University of St.\ Andrews, Fife, KY16 9SS Scotland, UK}

\author[0000-0001-6947-5846]{Luis~C.~Ho}
\affiliation{Kavli Institute for Astronomy and Astrophysics,
Peking University, Beijing 100871, China}
\affiliation{Department of Astronomy, School of Physics, Peking University,
Beijing 100871, China}

\author[0000-0002-0957-7151]{Y.~Homayouni}
\affiliation{Department of Physics, University of Connecticut, 2152 Hillside Rd., Unit 3046,
Storrs, CT 06269-3046, USA}

\author[0000-0003-1728-0304]{Keith~Horne}
\affiliation{SUPA Physics and Astronomy, University of St.\ Andrews, Fife, KY16 9SS Scotland, UK}

\author[0000-0001-6017-2961]{C.~S.~Kochanek}
\affiliation{Department of Astronomy, The Ohio State University, 140 W 18th Ave, Columbus, OH 43210, USA}
\affiliation{Center for Cosmology and AstroParticle Physics, 191 Woodruff Ave., Columbus, OH 43210, USA}

\author[0000-0002-0311-2812]{Jennifer~I-Hsiu~Li}
\affiliation{Department of Astronomy, University of Illinois
at Urbana--Champaign, Urbana, IL 61801, USA}

\author[0000-0001-6890-3503]{L.~Morelli}
\affiliation{Instituto de Astronom\`ia y Ciencias Planetarias, Universidad de Atacama, Copayapu 485, Copiap\`o, Chile}
\affiliation{INAF-Osservatorio Astronomico di Padova, Vicolo dell'Osservatorio 5 I-35122, Padova, Italy}

\author[0000-0001-9585-417X]{A.~Pizzella}
\affiliation{Dipartimento di Fisica e Astronomia ``G. Galilei,'' Universit\`{a} di Padova, Vicolo dell'Osservatorio 3, I-35122 Padova, Italy}
\affiliation{INAF-Osservatorio Astronomico di Padova, Vicolo dell'Osservatorio 5 I-35122, Padova, Italy}

\author[0000-0003-1435-3053]{R.~W.~Pogge}
\affiliation{Department of Astronomy, The Ohio State University, 140 W 18th Ave, Columbus, OH 43210, USA}
\affiliation{Center for Cosmology and AstroParticle Physics, 191 Woodruff Ave., Columbus, OH 43210, USA}

\author[0000-0001-7240-7449]{D.~P.~Schneider}
\affiliation{Department of Astronomy and Astorphysics, Eberly College of Science, The Pennsylvania State
University, 525 Davey Laboratory, University Park, PA 16802, USA}
\affiliation{Institute for Gravitation and the Cosmos, The Pennsylvania State
University, University Park, PA 16802, USA}

\author[0000-0003-1659-7035]{Yue~Shen}
\altaffiliation{Alfred P. Sloan Research Fellow}
\affiliation{Department of Astronomy, University of Illinois
at Urbana--Champaign, Urbana, IL 61801, USA}
\affiliation{National Center for Supercomputing Applications, University of Illinois at Urbana-Champaign, Urbana, IL 61801, USA}

\author[0000-0002-1410-0470]{J.~R.~Trump}
\affiliation{Department of Physics, University of Connecticut, 2152 Hillside Rd., Unit 3046,
Storrs, CT 06269, USA}

\author[0000-0001-9191-9837]{Marianne~Vestergaard}
\affiliation{DARK, Niels Bohr Institute, University of Copenhagen, Lyngbyvej 2, DK-2100 Copenhagen, Denmark}
\affiliation{Steward Observatory, University of Arizona, 933 North Cherry Avenue, Tucson, AZ 85721, USA}

\begin{abstract}
It is well known that reverberation mapping of active galactic nuclei
(AGN) reveals a relationship between AGN luminosity and the size of
the broad-line region, and that use of this relationship, combined
with the Doppler width of the broad emission line, enables an estimate
of the mass of the black hole at the center of the active nucleus
based on a single spectrum. An unresolved key issue is the choice of
parameter used to characterize the line width, either FWHM or line 
dispersion \sigl\ (the square root of the second moment of the line profile). We argue here that 
use of FWHM introduces a bias, stretching the mass scale such that high masses are
overestimated and low masses are underestimated. Here we investigate
estimation of black hole masses in AGNs based on individual or ``single epoch''
observations, with a particular emphasis in comparing mass estimates
based on line dispersion and FWHM. We confirm the recent findings that, in
addition to luminosity and line width, a third parameter is required
to obtain accurate masses and that parameter seems to be Eddington
ratio. We present simplified empirical formulae for estimating
black hole masses from the \hb\,$\lambda4861$ and \civ\,$\lambda1549$
emission lines. While the AGN continuum luminosity at 5100\,\AA\ is usually used
to predict the \hb\ reverberation lag, we show that the luminosity 
of the \hb\ broad component can be used instead without any loss of
precision, thus eliminating the difficulty of accurately accounting
for the host-galaxy contribution to the observed luminosity.
\end{abstract}

\newpage

\section{Introduction}
\label{section:intro}

\subsection{Reverberation-Based Black Hole Masses}

The presence of emission lines with Doppler widths of thousands of
kilometers per second is one of the defining characteristics of active
galactic nuclei \citep{Burbidge67,Weedman76}. It was long suspected
that the large line widths were due to motions in a deep gravitational
potential and this implied very large central masses
\citep[e.g.,][]{Woltjer59}, as did the Eddington limit
\citep{Tarter73}. Under a few assumptions, the central mass is $M
\propto V^2 R$, where $V$ is the Doppler width of the line and $R$ is
the size of the broad-line region (BLR). It is the latter quantity
that is difficult to determine. An early attempt
to estimate $R$ by \cite{Dibai80} was
based on the assumption of constant emissivity per unit
volume, but led to an incorrect dependence on luminosity as
in this case, luminosity is proportional to volume, so $R \propto L^{1/3}$.
\cite{Wandel85} inferred the BLR size from the
\hb\ luminosity. Other attempts were based on photoionization physics
\citep[see][]{Ferland85,Osterbrock85}.
\cite{Davidson72} found that the relative strength of emission lines
in ionized gas could be characterized by an ionization parameter
\begin{equation}
U= \frac{Q({\rm H})}{4\pi R^2 c n_{\rm H}},
\label{eq:Udef}
\end{equation}
where $Q({\rm H})$ is the rate at which H-ionizing photons are emitted
by the central source and $n_{\rm H}$ is the particle density of the
gas. The ionization parameter $U$ is proportional to the ratio of
ionization rate to recombination rate in the BLR clouds. The similarity of
emission-line flux ratios in AGN
spectra over orders of magnitude in luminosity suggested that $U$ is
constant, and the presence of C\,{\sc iii}]\,$\lambda1909$ sets an upper
limit on the density $n_{\rm H} \la 10^{9.5}$\,\cc\ \citep{Davidson79}.
Since $L \propto Q({\rm H})$, this naturally
led to the prediction that the BLR radius would scale with luminosity as
$R \propto L^{1/2}$. Unfortunately, best-estimate values for $Q({\rm
H})$ and $n_{\rm H}$ led to a significant overestimate of the BLR
radius \citep{Peterson85} as a consequence of the
simple but erroneous
assumption that all the broad lines arise cospatially (i.e.,
models employed a single representative BLR cloud).

With the advent of reverberation mapping
\citep[hereafter RM;][]{Blandford82,Peterson93}, direct measurements of $R$ enabled
improved black hole mass determinations. Attempts to estimate
black hole masses based on early RM results and the
$R \propto L^{1/2}$ prediction included those of \cite{Padovani88},
\cite{Koratkar91}, and \cite{Laor98}.  The first multiwavelength
RM campaigns demonstrated ionization stratification
of the BLR \citep{Clavel91,Krolik91,Peterson91} and this eventually led to
identification of the virial relationship, $R \propto V^{-2}$
\citep{PetersonWandel99,PetersonWandel00,
Onken02,Kollatschny03,Bentz10}, that gave
reverberation-based mass measurements higher levels of credibility.
Of course, the virial relationship demonstrates only that the
central force has a $R^{-2}$ dependence, which is also characteristic of
radiation pressure; whether or not radiation pressure from the
continuum source is important has not been clearly established
\citep{Marconi08,Marconi09,Netzer10}. If radiation pressure 
in the BLR turns out
to be important, then the black hole masses, as we discuss them here, 
are underestimated.

Masses of AGN black holes are computed as
\begin{equation}
\label{eq:masseqn}
M_{\rm BH} = f \left(\frac{V^2 R}{G}\right),
\end{equation}
where $V$ is the line width,
$R$ is the size of the BLR from the reverberation lag,
and $G$ is the gravitational constant.
The quantity in parentheses is often referred to as the
virial product $\mu$; it incorporates the two observables in RM,
line width and time delay $\tau = R/c$, and is in units of mass.
The scaling factor $f$
is a dimensionless quantity of order unity that
depends on the geometry, kinematics, and inclination of the AGN.
Throughout most of this work, we ignore $f$ (i.e., set it to unity) and work
strictly with the virial product. 

While reverberation mapping has emerged as the most effective
technique for measuring the black hole masses in AGNs
\citep{Peterson14}, it is resource intensive, requiring many
observations over an extended period of time at fairly high
cadence. Fortunately, observational confirmation of the $R$--$L$
relationship \citep{Kaspi00,Kaspi05,Bentz06a,Bentz09a,Bentz13} enables
``single-epoch'' (SE) mass estimates because, in principle, a single
spectrum could yield $V$ and also 
$R$, through measurement of $L$ 
\citep[e.g.,][]{Wandel99,McLure02,Vestergaard02,Corbett03,Vestergaard04,
Kollmeier06,Vestergaard06,Fine08,Shen08a,Shen08b,Vestergaard08}.
Of the three strong emission lines generally used 
to estimate central black hole
masses, the $R$--$L$ relationship is only well-established
for \hb\,$\lambda4861$
\citep[][and references therein, but see the discussion in
\S\ref{section:hbpredictor}]{Bentz13}. Empirically
establishing the $R$--$L$ relationship for
\mgii\,$\lambda2798$  \citep{Clavel90,Clavel91,Reichert94,Metzroth06,
Cackett15,Shen16b,Lira18,Czerny19,Zajacek20,Homayouni20}
as well as for  \civ\,$\lambda1549$ \citep{CWG89,Clavel90,Clavel91,Reichert94,
Korista95,Rod97,Wanders97,Obrien98,
Peterson05,Metzroth06,Kaspi07,
Trevese14,DeRosa15,Lira18,Grier19,Hoormann19}
has been difficult because of the 
nature of the UV line variability and the
high level of competition for suitable facilities.

Masses based on the \civ\,$\lambda1549$ emission line, in particular,
have been somewhat
controversial. Some studies claim that there is good agreement
between masses based on \civ\ and those measured from other lines
\citep{Vestergaard06,Greene10b,Assef11}. On the other hand, there are
several claims that there is inadequate agreement with masses based
on other emission lines
\citep{Baskin05,Netzer07,Sulentic07,Shen08b,Shen12,Trak12}. \cite{Denney09a}
and \cite{Denney13}, however, note that there are a number of biases
that can adversely affect single-epoch mass estimates, with low
$S/N$ ``survey quality'' data being an important problem with some
of the studies for which poor agreement between \civ\ and other
lines is found. It has also been argued, however, that some fitting
methodologies are more affected by this than others
\citep{Shen19}.
There have been more recent papers that attempt to
correct \civ\ mass determinations to better agree with those based on
other lines 
\citep[e.g.,][]{Bian12,Runnoe13a,Brotherton15a,Coatman17,Mejia18,Marziani19}.

\subsection{Characterizing Line Widths}

As first shown by \cite{Denney09a} and \cite{Denney12}
the apparent difficulties with \civ-based
masses trace back not only to the $S/N$ issue, but also
to how the line widths are characterized.  It has been
customary in AGN studies to characterize line widths by one of two
parameters, either FWHM or the line dispersion \sigl, which is defined by
\begin{equation}
\sigl = \left[ \frac{ \int (\lambda - \lambda_0)^2 P(\lambda)\,d\lambda}{\int P(\lambda)\,d\lambda} \right]^{1/2},
\label{eq:Defsigl}
\end{equation}
where $P(\lambda)$ is the emission-line
profile as a function of wavelength and $\lambda_0$ is the line centroid,
\begin{equation}
\lambda_0 = \frac{\int \lambda P(\lambda)\,d\lambda}{\int P(\lambda)\,d\lambda}.
\label{eq:Deflambda0}
\end{equation}
While both FWHM and \sigl\ have
been used in the virial equation to estimate AGN
black hole masses, they are not interchangeable. It is well known that
AGN line profiles depend on the line width \citep{Joly85}: broader
lines have lower kurtosis, i.e., they are ``boxier'' rather than
``peakier.''  Indeed, for AGNs, the ratio ${\rm FWHM}/\sigl$
has been found to be a simple but useful characterization of the line
profile \citep{Collin06,Kollatschny13}.

Each line-width measure has practical strengths and weaknesses
\citep{Peterson04,Wang20}.
The line dispersion
\sigl\ is more physically intuitive, but  it is
sensitive to the line wings, which are
often badly blended with other features. All
three of the strong lines usually used to estimate masses ---
\hb\,$\lambda4861$, \mgii\,$\lambda2798$, and
\civ\,$\lambda1549$ --- are blended with other features:
the \feii\,$\lambda4570$ and \feii\,$\lambda\lambda$5190, 5320 complexes
\citep{Phillips78} and \heii\,$\lambda4686$ in
the case of \hb, the UV \feii\ complexes in the case of \mgii, and
\heii$\,\lambda1640$ in the case of \civ. 
The red wing of the  \hb\ line is also 
blended with \oiii\,$\lambda\lambda4959$, 5007,
though because they do not vary on short timescales
these narrow lines disappear in the rms spectrum 
(defined below) and, on account of
their narrowness, can usually be removed from mean or single spectra
as we note below.
The FWHM can usually be
measured more precisely than \sigl\ (although \citealt{Peterson04}
note that the opposite is true for the rms spectra, which are
sometimes quite noisy), but it is not clear that FWHM yields more
{\em accurate} mass measurements. In practice,
FWHM is used more often than \sigl\ because
it is relatively simple to measure and can be measured more precisely,
while \sigl\ often requires deblending or modeling the emission features,
which does not necessarily yield unambiguous results.

There are, however,
a number of reasons to prefer \sigl\ to FWHM as the line-width
measure for estimating AGN black hole masses.
Certainly for radio-loud AGNs where inclination can be 
estimated from radio jets, core versus lobe dominance, or
radio spectral index, it is well-known that
FWHM correlates with inclination 
\citep{Wills86,Runnoe13b,Brotherton15b}.
\cite{Fromerth00} point out
that \sigl\ better characterizes an arbitrary or irregular
line profile.
\cite{Peterson04} note that
\sigl\ produces a tighter virial relationship than FWHM, and
\cite{Denney13} find
better agreement between \civ-based and \hb-based
mass estimates by using \sigl\ rather
than FWHM (these latter two are essentially the same argument).
In the case of NGC 5548, for which there are multiple reverberation-based
mass measures, a possible correlation with luminosity
is stronger for FWHM-based masses than for \sigl-based masses,
suggesting that the former are biased as the same mass should
be recovered regardless of the luminosity state of the AGN
\citep{Collin06,ShenKelly12}.
A possibly more compelling argument for using \sigl\ instead of FWHM
is bias in the mass scale that is
introduced by using FWHM as the line width. \cite{Steinhardt10} used
single-epoch masses for more than 60,000 SDSS quasars \citep{Shen08b} with
masses computed using FWHM. They found that, in any redshift bin,
if one plots the distribution of mass versus luminosity, the higher mass
objects lie increasingly below the Eddington limit; they refer to this
as the ``sub-Eddington boundary.'' There is no physical basis for this.
\cite{Rafiee11} point out, however,
that if the quasar masses are computed using \sigl\ instead of FWHM, the
sub-Eddington boundary disappears: the distribution of quasar black hole
masses approaches the Eddington limit at all masses. Referring to Figure~1
of \cite{Rafiee11}, the distribution of quasars in the mass vs.\ luminosity
diagram is an enlongated cloud of points whose axis is roughly parallel to the
Eddington ratio when \sigl\ is used to characterize the line width. However,
when FWHM is used, the axis of the distribution rotates as the higher masses
are underestimated and the lower masses are overestimated. However,
the apparent rotation of the mass distribution is in the same
sense that is expected from the Malmquist bias and a bottom heavy
quasar mass function \citep{Shen13}.
Unfortunately, these arguments are not statistically compelling.
Examination of the \mbh--$\sigma{*}$ relation using FWHM-based
and \sigl-based masses is equally unrevealing \citep{Wang19}.

In reverberation mapping, a further distinction among line-width measures
must be drawn since either FWHM or \sigl\ can be measured in the mean
spectrum,
\begin{equation}
\overline{F}(\lambda) =\frac{1}{N} \sum_{1}^{N} F_i(\lambda),
\label{eq:meanspec}
\end{equation}
where $F_i(\lambda)$ is the flux in the $i^{th}$ spectrum
of the time series at
wavelength $\lambda$ and $N$ is the number of spectra,
or they can be measured in the
rms residual spectrum
(hereafter simply ``rms spectrum''), which is defined as
\begin{equation}
\sigma_{\rm rms}(\lambda) = \left\{ \frac{1}{N-1}
\sum_{1}^{N}\left[ F_i(\lambda) -
\overline{F}(\lambda)\right]^2 \right\}^{1/2}.
\label{eq:rmsspec}
\end{equation}
In this paper, we will specifically refer to the
measurements of \sigl\ in the mean spectrum as  \sigm\ and in the
rms spectrum as \sigr. Similarly, \fwm\ refers to
FWHM of a line in the mean spectrum or a single-epoch spectrum and
\fwr\ is the FWHM in the rms spectrum.
It is common to use
\sigr\ as the line-width measure for determining black hole
masses from reverberation data --- it is
intuitatively a good choice as it isolates
the gas in the BLR that is actually responding to the continuum
variations. As noted previously, the strong and strongly
variable broad emission lines can be hard to isolate as they are
blended with other features. In the rms spectra, however, the
contaminating features are much less of a problem because they
are generally constant or vary either slowly or weakly and
thus nearly disappear in the rms spectra.

Since the goal is to
measure a black hole mass from a single (or a few) spectra, we must
use a proxy for \sigr. Here we will attempt to determine
if either \sigm\ or \fwm\ in a single or mean spectrum
can serve as a suitable proxy for \sigr; we know {\em a priori} that
there are good, but non-linear, correlations
between \sigr\ and both
\sigm\ and \fwm. It therefore seems likely that
either \sigm\ or \fwm\ could be used as
a proxy for \sigr.

Investigation of the relationship among the line-width measures motivated a
broader effort to produce easy-to-use prescriptions for computing
{\em accurate} black hole masses using \hb\ and \civ\ emission
lines and nearby continuum fluxes measurements for each line.
We do not discuss \mgii\ RM results
in this contribution as the present situation
has been addressed rather thoroughly by \cite{Bahk19},
\cite{Martinez20}, 
and \cite{Homayouni20}.
In \S{2}, the data used in this investigation are described.
In \S{3}, the relationship between the \hb\ reverberation lag
and different measures of the AGN luminosity are considered, and
we identify the physical parameters to lead to accurate black-hole
mass determinations.
In \S{4}, we will similarly discuss masses based on  \civ.  
In \S{5}, we
present simple empirical formulae for estimating black hole masses
from \hb\ or \civ;
we regard this as the most important result of this study.
The results of this investigation and our
future plans to improve this method are outlined in \S{6}.
Our results are briefly summarized in \S{7}.
Throughout this work, we assume $H_0 = 72$\,\kms\,Mpc$^{-1}$,
$\Omega_{\rm matter} =0.3$ and $\Omega_\Lambda = 0.7$.

\section{Observational Database and Methodology}

\subsection{Data}
\label{section:Data}

We use two high-quality databases for this investigation:
\begin{enumerate}

\item Spectra and measurements for previously reverberation-mapped AGNs,
for \hb\ (Table~A1) and
for \civ\ (Table~A2).
These are mostly taken from the literature
(see also \citealt{Bentz15} for a compilation\footnote{The database is
regularly updated at http://www.astro.gsu.edu/AGNmass}). Sources
without estimates of host-galaxy contamination to the optical luminosity
$L(5100\,{\rm \AA})$ have been excluded. This database provides
the fundamental $R$--$L$ calibration for the single-epoch mass scale.
In this contribution, we will refer to these collectively as the
``reverberation-mapping database (RMDB)''.

\item Spectral measurements from the Sloan Digital Sky Survey Reverberation
Mapping Program \citep[][hereafter ``SDSS-RM'' or more compactly
simply as ``SDSS'']{Shen15}.
We use both \hb\ (Table~A3)
and \civ\ (Table~A4) data from the 2014--2018 SDSS-RM
campaign \citep{Grier17b,Shen19,Grier19}.
Each spectrum is comprised of the average of the individual
spectra obtained for each of the 849 quasars in the SDSS-RM field.

\end{enumerate}

In addition, because \civ\ RM measurements remain rather scarce,
we augmented the \civ\ sample with measurements from
\cite{Vestergaard06} (hereafter VP06), who combined single-epoch
luminosity and line-width measurements from archival
UV spectra with \hb-based mass measurements
of the objects in Table~A1. The UV parameters are given
in Table~A5; we note, however, that we have
excluded 3C\,273 and 3C\,390.3 because they both have uncertainities
in their virial product larger than 0.5\,dex; the former was
a particular problem because there were far more measurements of
UV parameters for this source than for any other and the combination
of a large number of measurements and a poorly constrained virial
product conspired to disguise real correlations.

All SDSS-RM spectra have been reduced and processed as described by
\cite{Shen15} and \cite{Shen16b},
including post-processing with {\tt PrepSpec} (Horne,
in preparation). We note that only lags ($\tau$),
line dispersion in the rms spectrum ($\sigr$), and virial products 
($\murm = \sigr^2 c \tau/G$) are
taken from \cite{Grier17b} and \cite{Grier19};
all luminosities and other line-width measures are from
\cite{Shen19} (Tables~A3 and A4 are included here for the sake of clarity).

For each SDSS AGN, there are two determinations of
both \fwm\ and \sigm; one is the best-fit (BF) to the mean spectrum, and the
other is the mean of multiple Monte Carlo (MC) realizations. For each
MC realization, $N$ independent random selections of the $N$ spectra
are combined and the line width is measured for both \fwm\ and
\sigm. After a large number of realizations, the mean $\langle V
\rangle$ and rms $\Delta V$, for $V = \fwm$ and $V = \sigm$ are
computed, and the rms values are adopted as the uncertainties in each
line-width measure.

For the purpose of mass estimation, we need to establish relationships
based on the most reliable data.
Many of the SDSS average spectra are still quite noisy,
so we imposed quality cuts. Even though we are for the most
part restricting our attention to the SDSS-RM quasars for which
there are measured lags for \hb\ (44 quasars) or
\civ\ (48 quasars), we impose these cuts on the entire sample
for the sake of later discussion.
The first quality condition is that
\begin{equation}
\label{eq:minwidth}
V \geq 1000\,\kms
\end{equation}
for both $V = \fwm$ and $V =\sigm$,
since AGNs with lines narrower than 1000\,\kms\  are probably Type 2 AGNs;
there are some Type 1 AGNs with line widths narrower than this,
including several in Table A1,
but these are low-luminosity AGNs \cite[e.g.,][]{Greene07},
not SDSS quasars.
The second quality condition is
that the best fit value $V({\rm BF})$ must lie in the range
\begin{equation}
\label{eq:consistency}
\langle V \rangle - \Delta V \leq V({\rm BF}) \leq \langle V \rangle + \Delta V
\end{equation}
for both FWHM and \sigl. A third quality condition is
a ``signal-to-noise'' ($S/N$) requirement
that the line width must be significantly larger than
its uncertainty. Some experimentation showed that 
\begin{equation}
\label{eq:s2n}
\frac{V}{\Delta V} \geq 10
\end{equation}
is a good criterion 
for both $V = \fwm$ and $V = \sigm$
to remove the worst outliers from the line-width comparisons
discussed in \S{\ref{section:hblinewidths}}
and \S{\ref{section:civfundamental}}.

Finally, we removed quasars that were flagged by \cite{Shen19}
as having broad absorption lines (BALs),
mini-BALs, or suspected BALs in \civ. 

The effect of each quality
cut on the size of the database available for each emission line
is shown in Table~1. 
Of the 44 SDSS-RM quasars with measured \hb\ lags,
12 failed to meet at least one of the quality criteria,
usually the $S/N$ requirement, thus reducing the SDSS-RN \hb\ sample to 
32 quasars. Three quasars with
\civ\ reverberation measurements (RMID 362, 408, and 722) were
rejected for significant BALs, thus reducing the SDSS-RM
\civ\ reverberation sample to 45 quasars. As we will show
in \S{\ref{section:massformulae}}, another effect of imposing
quality cuts is, not surprisingly, that it removes some
of the lower luminosity sources from the sample.

\begin{deluxetable}{lcc}
\tablewidth{0pt}
\tablecaption{Effects of Quality Cuts on SDSS-RM Sample Size}
\tablehead{
\colhead{Criterion} &
\colhead{\hb}  &
\colhead{\civ}
}
\startdata
Original sample  & 221 & 540\\
(a) Minimum Line Width (eq.\ \ref{eq:minwidth})
& 199 & 520 \\
(b) Consistency (eq.\ \ref{eq:consistency}) & 194 & 368 \\
(c) $S/N$ (eq.\ \ref{eq:s2n}) & 121  & 462 \\
(a) + (b) & 174  & 352\\
(a) + (c) & 108  & 450\\
(b) + (c) & 107  & 309\\
(a) + (b) + (c) & 96 & 299 \\
All + BAL removal & 96 & 248
\enddata
\label{table:qcuts}
\end{deluxetable}


\subsection{Fitting Procedure}

Throughout this work, we use the
fitting algorithm described by \cite{Cappellari13}
that combines the Least Trimmed Squares technique of \cite{Rousseeuw06}
and a least-squares fitting algorithm which allows errors in all
variables and includes intrinsic scatter, as implemented by
\cite{DallaBonta18}.
Briefly, the fits we perform here are of the general form
\begin{equation}
\label{eq:powerlaw}
y = a + b\left(x - x_0 \right),
\end{equation}
where $x_0$ is the median value of the observed parameter $x$. The fit
is done iteratively with $5 \sigma$ rejection (unless stated otherwise)
and the best fit minimizes the quantity
\begin{equation}
\label{eq:chi2line}
	\chi^2=\sum_{i=1}^N \frac{[a+b (x_i-x_0) - y_i]^2}
	{(b \Delta x_i)^2 + (\Delta y_i)^2 + \varepsilon_y^2},
\end{equation}
where $\Delta x_i$ and $\Delta y_i$ are the errors on the variables
$x_i$ and $y_i$, and $\varepsilon_y$ is the sigma of the Gaussian
describing the distribution of intrinsic scatter in the $y$
coordinate; $\varepsilon_y$ is iteratively adjusted so that the
$\chi^2$ per degree of freedom $\nu=N-2$ has the value of unity
expected for a good fit. The observed scatter is
\begin{equation}
\label{eq:Deltadef}
\Delta = \left\{ \frac{1}{N-2} \sum_{i=1}^N
\left[y_i - a - b\left(x_i - x_0\right) \right]^2 \right\}^{1/2}.
\end{equation}
The value of $\varepsilon_y$ is added in quadrature when $y$ 
is used as a proxy for $x$.

The bivariate fits are intended to establish the physical relationships
among the various parameters and also to fit residuals.
The actual mass estimation equations that we
use will be based on multivariate fits of the general form
\begin{equation}
\label{eq:twoparameters}
z = a + b\left( x - x_0 \right) +c\left(y - y_0\right),
\end{equation}
where the parameters are as described above, plus an additional
observed parameter $y$ that has median value $y_0$.
Similarly to linear fits, the plane fitting minimizes the quantity
\begin{equation}
\label{eq:chi2plane}
	\chi^2=\sum_{i=1}^N \frac{[a + b (x_i-x_0) + c (y_i-y_0) - z_i]^2}
	{(b \Delta x_i)^2 + (c \Delta y_i)^2 + (\Delta z_i)^2 + \varepsilon_z^2},
\end{equation}
with $\Delta x_i$, $\Delta y_i$ and $\Delta z_i$ as the errors on the
variables $(x_i,y_i,z_i)$, and $\varepsilon_z$ as the sigma of the
Gaussian describing the distribution of intrinsic scatter in the $z$
coordinate; $\varepsilon_z$ is iteratively adjusted so that the $\chi^2$ per degrees of freedom $\nu=N-3$ has the value of unity expected for a good fit.
The observed scatter is
\begin{equation}
\label{eq:twoparmDeltadef}
\Delta = \left\{ \frac{1}{N-3} \sum_{i=1}^N
\left[{z_i} - a - b\left(x_i - x_0\right)
-c\left(y_i - y_0\right)\right]^2 \right\}^{1/2}.
\end{equation}

\section{Masses Based on \hb}

\subsection{The $R$--$L$ Relationships}

\begin{figure}
\begin{centering}
\includegraphics[scale=0.8]{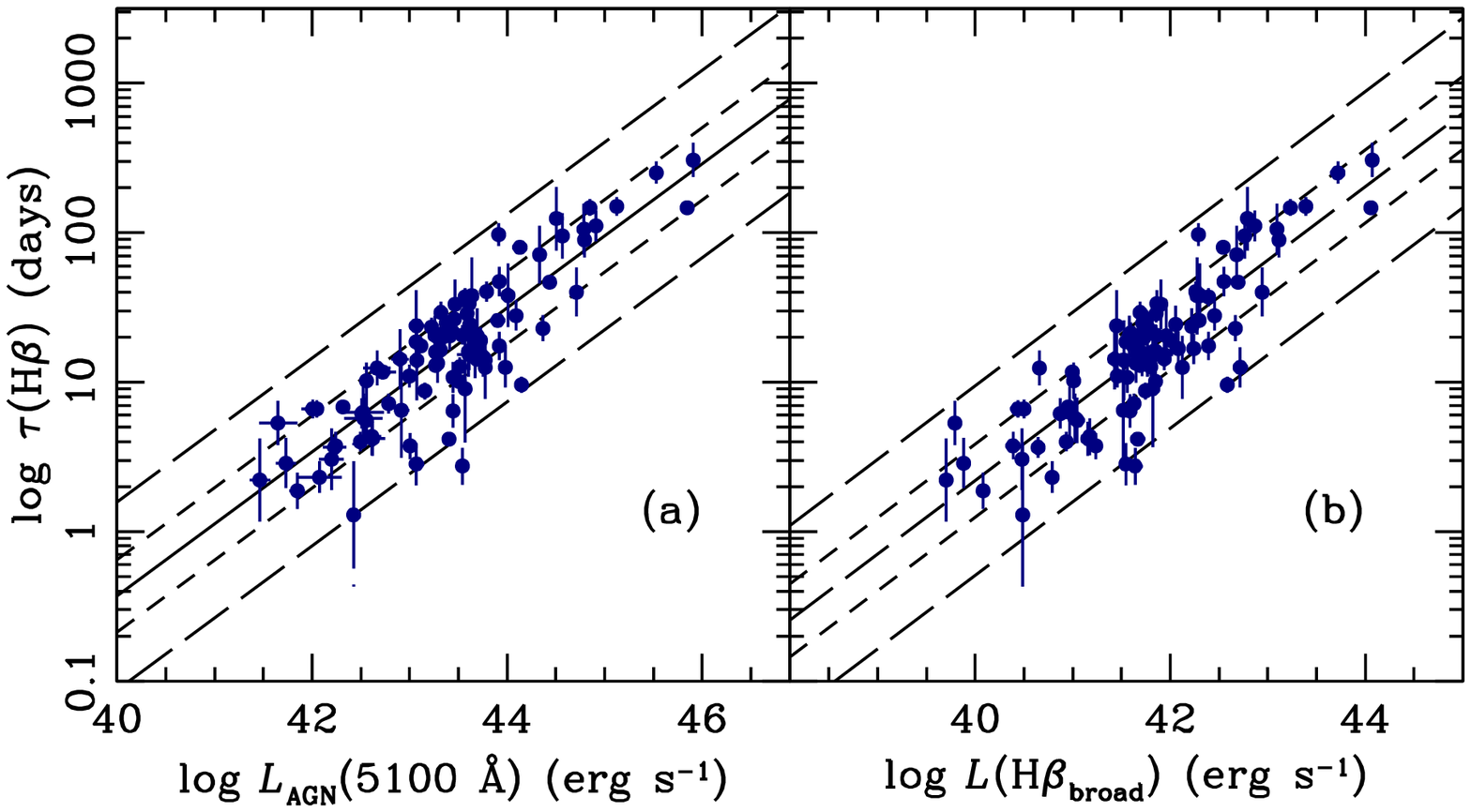}
\epsscale{0.20}
\caption{Panel (a):  The rest-frame \hb\ lag in days is
shown as a function of the AGN luminosity
$L_{\rm AGN}(5100\,{\rm \AA})$ in \ergsec.
The host-galaxy starlight contribution has been
removed by using unsaturated \hst\ images
\citep[see][]{Bentz13}. Panel (b):
The \hb\ lag in days is shown as a
function of the broad \hb\ luminosity $L(\hb_{\rm broad})$ in \ergsec.
The narrow component of \hb\ has been removed in each case
where it was sufficiently strong (i.e., easily identifiable) 
to isolate.
In both panels, the solid line shows the best-fit to the data
using equation (\ref{eq:powerlaw}) with coefficients given
in Table~2. The short dashed lines show the $\pm1\,\sigma$
uncertainty (equivalent to enclosing 68\% of the values for a Gaussian
distribution) and the long dashed lines show the $2.6\sigma$ uncertainties
(equivalent to enclosing 99\% of the values for a Gaussian distribution).
The Spearman rank correlation 
coefficient for the data in panel (a)
is $\rho = 0.797$ and the probability that the relationship arises by
chance is $P < 10^{-6}$, 
and for the data in panel (b), $\rho = 0.873$ 
with $P < 10^{-6}$. }
\label{Figure:HbRL}
\end{centering}
\end{figure}

In this section, we examine the calibration of the fundamental
\hb\ $R$--$L$ relationship using various luminosity measures.
The analysis in
this section is based only on the RMDB sample in
Table~A1 
because all these sources have been corrected
for host-galaxy starlight.
To obtain accurate masses from \hb, contaminating starlight from the
host galaxy must be accounted for in the luminosity measurement,
or the mass will be
overestimated. For reverberation-mapped sources, this
has been done by modeling unsaturated images of the AGNs
obtained with the {\em Hubble Space Telescope}
\citep{Bentz06a,Bentz09a,Bentz13}.
The AGN contribution was removed from each image by modeling the images
as an extended host galaxy plus a central point source representing
the AGN. The starlight contribution to the reverberation-mapping spectra
is determined by using simulated aperture photometry of the AGN-free image.
In panel (a) of Figure~\ref{Figure:HbRL}, 
we show the \hb\ lag as a function of the
AGN continuum with the host contribution removed in each case.
This essentially reproduces the result of \cite{Bentz13} as
small differences are due solely to improvements in the quality and
quantity of the RM database [cf.\ Table~A1]; 
we give the best-fit values to equation (\ref{eq:powerlaw}) in the
first line of Table~2. 

\begin{deluxetable}{lllcccccl}
\tablewidth{0pt}
\tablecaption{Radius--Luminosity and 
Luminosity--Luminosity Relations\tablenotemark{1}}
\tablehead{
\colhead{Line} &
\colhead{$x$} &
\colhead{$y$} &
\colhead{$a \pm \Delta a$} &
\colhead{$b \pm \Delta b$}  &
\colhead{$x_0$} &
\colhead{$\varepsilon_y$} &
\colhead{$\Delta$} &
\colhead{Figure} \\
\colhead{(1)} &
\colhead{(2)} &
\colhead{(3)} &
\colhead{(4)} &
\colhead{(5)} &
\colhead{(6)} &
\colhead{(7)} &
\colhead{(8)} &
\colhead{(9)} 
}
\startdata
1 & $\log L_{\scriptsize\rm AGN}(5100\,{\rm \AA})$ & $\log \tau(\hb)$ &
$1.228 \pm 0.025$ & $0.482 \pm 0.029$ & $43.444$ & $0.213 \pm 0.021$ & $0.241$
& 1a \\
2 & $\log L(\hb_{\scriptsize\rm broad})$ & $\log \tau(\hb)$ &
$1.200\pm 0.025$ & $0.492 \pm 0.030$ & $41.746$ & $0.218 \pm 0.022$ & $0.244$
& 1b\\
3 & $\log L(1350\,{\rm \AA})$ & $\log \tau(\civ)$ &
$1.915 \pm 0.047$ & $0.517 \pm 0.036$ & $45.351$ & $0.336 \pm 0.041$ & $0.361$
& 7 \\
4 &$\log L_{\scriptsize\rm AGN}(5100\,{\rm \AA})$ & $\log L(\hb_{\rm broad})$ &
$41.797 \pm 0.017$ & $0.960 \pm 0.020$ & $43.444$ & $0.158 \pm 0.014$ &$0.171$ & 2 \\
5 &$\log L(\hb_{\rm broad})$ &$\log L_{\scriptsize\rm AGN}(5100\,{\rm \AA})$ & 
$43.396 \pm 0.018$ & $1.003 \pm 0.022$ & $41.746$ & $0.161 \pm 0.015$ & $0.174$
& 2 
\enddata
\tablenotetext{1}{Continuum luminosities, $L(5100\,{\rm \AA})$ and
$L(1350\,{\rm \AA})$, and  line
luminosities, $L(\hb)$ and $L(\civ)$, are in units of \ergsec.
Time delays, $\tau(\hb)$ and $\tau(\civ)$, are in days.}
\label{table:RL}
\end{deluxetable}

\begin{figure}
\begin{centering}
\includegraphics[scale=0.8]{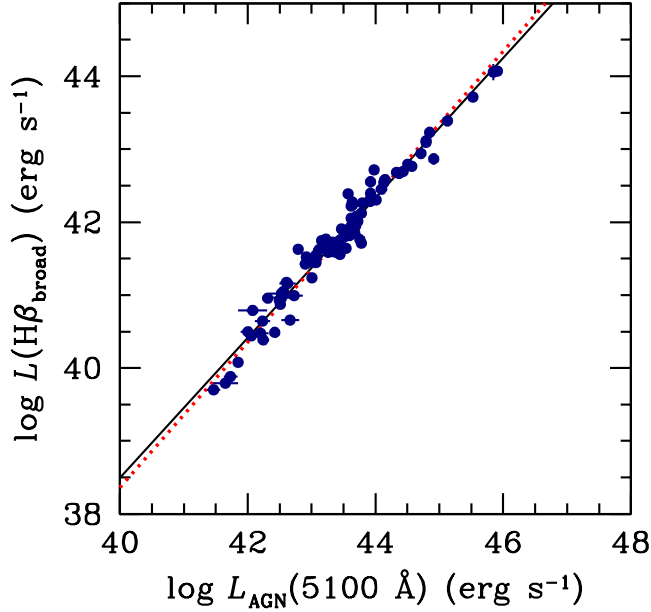}
\epsscale{0.20}
\caption{The relationship between the broad \hb\ emission line
luminosity and the starlight-corrected AGN luminosity
for the sources in Table A1. The black solid line is the 
regression of $L(\hb_{\rm broad})$ on $L_{\rm AGN}(5100\,{\rm \AA})$; 
the Spearman rank coefficient for this fit is
$\rho = 0.901$ with $P< 10^{-6}$.
The red
dotted line is the regression of 
$L_{\rm AGN}(5100\,{\rm \AA})$ on $L(\hb_{\rm broad})$,
which we use in equation (\ref{eq:LHbLAGN}); 
for this fit $\rho = 0.970$ and $P<10^{-6}$.
The coefficients for both fits are given in Table 2.}
\label{Figure:LAGNLHb}
\end{centering}
\end{figure}

Accounting for the host-galaxy contribution
in the same way for large number of AGNs, such as
those in SDSS-RM (not to mention the entire SDSS catalog), is simply not
feasible. It is well-known, however, that there is a tight correlation
between the AGN continuum luminosity and the luminosity of \hb\
\citep[e.g.,][]{Yee80,Ilic17}, and it has indeed been argued that the
\hb\ emission-line luminosity can be used as a proxy for the AGN 
continuum luminosity
for reverberation studies \citep{Kaspi05, Vestergaard06, Greene10a}.
However, 
in some of the reverberation-mapped sources, narrow-line \hb\
contributes significantly to the total \hb\ flux; NGC 4151 is an extreme
example \citep[e.g.,][]{Antonucci83,Bentz06a,Fausnaugh17}. Whenever
the narrow-line component can be isolated, it has been subtracted
from the total \hb\ flux. This also affects the line-width
measurements. In general, it is assumed that \oiii\,$\lambda5007$ can
be used as a template for narrow \hb. The template is shifted and scaled
to the largest flux that, when subtracted from the spectrum,
does not produce a depression at the
center of the remaining broad \hb\ component.
In Figure~\ref{Figure:LAGNLHb}, we show the tight relationship
between $L_{\rm AGN}(5100\,{\rm \AA})$ and $L(\hb_{\rm broad})$;
the best-fit coefficients for this relationship are given
in Table 2.

In panel (b) of Figure \ref{Figure:HbRL}, we show
the \hb\ lag
as a function of the luminosity of the broad component of \hb, with
the narrow component removed whenever possible.
We give the
best-fit values to the equation (\ref{eq:powerlaw}) in
the second row of Table~2, 
which shows that the
slope of this relationship is nearly identical to the slope of the
$R$--$L$ relationship using the AGN continuum.
The luminosity of the \hb\ broad component is thus an excellent
proxy for the AGN luminosity and requires only removal of the \hb\
narrow component (at least when it is significant) which is much
easier than estimating the starlight contribution to the continuum
luminosity at 5100\,\AA. Moreover, by using the line flux instead
of the continuum flux, we can include core-dominated radio
sources where the continuum may be enhanced by
the jet component \citep{Greene05}.
This is therefore the $R$--$L$ relationship we prefer
for the purpose of estimating single-epoch masses and we will
focus on this relationship through the remainder of this contribution.

\subsection{Line-Width Relationships}
\label{section:hblinewidths}

\begin{figure}
\begin{centering}
\includegraphics[scale=0.8]{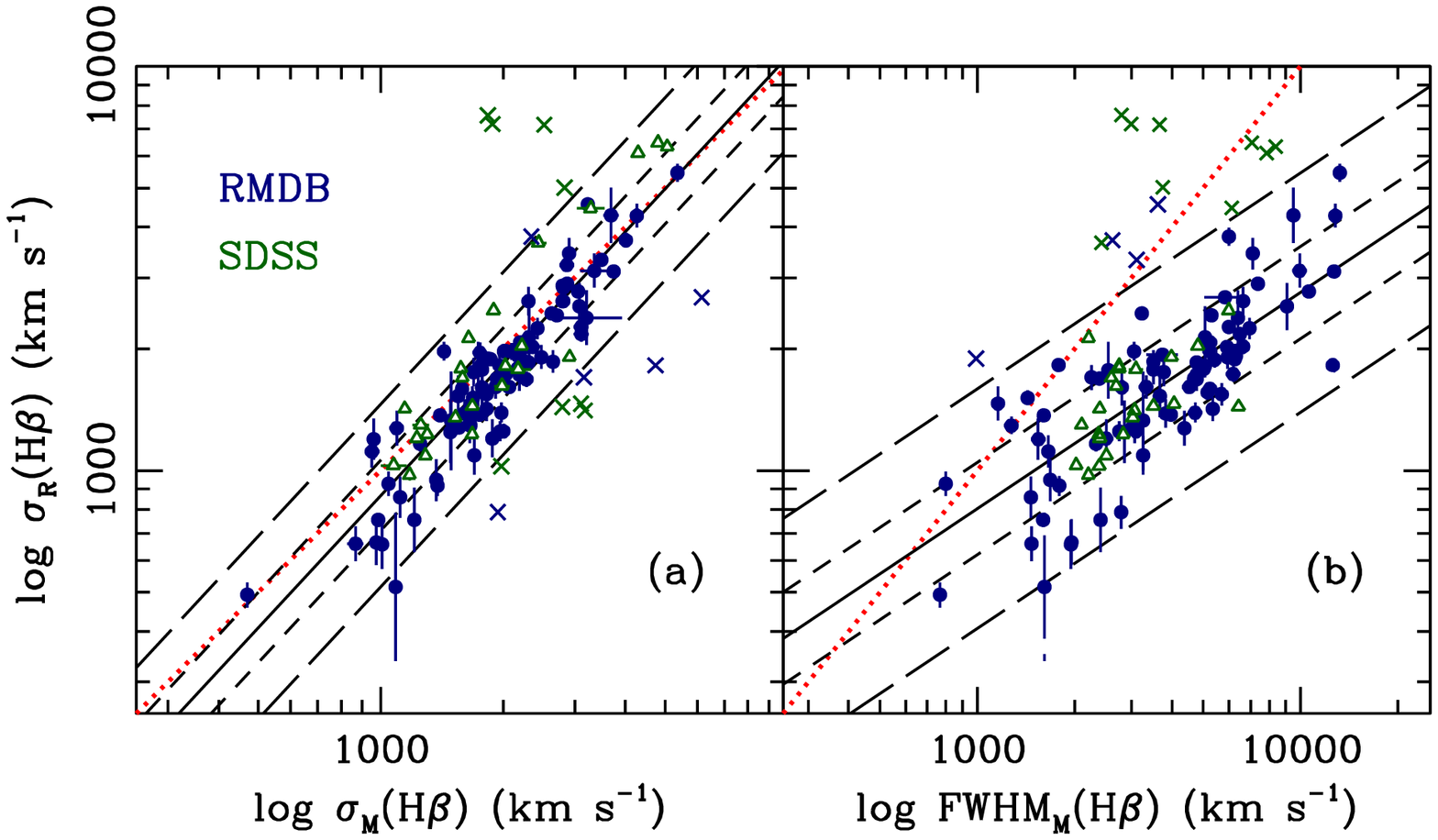}
\caption{The relationship between \hb\ line dispersion
in the rms $\sigr(\hb)$ and mean $\sigm(\hb)$ spectra
is shown in panel (a). The relationship between
\hb\ line dispersion in the rms spectrum \sigr(\hb) and
FWHM in the mean spectrum \fwm(\hb)
is shown in panel (b).
Blue filled circles are for the RMDB sample (Table~A1) and
open green triangles are for the SDSS sample (Table~A3).
The solid lines are best fits to equation (\ref{eq:powerlaw})
with coefficients in Table~3. The short dashed and long
dashed lines indicate the $\pm 1 \sigma$ and
$\pm 2.6 \sigma$ envelopes, respectively, and the
red dotted lines indicate where the two line-width measures are equal.
Crosses are points that were rejected
at the 2.6$\sigma$ (99\%) level and are colored-coded like the circles.
The relationship in panel (a)  is nearly linear (slope $= 1.085 \pm 0.045$)
and the scatter $\varepsilon_y$ is low (0.079\,dex).
The Spearman rank correlation coefficient for these data
is $\rho = 0.901$ and the probability of the correlation arising by
chance is $P < 10^{-6}.$
It is clear in panel (b) that \fwm(\hb) and
\sigr(\hb) are well-correlated, but the relationship
is significantly non-linear (slope $= 0.535 \pm 0.042$),
the scatter $\varepsilon_y$ is slightly larger (0.106\,dex), and there
are several significant outliers.
For these data, $\rho =0.786$ and $P<10^{-6}.$}
\label{Figure:windowhbwidths}
\end{centering}
\end{figure}

We now consider the use of \sigm\ and \fwm\ as proxies for \sigr\
\citep[cf.][]{Collin06,Wang19}.
Panel (a) of Figure~\ref{Figure:windowhbwidths} shows the relationship between
$\sigr(\hb)$, the \hb\ line dispersion in the rms spectrum,
and $\sigm(\hb),$ the \hb\ line dispersion in the
mean spectrum. The relationship is nearly 
linear (slope\ $ = 1.085\pm0.045$) and the
intrinsic scatter is small ($0.079$\,dex). The fit coefficients are given
in the first line of Table~3. 

\begin{deluxetable}{lllcccccl}
\tablewidth{0pt}
\tablecaption{Line-Width Relations\tablenotemark{1}}
\tablehead{
\colhead{Line} &
\colhead{$x$} &
\colhead{$y$} &
\colhead{$a \pm \Delta a$} &
\colhead{$b \pm \Delta b$}  &
\colhead{$x_0$} &
\colhead{$\varepsilon_y$} &
\colhead{$\Delta$} &
\colhead{Figure}\\
\colhead{(1)} &
\colhead{(2)} &
\colhead{(3)} &
\colhead{(4)} &
\colhead{(5)} &
\colhead{(6)} &
\colhead{(7)} &
\colhead{(8)} &
\colhead{(9)}
}
\startdata
1 &$\log \sigm(\hb)$ & $\log \sigr(\hb)$ &
$3.260 \pm 0.008$ & $1.085 \pm 0.045$ & 3.297 & $0.079 \pm 0.006$ & 0.087 &
3a \\
2 & $\log \fwm(\hb)$ & $\log \sigr(\hb) $ &
$3.205 \pm 0.011$ & $0.535 \pm 0.042$ & 3.559 & $0.106 \pm 0.001$ & 0.114 &
3b\\
3 & $\log \sigm(\civ)$ & $\log \sigr(\civ)$ &
$3.436 \pm 0.009$ &$0.822 \pm 0.059$ & 3.394 & $0.064 \pm 0.008$ & 0.067 &
8a\\
4 & $\log \fwm(\civ)$ & $\log \sigr(\civ)$ &
$3.447 \pm 0.016$ & $0.445 \pm 0.101$ & 3.580 & $0.121 \pm 0.014$ & 0.121 &
8b
\enddata
\tablenotetext{1}{All line widths are in \kms\ in the rest-frame of the AGN.}
\label{table:LW}
\end{deluxetable}

We also show in panel (a) of  
Figure~\ref{Figure:windowhbwidths} the relationship between
\sigr(\hb) and the FWHM of \hb\ in the mean spectrum, \fwm(\hb).
The fit coefficients are given in the second line
of Table~3. 
The relationship is far from linear (slope\ $= 0.535 \pm 0.042$),
and the scatter $\varepsilon_y$ is larger than it is for
the \sigr(\hb)--\sigm(\hb) relationship, even after
removal of the notable outliers.
The shallow slope of the relationship between \fwm\ and \sigr\
is why the mass distribution is stretched by using \fwm\ as the
line-width measure in equation (\ref{eq:masseqn}): for any given $R$,
the ratio $(\fwm/\sigr)^2$ is larger at the high-mass end of the
distribution than it is at the low-mass end. Use of \fwm\ in 
equation (\ref{eq:masseqn}) overestimates the high masses and underestimates
the low masses.
While it is clear that \sigm(\hb)  is an
excellent proxy for  \sigr(\hb), the value of
\fwm(\hb) is less clear, though the shallow slope
of the \fwm--\sigr\ relationship needs to be taken into account. 
We will fit
both versions in order to understand the relative
merits of each.

\subsection{Single-Epoch Predictors of the Virial Product}
\label{section:hbpredictor}

In the previous subsections, we have re-established the
correlations between $\tau(\hb)$ and $L(\hb_{\rm broad})$
and between \sigr(\hb) and both \sigm(\hb) and \fwm(\hb).
As a first approximation for a formula to estimate
single-epoch masses, we fit the following equations:
\begin{equation}
\log \murm(\hb) = a + b\left[\log L(\hb_{\rm broad}) - x_0\right]
+c\left[\log\sigm(\hb) - y_0\right],
\label{eq:Fit_139}
\end{equation}
and
\begin{equation}
\log \murm(\hb) = a + b\left[ \log L(\hb_{\rm broad}) - x_0\right]
+c\left[ \log \fwm(\hb) - y_0\right].
\label{eq:Fit_141}
\end{equation}

\begin{figure}
\begin{centering}
\includegraphics[scale=0.8]{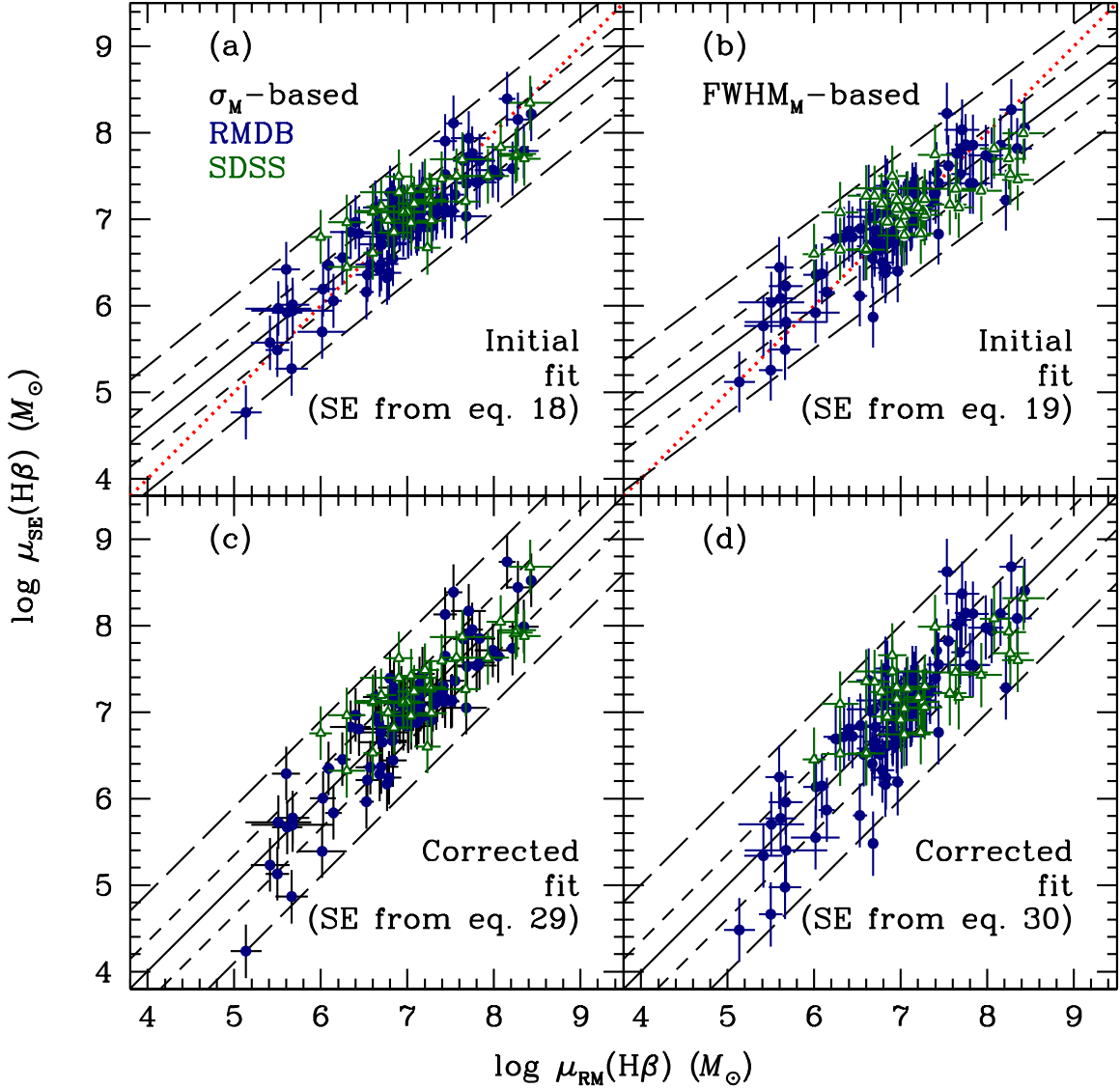}
\caption{Single-epoch \hb-based virial product predictions
using equations
(\ref{eq:SEsigm})
and (\ref{eq:SEfwm}) in panels (a) and (b), respectively,
compared with the
actual RM measurements for the same sources.
The coefficients and their uncertainties for these
initial predictors of $\log \muse(\hb)$ 
are presented in the first two lines of Table~4.
Blue filled circles represent RMDB data (Table~A1) and
green open triangles represent SDSS data (Table~A3). The solid line
shows the best fit to the data, and the red dotted line
shows where the two values are equal.
Coefficients for fits to the $\log \muse(\hb)$--$\log \murm(\hb)$
relationship are given in 
the first two lines of Table 5.
The short and
long dashed lines show the $\pm 1 \sigma$
and $\pm 2.6 \sigma$ envelopes, respectively.
It is clear that
this is an inadequate virial product predictor as
it systematically underestimates higher masses and overestimates
lower masses.
The panels (c) and (d) show the same relationship
after the empirical corrections as embodied in equations
(\ref{eq:SEsigmcorr}) and (\ref{eq:SEfwmcorr})
for \sigm\ and \fwm, respectively. In panels (c) and (d),
the best fit lines cover the
equality lines; results of these fits are given in lines
8 and 9 of Table 5. The intrinsic errors $\varepsilon_y$ have been
added in quadrature to the measurement uncertainties in $\log \muse(\hb)$.}
\label{Figure:Fit_139_141}
\end{centering}
\end{figure}

The results of these fits based on the combined RMDB data (Table~A1)
and SDSS data (Table~A3) 
are given in the first two lines of Table~4, 
and illustrated in panels (a) and (b) of Figure~\ref{Figure:Fit_139_141}.
Using these coefficients, we have initial 
of predictors of $\log \muse(\hb)$
using $\sigm$ as the line-width measure,
\begin{equation}
\log \muse(\hb) = 
6.975  + 0.566\left[ \log L(\hb_{\rm broad}) - 41.857\right]
+ 1.757\left[ \log \sigm(\hb) - 3.293\right],
\label{eq:SEsigm}
\end{equation}
and using \fwm\ as the line-width measure,
\begin{equation}
\log \muse(\hb) = 
6.981 + 0.587\left[ \log L(\hb_{\rm broad}) - 41.857\right]
+ 1.039\left[ \log \fwm(\hb) - 3.599\right].
\label{eq:SEfwm}
\end{equation}

The luminosity coefficient $b$
and the line-width coefficient $c$ are roughly as expected from the
virial relationship and the $R$--$L$ relationship,
and we note that the line-width coefficient for \fwm\ ($ c = 1.039$)
is much smaller than that of \sigm\ ($c = 1.757$),
as expected from Figure~\ref{Figure:windowhbwidths}.
It is clear that both equations (\ref{eq:SEsigm}) and (\ref{eq:SEfwm})
overestimate masses at the low end and underestimate them at the high end,
thus biasing the prediction. 
Coefficients based on fits to the 
relationship between $\log \muse(\hb)$ and $\log \murm(\hb)$
are given in the top two rows of Table 5, and the fits are
shown in panels (a) and (b) of 
Figure~\ref{Figure:Fit_139_141}.  
In both cases, the slopes are too shallow.
The failure of equations (\ref{eq:SEsigm}) and (\ref{eq:SEfwm})
to correctly recover $\log \murm(\hb)$
suggests that another parameter is required
for the single-epoch virial product prediction.

We investigated the possible importance of another parameter by
plotting the residuals $\Delta \log \mu =
\log \murm - \log \muse$ against other parameters,
specifically luminosity, mass (virial product),
Eddington ratio,
emission-line lag, and both line width and line-width ratio
${\rm FWHM}/\sigl$ for both mean and rms spectra.
The most significant correlation between the virial product residuals
and other parameters was for Eddington ratio, which has been
a result of other recent investigations
\citep{Du16,Grier17b,Du18,Du19,Martinez19,Alvarez20}. To determine the
Eddington ratio, we start with the Eddington luminosity
\begin{equation}
\label{eq:Eddingtonlimit}
L_{\rm Edd} = \frac{4 \pi G c m_e M}{\sigma_e}
= 1.257 \times 10^{38} \left(\frac{M}{\msun}\right),
\end{equation}
where $m_e$ is the electron mass and
$\sigma_e$ is the Thomson cross-section. The black hole
mass is $\log M = \log f + \log \mu$ and, as explained in the Appendix, we assume
$\log f = 0.683 \pm 0.150$ \citep{Batiste17} so the Eddington luminosity is
\begin{equation}
\label{eq:Ledd}
\log L_{\rm Edd} = \log f + 38.099 + \log \murm = 38.782 + \log \murm.
\end{equation}
The bolometric luminosity can be obtained from the observed 5100\,\AA\ AGN
luminosity plus a bolometric correction
\begin{equation}
\label{eq:bolometriclum}
\log L_{\rm bol} = \log L_{\rm AGN}(5100\,{\rm \AA})+ \log k_{\rm bol} .
\end{equation}

\begin{longrotatetable}
\begin{deluxetable}{llllccccccc}
\tablewidth{0pt}
\tablecaption{Multivariate Fits\tablenotemark{1}}
\tablehead{
\colhead{Line} &
\colhead{$x$} &
\colhead{$y$} &
\colhead{$z$}&
\colhead{$a \pm \Delta a$} &
\colhead{$b \pm \Delta b$} &
\colhead{$c \pm \Delta c$} &
\colhead{$x_0$} &
\colhead{$y_0$} &
\colhead{$\varepsilon_z$} &
\colhead{$\Delta$} \\
\colhead{ } &
\colhead{(\ergsec)}&
\colhead{(\kms)} &
\colhead{(\msun)}&
\colhead{ } &
\colhead{ } &
\colhead{ } &
\colhead{(\ergsec)} &
\colhead{(\kms)}&
\colhead{ } &
\colhead{ } \\
\colhead{(1)} &
\colhead{(2)} &
\colhead{(3)} &
\colhead{(4)} &
\colhead{(5)} &
\colhead{(6)} &
\colhead{(7)} &
\colhead{(8)} &
\colhead{(9)} &
\colhead{(10)}&
\colhead{(11)}
}
\startdata
1 & $\log L(\hb_{\rm broad})$ &
$\log \sigm(\hb)$ &
$\log \murm(\hb)$&
$6.975 \pm 0.029$ &
$0.566 \pm 0.035$ &
$1.757 \pm 0.160$ &
$41.857$ & $3.293$ & $0.273 \pm 0.025$ & $0.314$ \\
2 & $\log L(\hb_{\rm broad})$ &
$\log \fwm(\hb)$ &
$\log \murm(\hb)$ &
$6.981 \pm 0.033$ &
$0.587 \pm 0.040$ &
$1.039\pm 0.128$ &
$41.857$ & $3.559$ & $0.323 \pm 0.028$ & $0.352$\\
3 & $\log L(1350\,{\rm \AA})$ &
$\log \sigm(\civ)$ &
$\log \murm(\civ)$ &
$7.664 \pm 0.039$ &
$0.599 \pm 0.033$ &
$1.014 \pm 0.265$ &
$44.706$ & $3.502$ & $0.364 \pm 0.033$ & $0.397$
\enddata
\tablenotetext{1}{All values of \murm are in solar masses.}
\label{table:mfits}
\end{deluxetable}
\end{longrotatetable}

\begin{deluxetable}{lllccccccl}
\tablewidth{0pt}
\tablecaption{Initial, Residual, and Final Fits}
\tablehead{
\colhead{Line} &
\colhead{Data Set} &
\colhead{$x$} &
\colhead{$y$} &
\colhead{$a \pm \Delta a$} &
\colhead{$b \pm \Delta b$} &
\colhead{$x_0$} &
\colhead{$\varepsilon_y$} &
\colhead{$\Delta$}&
\colhead{Figure}  \\
\colhead{(1)} &
\colhead{(2)} &
\colhead{(3)} &
\colhead{(4)} &
\colhead{(5)} &
\colhead{(6)} &
\colhead{(7)} &
\colhead{(8)} &
\colhead{(9)} & 
\colhead{(10)} 
}
\startdata
& \multicolumn{3}{l}{Initial:} \\
1 & \hb\ &
$\log \murm(\sigm)$ &
$\log \muse(\sigm)$ &
$7.025 \pm 0.025$ &
$0.805 \pm 0.038$ &
$7.041$ &
$0.249 \pm 0.021$ &
$0.279$ & 4a\\
2 & \hb\ &
$\log \murm(\fwm)$ &
$\log \muse(\fwm)$ &
$7.012 \pm 0.028$ &
$0.749 \pm 0.042$ &
$7.007$ &
$0.278 \pm 0.023$ &
$0.290$ & 4b\\
3 & \civ\ &
$\log \murm(\civ) $ &
$\log \muse(\civ) $ &
$7.483 \pm 0.033$ &
$0.787 \pm 0.041$ &
$7.481$ &
$0.321 \pm 0.028$ &
$0.347$ & 9a \\
& \multicolumn{3}{l}{Residual:}\\
4 & \hb\ &
$\log \dot{m}$ &
$\Delta \log \mu (\sigm)$ &
$-0.010 \pm 0.022$ &
$-0.422 \pm 0.045$ &
$-0.951$ &
$0.187 \pm 0.021$ &
$0.246$ & 5a \\
5 & \hb\ &
$\log \dot{m}$ &
$\Delta \log \mu (\fwm)$ &
$-0.007 \pm 0.023$ &
$-0.543 \pm 0.046$ &
$-0.951$ &
$0.191 \pm 0.021$ &
$0.247$ & 5b\\
6 & \civ\ &
$\log \dot{m}$ &
$\Delta \log \mu$ &
$-0.049 \pm 0.026$ &
$-0.557 \pm 0.048$ &
$-1.155$ &
$0.213 \pm 0.027$ &
$0.282$ & 10a \\
7 & \civ\ &
$\log \murm$ &
$\Delta \log \mu$ &
$-0.012 \pm 0.026$ &
$0.297 \pm 0.024$ &
$7.481$ &
$0.000 \pm 0.000$ &
$0.139$ & 10d\\
& \multicolumn{3}{l}{Final:} \\
8 & \hb\ &
$\log \murm(\sigm)$ &
$\log \muse(\sigm)$ &
$7.040 \pm 0.031 $ &
$0.999 \pm 0.047 $ &
$7.041$ &
$0.309 \pm 0.027$ &
$0.346$ & 4c\\
9 & \hb\ &
$\log \murm(\fwm)$ &
$\log \muse(\fwm)$ &
$7.007 \pm 0.037 $ &
$1.000 \pm 0.055 $ &
$7.007$ &
$0.371 \pm 0.030$ &
$0.387$ & 4d\\
10 & \civ\ &
$\log \murm(\civ) $ &
$\log \muse(\civ)$ &
$7.485 \pm 0.041$ &
$0.963 \pm 0.006$ &
$7.481$ &
$0.408 \pm 0.035$ &
$0.439$ & 9b
\enddata
\label{table:residfits}
\end{deluxetable}
We ignore inclination effects and, following \cite{Netzer19},
the bolometric correction we use is
\begin{equation}
\label{eq:bolometriccorrection}
\log k_{\rm bol} = 10 - 0.2\log L_{\rm AGN}(5100\,{\rm \AA}).
\end{equation}
Since we are using $L(\hb_{\rm broad})$
as a proxy for $L_{\rm AGN}(5100\,{\rm \AA})$, we substitute
$L(\hb_{\rm broad})$ for $L_{\rm AGN}(5100\,{\rm \AA})$ by
fitting the luminosities in Table~A1, yielding (see Table 2)
\begin{equation}
\log L_{\rm AGN}(5100\,{\rm \AA}) = 43.396 
+ 1.003\left[\log L(\hb_{\rm broad}) - 41.746\right],
\label{eq:LHbLAGN}
\end{equation}
so we can write the bolometric luminosity as 
\begin{equation}
\log L_{\rm bol} = 44.717 + 0.802\left[\log(\hb_{\rm broad}) - 41.746 \right].
\label{eq:boldef}
\end{equation}
The Eddington ratio \mdot\ is given by\footnote{Strictly speaking,
the Eddington ratio is defined as $\mdot = \dot{M}/\dot{M}_{\rm Edd}$.
Since $\dot{M} = L_{\rm bol}/\eta c^2$, 
$\mdot = L_{\rm bol}/L_{\rm Edd}$ as long as the efficiency $\eta$ is 
constant and not a function of the accretion rate, which we will
assume for simplicity.}
\begin{equation}
\log \mdot = \log L_{\rm bol} - \log L_{\rm Edd}.
\label{eq:Eddratio}
\end{equation}
Using equations (\ref{eq:boldef}) and (\ref{eq:Ledd}), the
Eddington ratio can then be written as
\begin{equation}
\log \dot{m} = 5.935 + 0.802 \left[ \log L(\hb_{\rm broad}) - 41.746\right]
- \log \murm.
\label{eq:Eddalt}
\end{equation}

\begin{figure}
\begin{centering}
\includegraphics[scale=0.8]{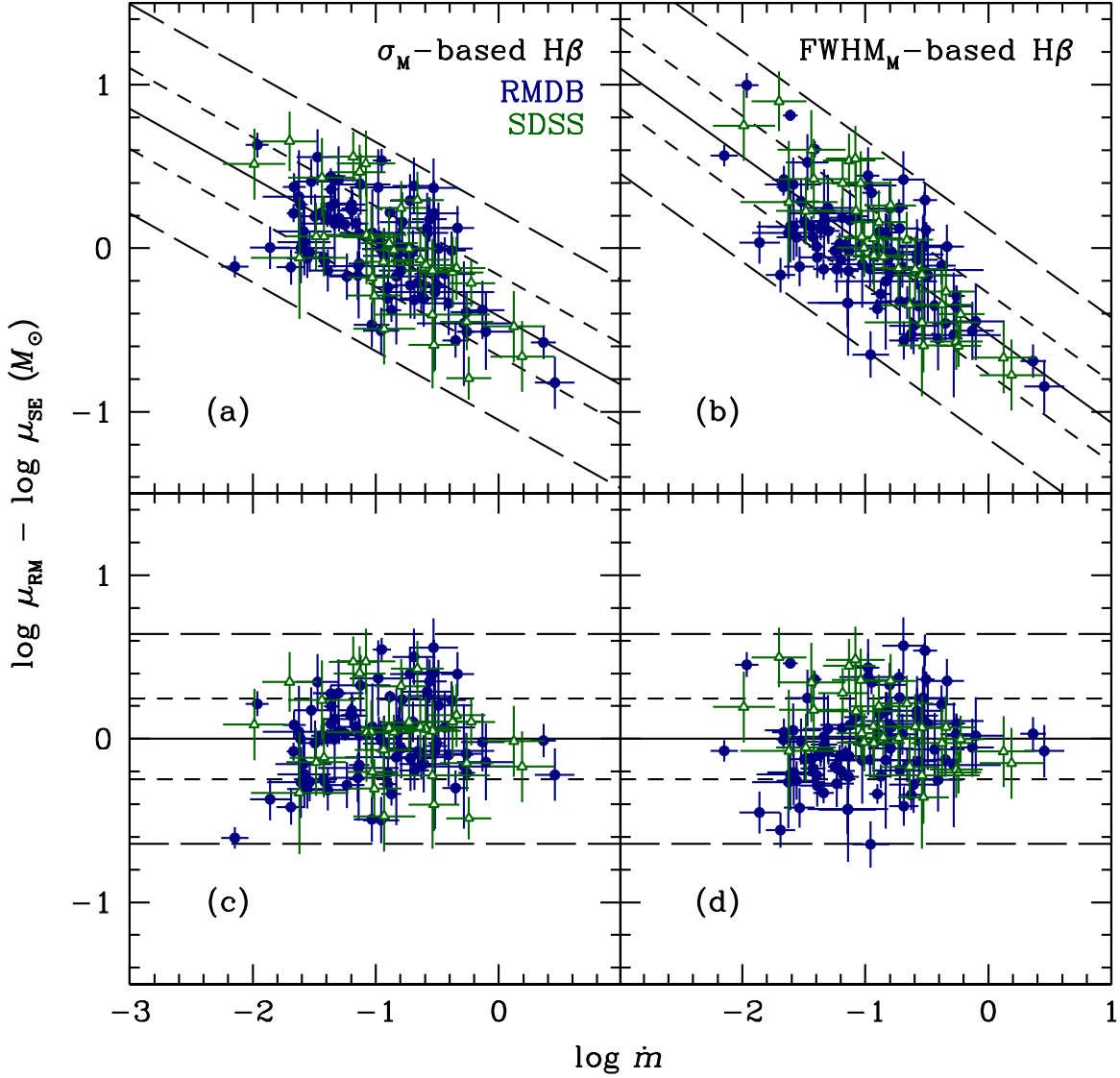}
\caption{Panel (a): The mass residuals (equation \ref{eq:residuals})
are the difference between the measured reverberation virial products
and those predicted by equation (\ref{eq:SEsigm}). The residuals are
plotted vs.\ Eddington ratio $\dot{m}$ (equation \ref{eq:Eddalt})
for single-epoch virial products based on \sigm(\hb).
The Spearman rank correlation coefficient is $\rho = -0.577$
with the probability that the correlation arises by chance
$P < 10^{-6}.$
Panel (b): The mass residuals (equation \ref{eq:residuals})
are the difference between the measured reverberation virial products
and those predicted by equation (\ref{eq:SEfwm}). The residuals are
plotted vs.\ Eddington ratio $\dot{m}$ 
for single-epoch virial products based on \fwm(\hb). 
For these data, $\rho = -0.679$ with $P < 10^{-6}.$
Panels (c) and (d) show residuals after subtraction of the best fit 
in panels (a) and (b), respectively. 
The $\varepsilon_y$ scatter in the residuals
is 0.197\,dex for the \sigm-based virial products and
0.204\,dex for the \fwm-based virial products.
In all panels, the solid blue circles represent RMDB data (Table~A1)
and the open green triangles represent SDSS data (Table~A3).
The solid line shows the best fit to the data. The short dashed and
long dashed lines are the $\pm 1 \sigma$ and $\pm 2.6 \sigma$
envelopes, respectively.
The coefficients of the fits are given in Table~5.
Error bars on the residuals are measurement uncertainties only, without
systematic errors.}
\label{Figure:windowhbresiduals}
\end{centering}
\end{figure}

To correct the single-epoch masses for Eddington ratio,
we fit the equation
\begin{equation}
\Delta \log \mu = \log \murm - \log \muse =
a + b(\log \dot{m} - x_0),
\label{eq:residuals}
\end{equation}
and use this as a correction to our initial fits, 
equations (\ref{eq:SEsigm}) and (\ref{eq:SEfwm}).
The best-fit parameters for comparison of the \sigm\ and \fwm-based
predictors of \muse\ with the reverberation measurements \murm\
are given in lines 4 and 5 of Table~5 and
shown in panels (a) and (b) of Figure~\ref{Figure:windowhbresiduals}.
Combining the correction equation (\ref{eq:residuals}) with
the best-fit coefficients in Table 5 and equations (\ref{eq:SEsigm})
and (\ref{eq:SEfwm}) yields the corrected single-epoch masses
\begin{equation}
\log \muse(\hb)  = 
6.965  + 0.566\left[ \log L(\hb_{\rm broad}) - 41.857\right]
+ 1.757\left[ \log \sigm(\hb) - 3.293\right]
- 0.422\left[ \log \mdot + 0.951 \right],
\label{eq:SEsigmcorr}
\end{equation}
and 
\begin{equation}
\log \muse(\hb)  = 
6.974 + 0.587\left[ \log L(\hb_{\rm broad}) - 41.857\right]
+ 1.039\left[ \log \fwm(\hb) - 3.599\right] 
- 0.543 \left[\log \mdot +0.951 \right],
\label{eq:SEfwmcorr}
\end{equation}
for \sigm\ and \fwm, respectively. 

Once the dependence on Eddington ratio is removed
(panels c and b of Figure~\ref{Figure:Fit_139_141}), the residuals do
not appear to correlate with other properties. 
We can now use equations (\ref{eq:SEsigmcorr}) and
(\ref{eq:SEfwmcorr}) to make single-epoch mass predictions
and we plot these versus the reverberation measurements in 
panels (c) and (d) of Figure~\ref{Figure:Fit_139_141}.
The quality of the correction can be tested by fitting
these relationships.
The best-fit coefficients for the corrected
$\log \muse(\hb)$--$\log \murm(\hb)$ relationship
are given in lines 8 and 9 of Table 5.

\section{Masses Based on \civ}

\subsection{Fundamental Relationships}
\label{section:civfundamental}

\begin{figure}
\begin{centering}
\includegraphics[scale=0.8]{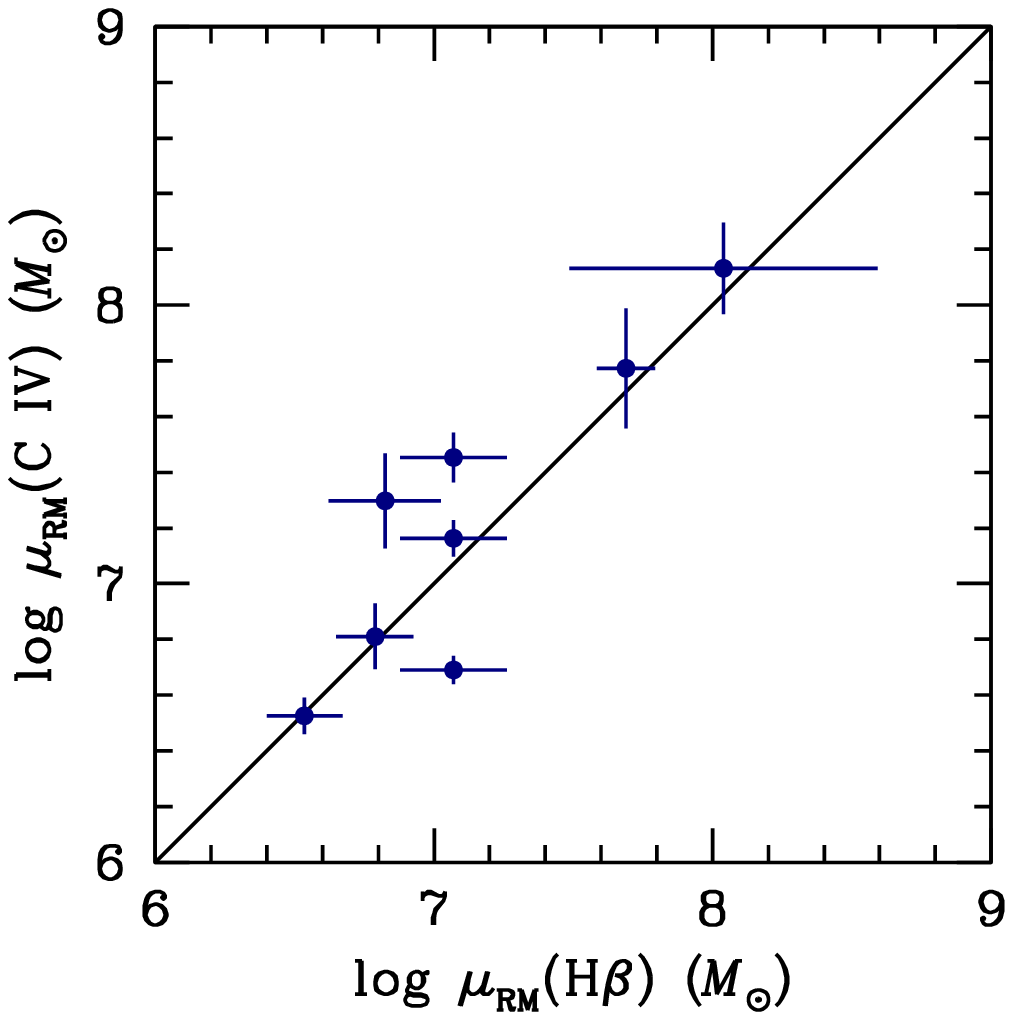}
\caption{Virial products based on \civ\
and \hb\ for the few cases in the RMDB sample
for which both are available. The solid line is
the locus where the two virial products are equal.
The values are weighted means of $\murm(\hb)$
and $\murm(\civ)$ for individual AGNs
that appear in both Tables~A1 and A2.
The Spearman rank coefficient for these data is
$\rho = 0.805$ and the probability that the
correlation arises by chance is $P = 0.016$.}
\label{Figure:RMcompare}
\end{centering}
\end{figure}

As noted in \S\ref{section:intro}, the veracity of
\civ-based mass estimates is unclear and
remains controversial. The ideal situation would be
to have a large number of AGNs with both \civ\
and \hb\ reverberation measurements to effect a direct comparison.
There are, unfortunately, very few AGNs that have
both; indeed Table~A2 
of the Appendix includes all \civ\ results  for which there are
corresponding \hb\ measurements in Table A1. For the
few sources with both \civ\ and \hb\ reverberation
measurements, we plot the virial products
$\mu_{\rm RM}(\civ)$ and $\mu_{\rm RM}(\hb)$
in Figure~\ref{Figure:RMcompare}; these
are in each case a weighted mean value of
\begin{equation}
\label{eq:mudef}
\mu_{\rm RM} = \left( \frac{c \tau \sigr^2}{G}\right)
\end{equation}
for each of the observations of \hb\ and \civ\ for the
AGNs that appear in both Tables~A1 and A2.
The close agreement of these values reassures us that
the \civ-based RM masses can be trusted, at least over the
range of luminosities sampled.

\begin{figure}
\begin{centering}
\includegraphics[scale=0.8]{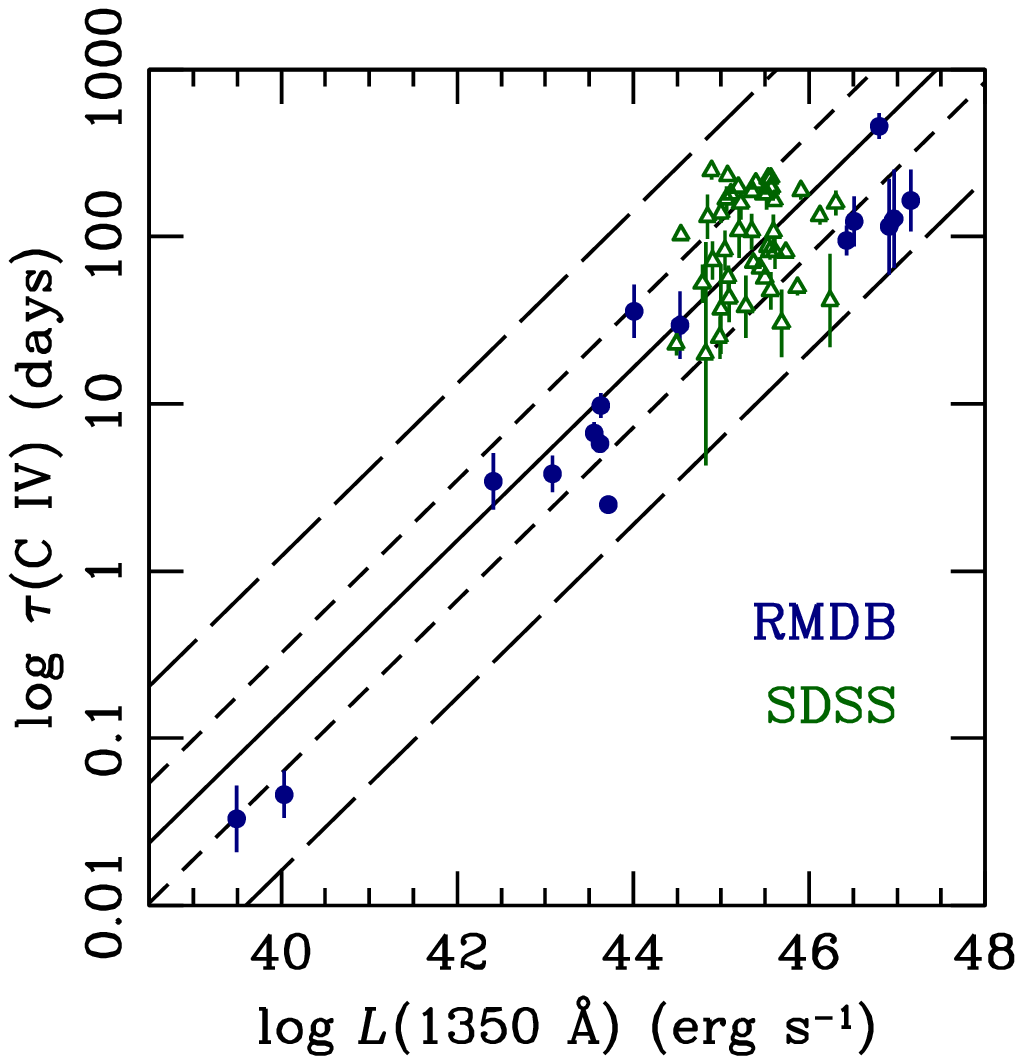}
\caption{Relationship between the \civ\ rest-frame emission-line
lag $\tau(\civ)$ and the continuum luminosity at 1350\,\AA.
Blue filled circles represent RMDB data (Table~A2) and
green open triangles represent SDSS data (Table~A4).
The solid line is the best fit to the data using
equation (\ref{eq:powerlaw}) with coefficients given in
Table~2. The short dashed and long dashed lines
are the $\pm 1 \sigma$ and $\pm 2.6 \sigma$ envelopes, respectively.
The Spearman rank coefficient for these data is
$\rho = 0.503$ with a probability $P = 1.1 \times 10^{-5}$
that the correlation arises by chance. If the two lowest luminosity
points (both measurements of the dwarf Seyfert NGC\,4395)
are omitted, the Spearman rank coefficient is decreased
to $\rho = 0.481$ with $P = 1.1 \times 10^{-4}$.}
\label{Figure:civradlum}
\end{centering}
\end{figure}

We now need to consider whether or not luminosities and mean line
widths are suitable proxies for emission-line lag and rms line widths
in the case of \civ.  In Figure~\ref{Figure:civradlum}, we show the
relationship between the UV continuum luminosity $L(1350\,{\rm \AA})$
and the \civ\ emission-line lag $\tau(\civ)$ based on the \civ\ data
in Table~A2, plus the SDSS-RM \civ\ data in Table~A4. The coefficients
of the fit are given in line 3 of Table~2.  We note again that we have removed
from the \cite{Grier19} sample in Table~A4 three quasars with BALs,
thus reducing the sample size from 48 to 45.  The slope of the
\civ\ $R$--$L$ relation ($0.517$) is consistent with that of
\hb\ ($0.492$), though the $\varepsilon_y$ scatter is substantially
greater ($0.336$\,dex for \civ\ compared to $0.213$\,dex for
\hb). Definition of the relationship does not depend on the two
separate measurements of very short \civ\ lag measurements for the
dwarf Seyfert NGC 4395 \citep{Peterson05}.  Thus it seems clear that
we can use $L(1350\,{\rm \AA})$ as a reasonable proxy for
$\tau(\civ)$.

\begin{figure}
\begin{centering}
\includegraphics[scale=0.8]{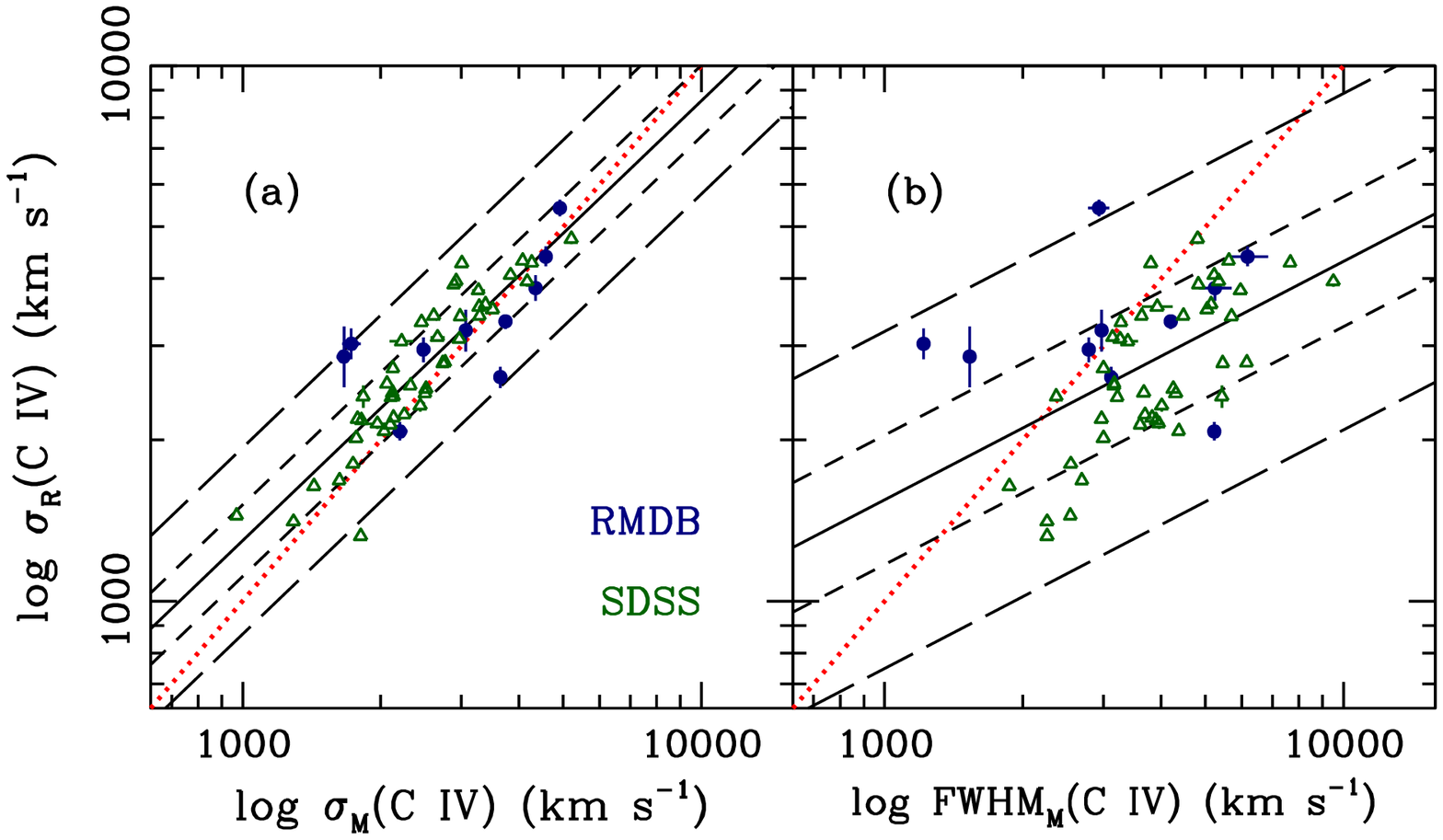}
\caption{Panel (a): Relationship between \civ\
line dispersion in the mean and rms spectra
of reverberation-mapped AGNs.  The Spearman
rank coefficient is $\rho = 0.873$ with a probability
of $P < 10^{-6}$ that the correlation arises by chance.
Panel (b): Relationship between \fwm(\civ) and
\sigr(\civ) for
reverberation-mapped AGNs. The Spearman
rank coefficient for these data is
$\rho = 0.524$ with $P=3.96 \times 10^{-5}$.
In both panels,
blue filled circles represent RMDB sources
in Table~A2 and green open triangles represent
SDSS-RM sources in Table~A4. The red dotted line
shows the locus where the two line-width measures are equal.
The solid line is the best fit to equation
(\ref{eq:powerlaw}) and the coefficients are
given in Table~3. The short dashed and long dashed
lines show the $\pm 1 \sigma$ and $\pm 2.6 \sigma$
envelopes, respectively.}
\label{Figure:windowcivwidths}
\end{centering}
\end{figure}

We show the relationship between the \civ\ line dispersion measured
in the rms spectrum \sigr(\civ) and the line dispersion
in the mean spectrum \sigm(\civ) in Figure~\ref{Figure:windowcivwidths}.
The best-fit coefficients are given in line 3 of Table~3.
The correlation is good. However, the correlation between
\fwm(\civ) and \sigr(\civ), 
also shown in Figure~\ref{Figure:windowcivwidths}
with coefficients in line 4 of Table~3,
is rather poor \citep[see also][]{Wang20}
and  demonstrates that
\fwm(\civ) is a dubious proxy for \sigr(\civ).  Measurement
of \fwm(\civ) is clearly a much less reliable predictor of 
\sigr(\civ) than is \sigm(\civ), 
so we will not consider \fwm(\civ) further.

\subsection{Single-Epoch Masses}
\label{section:civ}

\begin{figure}
\begin{centering}
\includegraphics[scale=0.8]{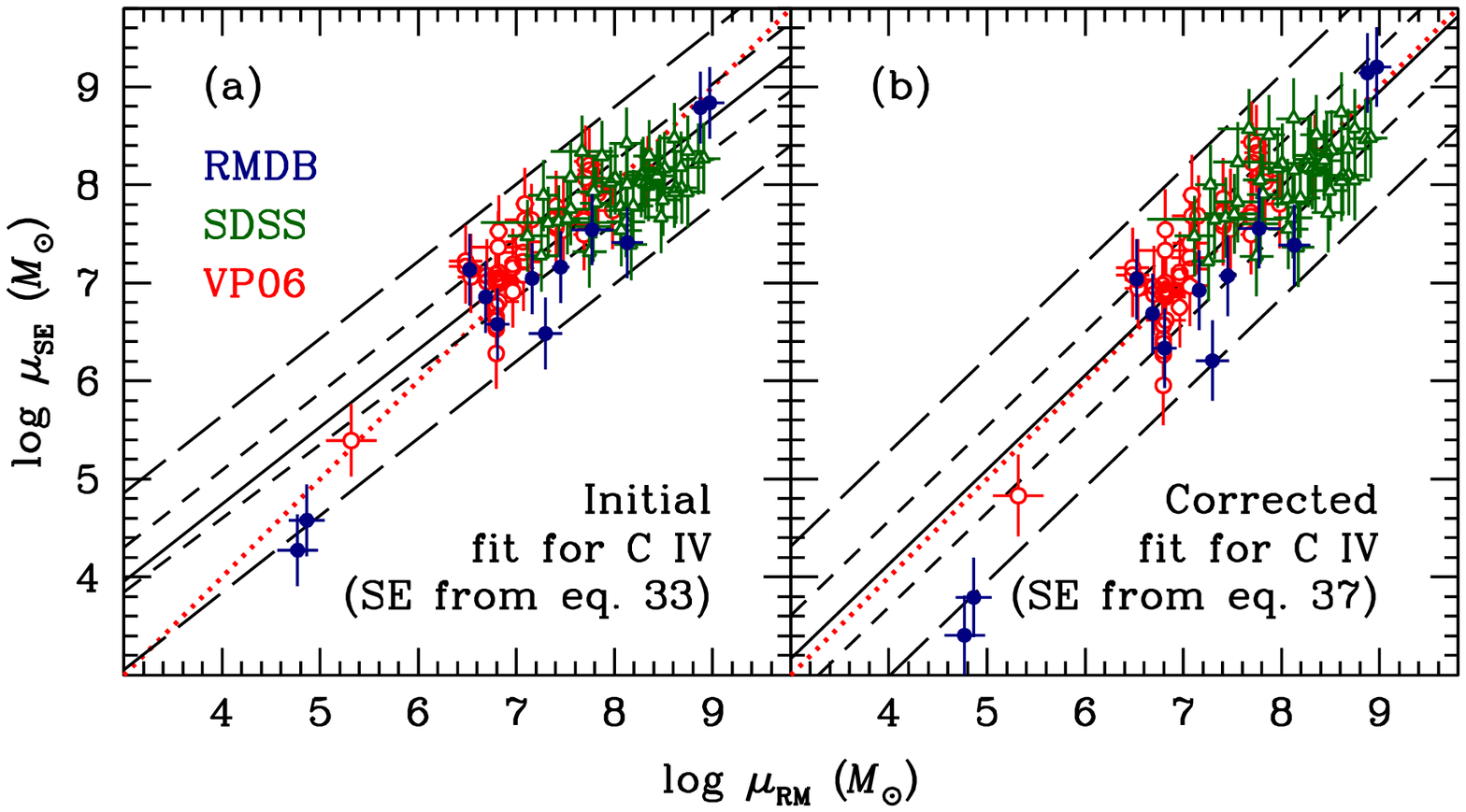}
\caption{Panel (a): Comparison of single-epoch virial products \muse(\civ) and
reverberation measurements \murm(\civ) for the data in Table~A2
(blue filled circles),  the SDSS-RM \civ\ reverberation data from
Table~A4 (green open triangles), and data from Table~A5
(red open circles).
The solid line is the best fit to the data and has
slope $0.787 \pm 0.041$. As was the case with \hb, 
masses are overestimated at the
low end and underestimated at the high end, excepting the three
very low mass measurements.
Panel (b): Comparison of single-epoch virial products after empirical
correction as given in equation (\ref{eq:predictciv}).
In both panels, the solid line is the best fit to the
relationship between $\log \muse(\civ)$ and $\log \murm(\civ)$.
The short dashed and long dashed
lines define the $\pm 1 \sigma$ and $\pm 2.6 \sigma$ envelopes,
respectively.
The diagonal red dotted line is the locus where \murm\ and \muse\
are equal. 
Coefficients for both fits are given in Table~5,
in line 3 for panel (a)  and in line 10 for panel (b).
In both panels, the intrinsic errors $\varepsilon_y$ have been
added in quadrature to the measurement errors in $\log \muse(\civ)$.}
\label{Figure:civvps}
\end{centering}
\end{figure}

Following the same procedures as with \hb,
we use the RMDB data (Table~A2) and the
SDSS-RM data (Table~A4) to fit the equation
\begin{equation}
\log \murm = a + b\left[ \log L(1350\,{\rm \AA}) - x_0 \right]
+ c \left[ \log  \sigm(\civ) - y_0 \right].
\label{eq:Fit_244}
\end{equation}
The resulting fit is shown in Figure~\ref{Figure:civvps}
and the best-fit coefficients are given in line 3 of Table~4. 
Thus our initial single-epoch virial product prediction is
\begin{equation}
\label{eq:SEciv}
\log \muse(\civ)  =  7.664
+ 0.599\left[ \log L(1350\,{\rm \AA}) - 44.706 \right]
+ 1.014\left[ \log  \sigm(\civ) - 3.502 \right].
\end{equation}
Single-epoch virial product 
estimates based on equation (\ref{eq:SEciv}) are plotted
against the actual reverberation measurements in 
Figure~\ref{Figure:civvps} and the results of a fit to
these data are given in line 3 of Table 5.
As was the case with \hb, the slope of this relationship
is too shallow, indicating that equation (\ref{eq:SEciv}) is
too simple a prescription and suggesting that another parameter is required.

\begin{figure}
\begin{centering}
\includegraphics[scale=0.8]{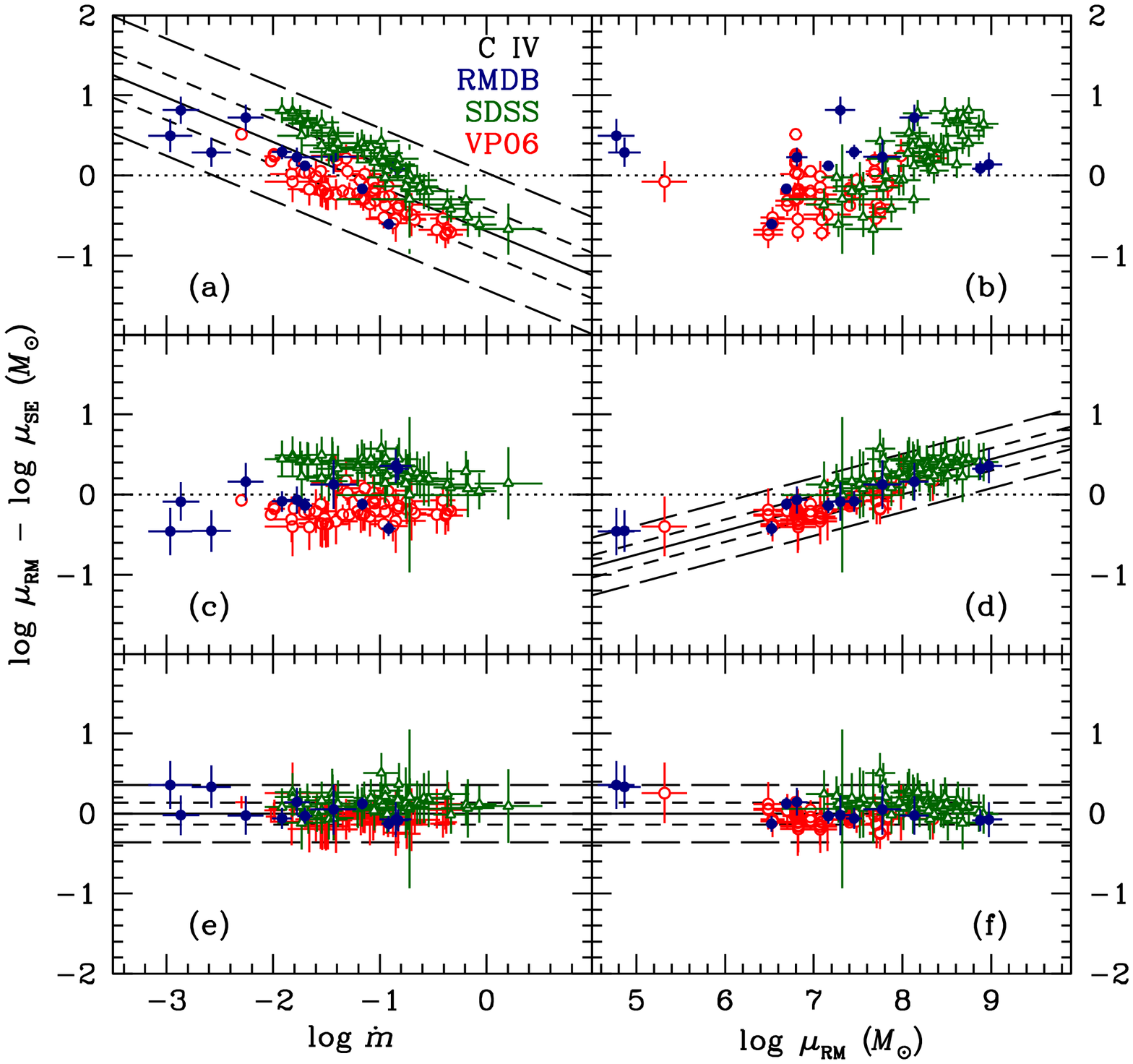}
\caption{Mass residuals $\Delta \log \mu =\log \murm - \log \muse$
versus Eddington rate \mdot\ (left column) and
virial product \murm\ (right column) for \civ.
Panel (a) shows the
residuals between \murm(\civ) and \muse(\civ)
versus Eddington ratio \mdot\ (equation \ref{eq:Eddratio}).
The fit to these data has Spearman rank coefficient
$\rho = -0.693$ with a probability that the correlation 
arises by chance $P < 10^{-6}.$
Panel (b) shows the residuals versus
virial product \murm.
Panels (c) and (d) show the residuals
versus \mdot\ and \murm\
after subtracting the fit in panel (a).
Panel (d) also shows a best fit to
the residuals versus mass; coefficients are
given in line 7 of Table~5. Note that there is no intrinsic
scatter in this relationship because the error bars are so large. 
For these data, $\rho = 0.883$ and $P < 10^{-6}.$
Panels (e) and (f)
show the mass residuals versus \mdot\
and \murm\ after subtracting
the fit in panel (d).
The scatter in panels (e) and (f)  is 0.138\,dex.
In all panels, the blue filled circles represent RMDB data (Table~A2),
the green open triangles are SDSS data (Table~A4), and the red open
circles are VP06 data (Table~A5). Best fits are shown as solid lines
and the short dashed and long dashed lines indicate the
$\pm 1 \sigma$ and $\pm 2.6 \sigma$ envelopes, respectively.
}
\label{Figure:window244_247}
\end{centering}
\end{figure}

Guided by our result for \hb, we
plot the residuals in $\log \murm - \log \muse$
versus Eddington ratio \mdot\ in 
panel (a) of Figure~\ref{Figure:window244_247}.
The Eddington ratio for the UV data is
\begin{equation}
\log \mdot = -33.737 + 0.9 \log L(1350\,{\rm \AA}) - \log \mu_{\rm RM},
\label{eq:Edddefciv}
\end{equation}
where again we have used a bolometric correction from $L(1350\,{\rm \AA})$
from \cite{Netzer19},
\begin{equation}
\log k_{\rm bol} = 5.045 - 0.1 \log L(1350\,{\rm \AA}).
\label{eq:bolometricciv}
\end{equation}
We fitted equation (\ref{eq:residuals}) to the \civ\ 
mass residuals and Eddington ratio
and the results are given in line 6 of Table~5
and also plotted in panel (a) 
of Figure~\ref{Figure:window244_247}.

\begin{figure}
\begin{centering}
\includegraphics[scale=0.5]{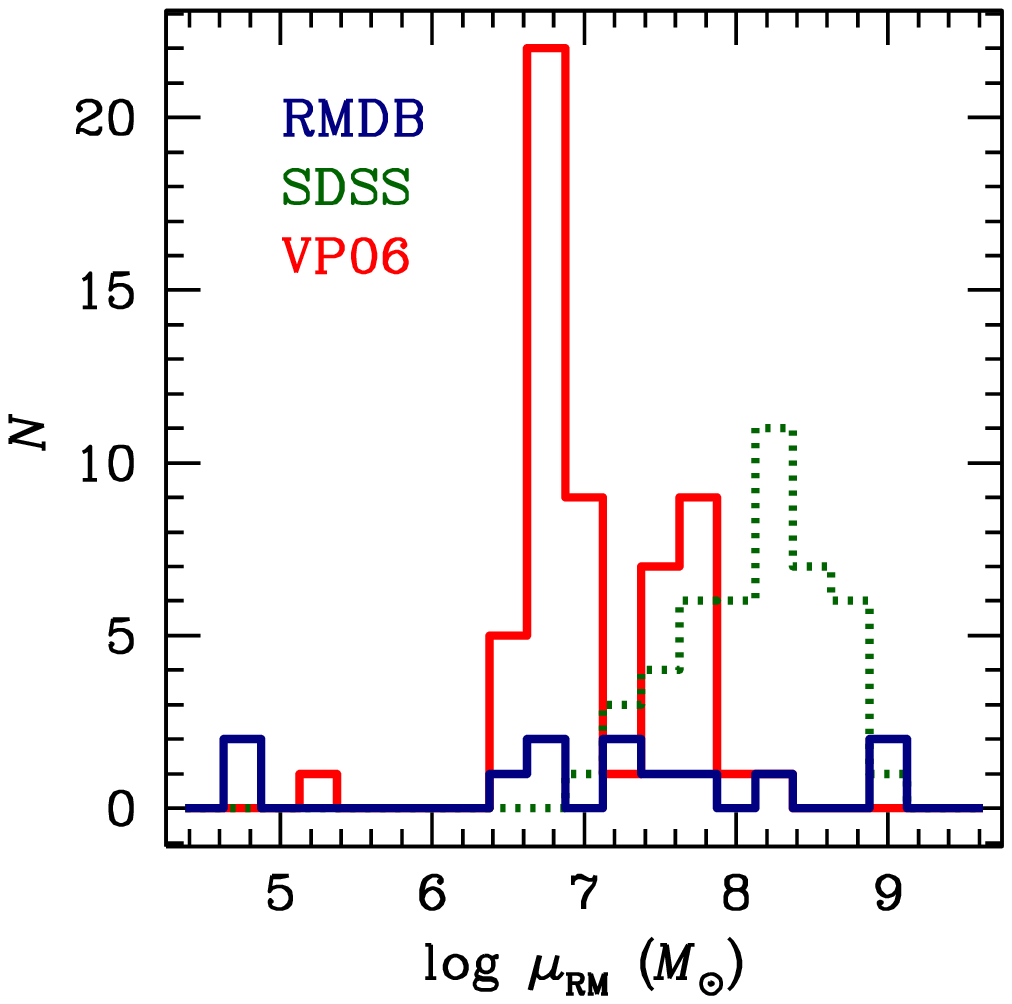}
\caption{Distribution in virial
product \murm\ for the
RMDB (Table~A2, blue solid line),
SDSS (Table~A3, green dotted line), and
VP06 (Table~A4, red solid line) samples.
The VP06 sample is a subset of the RMDB sample,
which is dominated by the relatively low-mass
Seyfert galaxies that were the first sources
studied by reverberation mapping. The SDSS
quasars are comparatively more massive and more
luminous.
}
\label{Figure:masshist}
\end{centering}
\end{figure}

The offset between the residuals in 
the panel (a) of Figure~\ref{Figure:window244_247}
between the RMDB and VP06 data on one hand
and the SDSS data on the other might seem to be problematic and we
were initially concerned that this might be a data integrity issue.
However, upon examining the distribution of mass
and luminosity for these
three samples as seen in Figure~\ref{Figure:masshist}, we see clearly
that the mass distribution of the SDSS sources is skewed toward much
higher values than for the RMDB and VP06 sources, which are
relatively local and less luminous than the SDSS quasars. We will
thus proceed by examining mass residuals versus
both Eddington ratio and \murm.

Figure~\ref{Figure:window244_247} illustrates the process by which we
eliminate the mass residuals in successive iterations. We compute the
mass residuals $\Delta \log \mu = \log \murm - \log \muse$ from
equation (\ref{eq:SEciv}); these are shown versus \mdot\ (left
column) and $\murm$ (right column).  We fit these residuals versus
\mdot\ (panel a) and subtract the best fit to get the corrected
residuals shown in the panels (c) and (d).  Examination of these residuals as a
function of other parameters revealed that they are still correlated
with \murm\ (panel d), suggesting that the importance of the
Eddington ratio depends on the black hole mass.  We therefore fit the
residuals a second time, this time as
\begin{equation}
\Delta \log \mu = a + b(\log \murm\ - x_0).
\label{eq:massresiduals}
\end{equation}
The best fit to this equation is shown in panel (d)
and the coefficients are given in Table~5. Subtraction of the best
fit yields the residuals shown in panels (e) and (f).
We would under most circumstances consider this
procedure with some trepidation from a statistical
point of view, since  $\murm$ appears explicitly in one correction
and is implicitly in the Eddington ratio. A generalized
solution would have multiple degeneracies as both
mass and luminosity appear in multiple terms. However,
the residual corrections are physically motivated; several
previous investigations have also concluded that 
Eddington ratio is correlated with the deviation from the
\cite{Bentz13} $R$--$L$ relationship, and panels (c) and (d) of 
Figure \ref{Figure:window244_247} suggests that the impact
of Eddington ratio varies slightly with mass. Nevertheless,
one would prefer to work with parameters that are
correlated with or indicators of \mdot\ and \murm,
as we will discuss in \S{\ref{section:discussion}}. 

Combining the original fit (equation \ref{eq:SEciv}) with the two
corrections (equations \ref{eq:Edddefciv} and \ref{eq:massresiduals})
yields a corrected single-epoch virial product predictor,
\begin{equation}
\label{eq:SEcivcorr}
\log \muse(\civ) =  7.714 
+ 0.761\left[ \log L(1350\,{\rm \AA}) - 44.706 \right]
+ 1.289\left[ \log  \sigm(\civ) - 3.502 \right].
\end{equation}
Single-epoch virial products for all three samples are
compared with the reverberation measurements in the right
panel of Figure~\ref{Figure:civvps}. The coefficients of the
best fit to these data are given in line 10 of Table~5.

It is worth noting in passing that after correcting for Eddington ratio
(Figure~\ref{Figure:windowhbresiduals}), the residuals in the \hb-based
mass estimates show no correlation with either mass or luminosity.

\section{Computing Single-Epoch Masses}
\label{section:massformulae}

To briefly reiterate our approach so far, we started with the assumption
that $\muse= f(R,L)$ only. This proved to be inadequate, so we examined the
residuals in the $\log \muse$--$\log \murm$ relationship 
and found that these correlated
best with Eddington ratio \mdot: fundamentally, at increasing \mdot,
the \cite{Bentz13} $R$--$L$ relationship overpredicts the size of the 
BLR $R$ \citep{Du19}. In the case of \civ, 
we found additional residuals that correlated 
with \murm, although we cannot definitively demonstrate that some part of this
is not attributable to inhomogeneities in the data base (a point that will be
pursued in the future). While we believe this analysis identifies the
physical parameters that affect the mass estimates, there are multiple 
degeneracies, with both mass and luminosity appearing in more than one term.

Instead of trying to fit coefficients to all the physical parameters that
have been identified, we  can do a purely empirical correction to 
equations (\ref{eq:Fit_139}), (\ref{eq:Fit_141}),
and (\ref{eq:Fit_244}) since the
residuals in the $\log \murm$--$\log \muse$ relationships
(upper panels in Figure~\ref{Figure:Fit_139_141}
and left panel of Figure~\ref{Figure:civvps}) are rather small. We can combine
the basic $R$--$L$ fits (equations \ref{eq:Fit_139}, \ref{eq:Fit_141},
and \ref{eq:Fit_244}) with the residual fits (equations
\ref{eq:residuals} and \ref{eq:massresiduals}) to obtain
prescriptions that work over the mass range sampled. Renormalizing
for convenience, we can estimate single-epoch masses based on
\hb(\sigm) from
\begin{eqnarray}
\log M_{\rm SE} & = & \log f + 7.530  
+ 0.703 \left[ \log L(\hb)-42 \right]
 + 2.183  \left[\log \sigm(\hb) - 3.5 \right],
\label{eq:predictHbsigm}
\end{eqnarray}
with associated uncertainty
\begin{equation}
\label{eq:predictHbsigmerror}
\Delta \log M_{\rm SE} = \left\{ (\Delta \log f)^2 +
\left[0.703\ \Delta \log L(\hb) \right]^2 +
\left[2.183\ \Delta \log \sigm(\hb)\right]^2 \right\}^{1/2}
\; .
\end{equation}
Here $f$ is the scaling factor which is discussed briefly in the Appendix,
and $\Delta \log P$ is the uncertainty in the parameter $\log P$. The intrinsic
scatter in this relationship is $0.309$\,dex,  and this must be added in
quadrature to the random error.
For the case of \hb(\fwm), a single-epoch mass estimate is obtained from
\begin{eqnarray}
\log M_{\rm SE} & = & \log f + 7.015  + 
0.784 \left[ \log L(\hb)-42 \right]
 + 1.387  \left[\log \fwm(\hb) - 3.5 \right],
\label{eq:predictHbFWHM}
\end{eqnarray}
with associated uncertainty
\begin{equation}
\label{eq:predictHbFWHMmerror}
\Delta \log M_{\rm SE} = \left\{ (\Delta \log f)^2 +
\left[0.784\ \Delta \log L(\hb) \right]^2 +
\left[1.387\ \Delta \log \fwm(\hb)\right]^2 \right\}^{1/2} \;
.
\end{equation}
In this case, the intrinsic scatter is $0.371$\,dex.

A comparison of the reverberation-based virial products \murm(\hb)
and the single-epoch masses \muse(\hb) based on equations
(\ref{eq:predictHbsigm}) and (\ref{eq:predictHbFWHM})
is shown in panels (c) and (d)  of
Figure~\ref{Figure:Fit_139_141}.

Similarly, single-epoch masses based on \civ\ can be computed from
\begin{eqnarray}
\log M_{\rm SE} &  = & \log f +7.934  
+ 0.761 \left[ \log L(1350\,{\rm \AA})-45 \right]
+ 1.289  \left[\log \sigm(\civ) - 3.5 \right],
\label{eq:predictciv}
\end{eqnarray}
with associated uncertainty
\begin{equation}
\label{eq:predictciverror}
\Delta \log M_{\rm SE} = \left\{ (\Delta \log f)^2+
\left[0.761\ \Delta \log L(1350\,{\rm \AA} \right]^2 +
\left[1.289\ \Delta \log \sigm(\civ)\right]^2  \right\}^{1/2}\;
.
\end{equation}
The intrinsic scatter in this relationship is 0.408\,dex. Single-epoch predictions
and reverberation-based masses for the AGNs in Tables~A2, A4, and A5
are compared in panel (b) of Figure~\ref{Figure:civvps}.
Coefficients for this fit are given in line 10 of Table~5.

\begin{figure}
\begin{centering}
\includegraphics[scale=0.5]{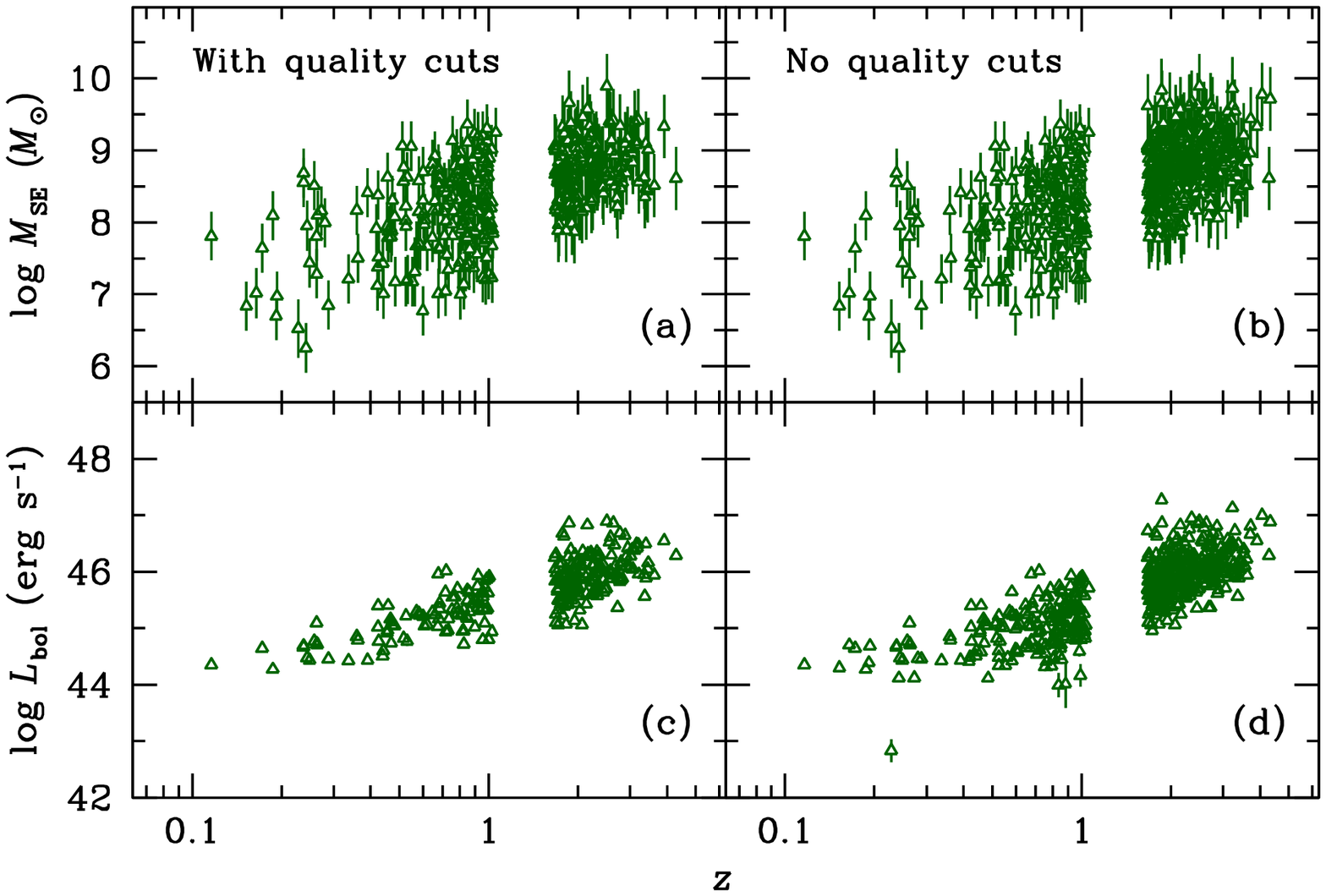}
\caption{Distribution of masses (panels a and b)
and bolometric luminosities (panels c and d) for the entire
SDSS-RM sample for which \hb\ or \civ\ single-epoch
masses can be computed using
equations (\ref{eq:predictHbsigm}) and
(\ref{eq:predictciv}). Here we assume
$f=4.28$ \citep{Batiste17}.
Bolometric corrections were made using
equations (\ref{eq:bolometriccorrection})
and (\ref{eq:bolometricciv}).
In the left column,
the quality cuts of \S\ref{section:Data} have been
imposed. In the right column, no quality 
cuts have been made.
}
\label{Figure:zdistsdssselect}
\end{centering}
\end{figure}

In Figure \ref{Figure:zdistsdssselect}, we show the distribution in bolometric
luminosity and black hole mass  based on our prescriptions for the entire sample of SDSS-RM
quasars for which \hb\ or \civ\ single-epoch masses can be estimated.

\section{Discussion}
\label{section:discussion}

\subsection{Single-Epoch Masses}

Our primary goal has been to find simple, yet unbiased, prescriptions for
estimating the masses of the black holes that power AGNs. Our underlying assumption
has been that the most accurate measure of the virial product is
given by using the emission-line lag $\tau$ and line width in the
rms spectrum \sigr\ (e.g., equation \ref{eq:explicitmu} in the Appendix)
as that quantity produces, upon adjusting by the scaling factor $f$,
an $M_{\rm BH}$--$\sigma_*$ relationship for AGNs that is in good
agreement with that for quiescent galaxies. Given that both
$\tau$ and \sigr\ average over structure in a complex system 
\citep[cf.][]{Barth15}, it is
somewhat surprising that this method of mass estimation works as well as it does.

Here we have shown that the luminosity of the broad component of the \hb\ emission line
is a good proxy for the starlight-corrected AGN luminosity (Figure
\ref{Figure:HbRL}).  This is useful since it eliminates the difficult
task of accurately modeling the host-galaxy starlight contribution to
the continuum luminosity.  
Moreover, the line luminosity and \sigr\ reflect the BLR
state at the same time; a measurement of the continuum luminosity, by contrast,
better represents the state of the BLR at a time $\tau$ in the future
on account of the light travel-time delay within the system
\citep{Pogge92,Gilbert03,Barth15}; this is, however, generally a 
very small effect.
For the sake of completeness, we also note that there is a small, but
detectable, lag between continuum variations at shorter wavelengths and 
those at longer wavelengths \citep{McHardy14,Shappee14,Edelson15,
Fausnaugh16, Edelson17,McHardy18,Edelson19}.

We have also confirmed that, for the case of \hb, 
both \sigm\ and \fwm\ are reasonable proxies for
\sigr, though \sigm\ is somewhat better than \fwm.

On the other hand, the case of \civ\ remains problematic, as it differs
in a number of ways from the other strong emission lines:
\begin{enumerate}
\item The equivalent width of \civ\ decreases with luminosity, which 
is known as the Baldwin Effect \citep{Baldwin77}; 
\civ\ is driven by higher-energy photons than, say, the Balmer lines
and the Baldwin Effect reflects a softening of the high-ionization
continuum. This could be due to higher Eddington ratio
\citep{Baskin04} or because
more massive black holes have cooler accretion disks
\citep{Korista98}.
\item The \civ\ emission line is typically blueshifted with respect
to the systemic redshift of the quasar, which is attributed to 
outflow of the BLR gas \citep{Gaskell82,Wilkes84,Wilkes86,Espey89,Wills93,
Richards02,Sulentic07,Richards11,Coatman16,Shen16a,Bisogni17,Vietri18}.
\item BALs in the short-wavelength wing of \civ, another signature of outflow, are
common \citep{Weymann91,Hall02,Hewett03,Allen11}.
We remind the reader that in \S{\ref{section:Data}} we removed
$\sim17$\% of our SDSS \civ\ sample because the presence of BALs precludes accurate
line-width measurements.
\item The pattern of ``breathing'' in \civ\ is the opposite of what
is seen in \hb\ \citep{Wang20}. Breathing refers to the
response of the emission lines, both lag and line width, to changes in
the continuum luminosity. In the case of \hb, an increase in luminosity
produces an increase in lag and a decrease in line width 
\citep{Gilbert03,Goad04,Cackett06}. In the case of \civ, however,
the line width increases when the continuum luminosity increases, contrary
to expectations from the virial theorem (equation \ref{eq:masseqn}).
\end{enumerate}
We must certainly be mindful that outflows can affect a mass measurement,
though the effect is small if the gas is at escape velocity. Notably,
in the cases studied to date there is good agreement between \hb-based
and \civ-based virial products (Figure \ref{Figure:civradlum}),
though, again, these are local Seyfert galaxies that are not
representative of the general quasar population.

The \civ\ breathing issue is addressed in detail by \cite{Wang20},
building on evidence for a non-reverberating narrow core or blue excess
in the \civ\ emission line presented by \cite{Denney12}. In this two-component model,
the variable part of the line is much broader
than the non-variable core.  As
the continuum brightens, the variable 
broad component increases in prominence,
resulting in a larger value of \sigm. As the broad
component reverberates in response to continuum variations, 
\sigm\ will track \sigr\ much better than \fwm, thus explaining the
breathing characteristics and why \fwm\ is a poor line-width measure for
estimating black hole masses. Physical interpretation of the non-varying
core remains an open question: \cite{Denney12} suggests that it might
be an optically thin disk wind or an inner extension of 
the narrow-line region.

\subsection{Eigenvector 1 and the Role of Eddington Ratio}

Aside from the Baldwin Effect \citep{Baldwin77}, the average
spectra of quasars show little dependence on luminosity 
\citep[e.g.,][]{Vandenberk04}. However, individual objects show
considerable spectral diversity or differences from the mean spectrum,
regardless of luminosity.
Many of these spectral differences 
show strong correlations and anticorrelations with other spectral
properties or physical parameters
as revealed by Principal Component Analysis (PCA), as
first shown by \cite{Boroson92}.
The strongest of these multiple correlations, \EVone,
is most clearly characterized by the anticorrelation
between (a) the strength of the \feii\,$\lambda4570$ and
\feii\,$\lambda\lambda5190$, 5320 complexes on either side of the
broad \hb\ complex and (b) the strength of 
the \oiii\,$\lambda\lambda 4959$, 5007 doublet.
The \feii\ strength is typically characterized 
by the ratio of the equivalent widths (EW) or fluxes of \feii\ to \hb,
i.e., $\RFe = {\rm EW(\feii)}/{\rm EW(\hb)}$.
\cite{Boroson92} speculate that the physical driver
behind \EVone\ is Eddington ratio as they are able to argue
against inclination effects.
\cite{Sulentic00} incorporate UV data into the PCA
and found that the magnitude of the
\civ\ emission-line blueshift,
a ubiquitous feature of AGN UV spectra \citep[e.g.,][]{Richards02},
is also an \EVone\ component, with larger blueshifts associated
with higher \RFe\ and lower \oiii\ strength. 
This has been confirmed in a number of subsequent studies
\citep{Baskin05,Coatman16,Sulentic17}.
\cite{Sulentic00}  also demonstrated that the 
``narrow-line Seyfert 1'' (NLS1) galaxies \citep{OsterbrockPogge85}, a subset of Type 1 AGNs
with particularly small broad-line widths
(${\rm FWHM}(\hb) < 2000\,\kms$),
lie at the strong \RFe--weak \oiii\ extreme of \EVone.
To see why this is so, if we combine the $R$--$L$ relation
with eq.\ (\ref{eq:masseqn}), the expected line width dependence is seen to be
\begin{equation}
\label{eq:linewidth}
V \propto\left( \frac{M}{L^{1/2}} \right)^{1/2} \propto 
\left( \frac{M}{\mdot} \right)^{1/4},
\end{equation}
where $\mdot \propto L/M$  is the Eddington ratio (eq.\ \ref{eq:Eddratio}).
Thus AGNs with the highest Eddington ratios have the smallest broad-line widths, and
many such sources are classified as NLS1s. \cite{Boroson02} argues that
the physical parameter driving \EVone\ is indeed Eddington ratio,
and that Eigenvector 2 is driven by accretion rate; these two
physical parameters, plus inclination, appear to account for most of the
spectral diversity among quasars. 
There is now, we believe, general consensus in the literature that \EVone\
is driven by Eddington ratio \citep[e.g.,][]{Shen14,Sun15,Marziani18},
and our own analysis supports this.

The necessity of including an Eddington ratio correction to
single-epoch mass estimators became an issue when poor argeement was found
between \hb\ and \mgii-based SE masses on one hand and
\civ-based masses on the other.
\cite{Shen08b} found that the offset between \mgii-based SE masses
and those based on \civ\ correlated with the \civ\ blueshift,
an \EVone\ parameter as already noted, thus enabling an empirical correction.
Similarly \cite{Runnoe13a} and \cite{Brotherton15a}
use the strength of the \siiv-\oiv\ blend, another \EVone\ parameter,
to effect an empirical correction.

The Super-Eddington Accreting Massive Black Holes (SEAMBH) collaboration has focused
on high-\mdot\ candidates in their reverberation-mapping program
\citep{Du14,Du16,Du18,Du19}. An important result from these studies, as we have
noted earlier, is that the \hb\ lags are smaller than predicted by
the current state-of-the-art $R$--$L$ relationship \citep{Bentz13}.
This implies that in these objects the ratio of hydrogen-ionizing photons
to optical photons is lower than in the lower \mdot\ sources; 
this is also consistent
with the relative strength of \RFe,
the weakness of high-ionization lines such as \oiii, and the
soft X-ray spectra \citep{Boller96} of high \mdot\  sources.
\cite{Du19} choose to make their correction 
to the BLR radius through adding a 
term that correlates with the deficiency of ionizing photons. 
In our approach, we absorb the correction directly into the
virial product computation.

The studies cited above have noted that an Eddington ratio correction
is required for single-epoch masses based on \hb. 
We find, as have others \citep{Shen08b,Bian12,Shen12,
Runnoe13a,Brotherton15a,Coatman17},
that a similar correction is required for
\civ-based masses as well.

\begin{figure}
\begin{centering}
\includegraphics[scale=0.5]{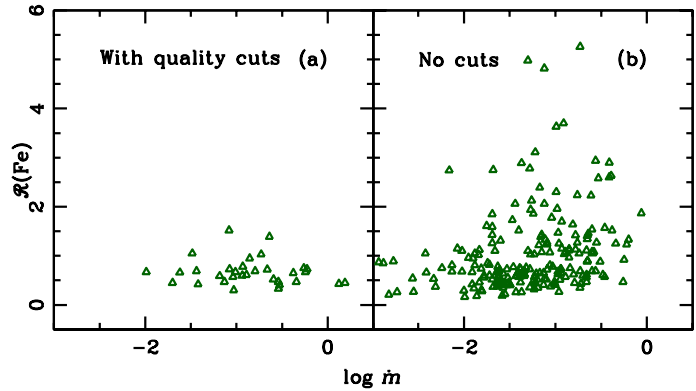}
\caption{
Correlation between ${\cal R}{\rm (Fe)}$ and
Eddington ratio \mdot\ for the subset of
SDSS-RM quasars selected for our study on the
basis of quality cuts (\S\ref{section:Data}) in
panel (a) and for all SDSS-RM quasars with measured
\feii\ equivalent widths in the compilation
of \cite{Shen19} in panel (b). The Eddington
ratio used by \cite{Shen19} differs slightly from
that used here.}
\label{Figure:rfemdot}
\end{centering}
\end{figure}

As noted in \S\ref{section:civ}, from a statistical point of
view, in the single-epoch mass equations 
it would be preferrable to replace the Eddington ratio with a parameter
strongly correlated with it.  
However, we find that the scatter in these relationships is so large that
any gain in the accuracy of black hole mass estimates
is offset by a large loss of precision. For example,
while the correlation between \RFe\ and Eddington ratio exists,
as shown for the SDSS-RM sample in Figure~\ref{Figure:rfemdot},
the scatter is so large that the correlation
has no real predictive power. We therefore
elect at this time to focus on the empirical formulae given
in \S\ref{section:massformulae}.

\subsection{Future Improvements}

While we believe our current single-epoch prescription for estimating
quasar black hole masses is more accurate than previous prescriptions,
we also recognize that there are additional improvements that can be
made to improve both accuracy and precision, some of which we became
aware of near the end of the current project. We intend to implement
these in the future. Topics that we will investigate in the future
include the following:
\begin{enumerate}
\item Replace those reverberation lag measurements made with the
interpolated cross-correlation function 
\citep{Gaskell87,White94,Peterson98b,Peterson04} with
lag measurements and uncertainties from {\tt JAVELIN} \citep{Zu11}.
Recent tests \citep{Li19,Yu20} show that while the {\tt JAVELIN}
and interpolation cross-correlation lags are generally consistent, 
the uncertainties predicted by
{\tt JAVELIN} are more reliable.
\item Utilize the expanded SDSS-RM database, which now extends
over six years, not only to make use of additional lag detections, but to
capitalize on the gains in $S/N$ that will
increase the overall quality of the lag and line-width measurements 
and result in fewer rejections of poor data.
\item Expand the database in Table~A1 with recent results and 
other previous results that we excluded because they did 
not have starlight-corrected continuum luminosities.
\item Update the VP06 database used to produce Table~A5. There are now
additional reverberation-mapped AGNs with archived {\em HST} UV spectra.
Some of the poorer data in Table~A5 can be replaced 
measurements based on higher-quality spectra.
\item Consider use of other line-width measures that
may correlate well with \sigl, but are less sensitive to blending
in the wings. Mean absolute deviation (MAD) is one such candidate;
indeed, \cite{Park17} have already demonstrated that \civ-based
masses are more consistent with those based on other lines if
either \sigl\ or MAD is used instead of FWHM to characterize the line width.
\item Improve line-width measurements. There appear to be some
systematic differences among the various data sets, probably due to different
processes for measuring \sigm; for example, panels (e) and (f) of
Figure \ref{Figure:window244_247} show that the SE mass estimates for
the VP06 sample are slightly higher than those from SDSS
(compare also the last two columns in Table~A5). Work on deblending 
alogrithms would aid
more precise measurement of \sigm, in particular.
\end{enumerate}

\section{Summary}
The main results of this paper are:
\begin{enumerate}
\item We confirm that the luminosity of the broad component
of the \hb\ emission line $L(\hb_{\rm broad})$ is an excellent 
substitute for the AGN continuum luminosity $L_{\rm AGN}(5100\,{\rm \AA})$
for predicting the \hb\ emission-line reverberation lag $\tau(\hb)$.
It has the advantage of being easier to isolate than
$L_{\rm AGN}(5100\,{\rm \AA})$, which requires an accurate
estimate of the host-galaxy starlight contribution to the
observed luminosity. 
The fact that there is no statistical penalty for
using $L(\hb)$ as the luminosity measure is, from a
practical point of view, one of the most important 
findings of this work because the high-quality unsaturated
space-based images that are used for host-galaxy
modeling \citep[see][and references therein]{Bentz13}
may not be so easily acquired in the future.
\item We confirm that 
the line dispersion of the \hb\ broad component \sigm(\hb)
and the full-width at half maximum for the 
\hb\ broad component \fwm(\hb) 
in mean, or single-epoch, spectra
are both reasonable proxies
for the line dispersion of \hb\ in the rms spectrum \sigm(\hb)
for computing single-epoch virial products \muse(\hb).
We find that \sigm(\hb) gives better results than \fwm(\hb),
but both are usable.
\item In the case of \civ, we find that 
the line dispersion of the \civ\ emission line \sigm(\civ)
in the mean, or single-epoch, spectrum is a good
proxy for the line dispersion in the rms spectrum
\sigr(\civ) for estimating single-epoch virial products
\muse(\civ). We find that \fwm(\civ), however, does not
track \sigr(\civ) well enough to be used as a proxy.
\item Although the $R$--$L$ relationship based
on the continuum luminosity $L(1350\,{\rm \AA})$ and
\civ\ emission-line reverberation lag $\tau(\civ)$ is not
as well defined as that for \hb, the relationship appears to
have a similar slope and it appears to be suitable 
for estimating virial products \muse(\civ).
\item We confirm for both \hb\ and \civ\ that combining the 
reverberation lag estimated from the luminosity with
a suitable measurement of the emission-line width
together introduces a bias where the high masses
are underestimated and the low masses are overestimated.
We confirm that the parameter that accounts for
the systematic difference between reverberation
virial product measurements \murm\ and those estimated
using only luminosity and line width is Eddington ratio.
Increasing Eddington ratio causes the reverberation
radius to shrink, suggesting a softening of the
hydrogen-ionizing spectrum.
\item While the virial product estimate from combining
luminosity and line width causes a systematic bias,
the relationship between the reverberation virial 
product \murm\ and the single-epoch estimate \muse\
is still a power-law, but with a slope somewhat
less than unity (upper panels of Figure~\ref{Figure:Fit_139_141},
left panel of Figure~\ref{Figure:civvps}).
We are therefore able to empirically
correct this relationship to an unbiased estimator
of \muse\ by fitting the residuals and essentially
rotating the power-law distribution to have a slope of
unity (lower panels of Figure~\ref{Figure:Fit_139_141},
right panel of Figure~\ref{Figure:civvps}).
We present these empirical estimators
for \muse(\hb) and \muse(\civ) in \S{\ref{section:massformulae}}.
On account of its potential utility, we regard 
this as the most important conclusion of
this study.

\end{enumerate}

\acknowledgements

EDB is supported by Padua University through grants DOR1715817/17,
DOR1885254/18, and DOR1935272/19 and by MIUR grant PRIN 2017
20173ML3WW\_001. EDB and BMP are grateful for the hospitality of STScI
early in this investigation.  JVHS and KH acknowledge support from
STFC grant ST/R000824/1.  YS acknowledges support from an Alfred
P.\ Sloan Research Fellowship and NSF grant AST-1715579. CJG, WNB,
JRT, and DPS acknowledge support from NSF grants AST-1517113 and
AST-1516784. KH acknowledges support from STFC grant ST/R000824/1. PBH
acknowledges support from NSERC grant 2017-05983.  YH acknowledges
support from NASA through STScI grant HST-GO-15650. SW, LJ, and LCH
acknowledge support from the National Science Foundation of China
(11721303, 11890693, 11991052) and the National Key R\&D Program of
China (2016YFA0400702, 2016YFA0400703).  MV gratefully acknowledges
support from the Independent Research Fund Denmark via grant number
DFF 8021-00130.

Funding for the Sloan Digital Sky Survey IV has been provided by the
Alfred P.\ Sloan Foundation, the U.S.\ Department of Energy Office of
Science, and the Participating Institutions. SDSS-IV acknowledges
support and resources from the Center for High-Performance Computing
at the University of Utah. The SDSS web site is www.sdss.org. SDSS-IV
is managed by the Astrophysical Research Consortium for the
Participating Institutions of the SDSS Collaboration including the
Brazilian Participation Group, the Carnegie Institution for Science,
Carnegie Mellon University, the Chilean Participation Group, the
French Participation Group, Harvard-Smithsonian Center for
Astrophysics, Instituto de Astrof\'isica de Canarias, The Johns
Hopkins University, Kavli Institute for the Physics and Mathematics of
the Universe (IPMU) / University of Tokyo, the Korean Participation
Group, Lawrence Berkeley National Laboratory, Leibniz Institut f\"ur
Astrophysik Potsdam (AIP), Max-Planck-Institut f\"ur Astronomie (MPIA
Heidelberg), Max-Planck-Institut f\"ur Astrophysik (MPA Garching),
Max-Planck-Institut f\"ur Extraterrestrische Physik (MPE), National
Astronomical Observatories of China, New Mexico State University, New
York University, University of Notre Dame, Observat\'ario Nacional /
MCTI, The Ohio State University, Pennsylvania State University,
Shanghai Astronomical Observatory, United Kingdom Participation Group,
Universidad Nacional Aut\'onoma de M\'exico, University of Arizona,
University of Colorado Boulder, University of Oxford, University of
Portsmouth, University of Utah, University of Virginia, University of
Washington, University of Wisconsin, Vanderbilt University, and Yale
University.

\clearpage

\appendix
\section*{Database of Reverberation-Mapped AGNs}
\setcounter{table}{0}
\renewcommand{\thetable}{A\arabic{table}}

\setcounter{equation}{0}
\renewcommand{\theequation}{A\arabic{equation}}

Reverberation-mapped AGNs provide the fundamental data that anchor the
AGN mass scale.  We selected all AGNs from the
literature (as of 2019 August)
for which unsaturated host-galaxy images acquired with \hst\
are available, since removal of the host-galaxy starlight contribution
to the observed luminosity is critical to this calibration, and
measurements of \hb\ time lags. It is worth noting, however, that
since our analysis shows that the broad \hb\ flux is a useful
proxy for the 5100\,\AA\ continuum luminosity, this criterion
is over-restrictive and we will avoid imposing it in future compilations.
In many cases, there is more than one
reverberation-mapping data set available in the literature. In a few
cases, the more recent data were acquired to replace, say, a more
poorly sampled data set or one for which the initial result was
ambiguous for some reason. In other cases, there are
multiple data sets of comparable quality for individual AGNs, and in
these cases we include them all. The particularly well-studied AGN NGC
5548 has been observed many times and in some sense has served as a
``control'' source that provides our best information about the
repeatability of mass measurements as the continuum and line widths
show long-term (compared to reverberation time scales) variations.

The final reverberation-mapped sample for \hb\ is given in Table~A1. 
It consists of 98 individual time series for 50 individual low-redshift ($z <
0.3$) AGNs. They span a range of AGN luminosity $41.46 \leq \log
L(5100\,{\rm \AA}) \leq 45.81$, in \ergsec.
Luminosities have been corrected for Galactic absorption
using extinction values on the NASA Extragalactic Database, which are
based on the \cite{Schlafly11} recalibration of the \cite{Schlegel98}
dust map. Line-width and time-delay measurements are in the rest-frame
of the AGNs. Luminosity distances are based on redshift, except the
cases noted by \cite{Bentz13}, for which the redshift-independent
distances quoted in that paper are used. For two of these sources, NGC
4051 and NGC 4151, we use preliminary Cepheid-based distances
(M.M.\ Fausnaugh, private communication), and for NGC 6814, we use the
Cepheid-based distance from \cite{Bentz19}.  Individual virial
products for these sources are easily computed using the \hb\ time
lags (Column 6) and line dispersion measurements (Column 12) and the
formula
\begin{equation}
\label{eq:explicitmu}
\mu = 0.1952\left( \frac{\tau(\hb)}{{\rm days}}\right)
\left( \frac{\sigr(\hb)} {{\rm km\,s}^{-1}} \right)^2\,\msun.
\end{equation}

Further conversion to mass requires multiplication by the virial
factor $f$, i.e. $\log M = \log f + \log \mu$, a dimensionless
factor that depends on the inclination, structure, and kinematics of
the broad-\hb-emitting region --- indeed, detailed modeling of 9 of
these objects \citep{Pancoast14,Grier17a} shows that $f$ depends most
clearly on inclination \citep{Grier17a}. Since such models are
available for only a very limited number of AGNs, it is more common to
use a statistical estimate of a mean value of $f$ based on a secondary
mass indicator, specifically the well-known $M_{\rm BH}$--$\sigma_*$
relationship \citep{Ferrarese00,Gebhardt00,Gultekin09}, where
$\sigma_*$ is the host-galaxy stellar bulge velocity dispersion. The
required assumption is that the AGN $M_{\rm BH}$--$\sigma_*$ is
identical to that of quiescent galaxies \citep{Woo13}. In fact, it is
found that the $\mu$--$\sigma_*$ has a slope consistent with the
$M_{\rm BH}$--$\sigma_*$ slope for quiescent galaxies \citep{Grier13},
and the zero points disagree by only a multiplicative factor, which is
taken to be $f$. Here we take $\langle \log f \rangle =
0.683 \pm 0.150$ \citep{Batiste17} where the error on the mean is
$\Delta \log f = 0.030$ --- this error must be propagated into the
mass measurement error when comparing AGN reverberation-based masses
to those based on other methods.

\begin{longrotatetable}
\begin{deluxetable}{lccccccccccc}
\tabletypesize{\footnotesize}
\tablewidth{0pt}
\tablecaption{Reverberation-Mapped AGNs (\hb)}
\label{table:RMDBhb}
\tablehead{
\colhead{Source} &
\colhead{Ref.} &
\colhead{JD Range} &
\colhead{$z$} &
\colhead{$D_L$} &
\colhead{$\tau(\hb)$} &
\colhead{$\log L_{\rm total}(5100)$} &
\colhead{$\log L_{\rm AGN}(5100)$} &
\colhead{$\log L(\hb_{\rm broad})$} &
\colhead{\fwm(\hb)} &
\colhead{\sigm(\hb)}&
\colhead{\sigr(\hb)} \\
\colhead{ } &
\colhead{ } &
\colhead{($-2400000$)}&
\colhead{ } &
\colhead{(Mpc)} &
\colhead{(days)} &
\colhead{(\ergsec)} &
\colhead{(\ergsec)} &
\colhead{(\ergsec)} &
\colhead{(\kms)} &
\colhead{(\kms)} &
\colhead{(\kms)}\\
\colhead{(1)} &
\colhead{(2)} &
\colhead{(3)} &
\colhead{(4)} &
\colhead{(5)} &
\colhead{(6)} &
\colhead{(7)} &
\colhead{(8)} &
\colhead{(9)} &
\colhead{(10)} &
\colhead{(11)} &
\colhead{(12)}
}
\startdata
Mrk335     &1 &  49156-49338&   0.02579&  109.5&   $  16.8^{+4.8}_{-4.2}$&$    43.802 \pm  0.010$ &$  43.703 \pm 0.013$  &$   42.083\pm  0.010$  &$1792     \pm  3 $&$ 1380	\pm 6$ &	$    917  \pm 52$ \\
Mrk335     &1 &  49889-50118&	0.02579&  109.5&   $  12.5^{+6.6}_{-5.5}$&$   43.861 \pm  0.010$ &$  43.777 \pm 0.013$   &$  42.124\pm	0.010$    &$1679	\pm 2  $&$ 1371	\pm 8$ &	$     948	\pm 113 $ \\
Mrk335	   &1 &  55431-55569&	0.02579&  109.5&   $  14.3^{+0.7}_{-0.7}$&$   43.791 \pm  0.007$ &$  43.683 \pm  0.061$  &$  41.940\pm    0.009$ &$1273	\pm 3  $&$1663 \pm 6$ &	$     1293	\pm 64  $ \\
Mrk1501	   &2 &  55430-55568&	0.08934&  402.5&   $  12.6^{+3.9}_{-3.9}$&$   44.314 \pm  0.011$ &$  43.980 \pm  0.053$  &$  42.719\pm    0.015$ &$3106	\pm 15 $&$3494 \pm35$ &	$     3321	\pm 107 $ \\
PG0026+129 &3 &  48545-51084&	0.14200&  653.1&   $  111.0^{+24.1}_{-28.3}$&$ 44.977 \pm  0.010$ &$ 44.911 \pm  0.011$  &$  42.867\pm    0.016$ &$2544	\pm 56 $&$1738 \pm 100$ &$     1773	\pm 285 $ \\
PG0052+251 &3 &  48461-51084&	0.15445&  751.9&   $  89.8^{+24.5}_{-24.1}$&$  44.964 \pm  0.013$ &$ 44.791 \pm  0.020$  &$  43.113\pm    0.016$ &$5008	\pm 73 $&$2167 \pm 30$ &	$     1783	\pm 86  $ \\
Fairall9   &4 &  49475-49743&	0.04702&  202.8&   $  17.4^{+3.2}_{-4.3}$&$   44.224 \pm  0.007$ &$  43.920 \pm  0.026$  &$  42.393\pm    0.007$ &$5999	\pm 60 $&$ 2347\pm 16$ &$     3787	\pm 197 $ \\
Mrk590	   &1 &  48090-48323&	0.02639&  112.1&   $  20.7^{+3.5}_{-2.7}$&$   43.842 \pm  0.010$ &$  43.544 \pm  0.029$  &$  41.855\pm    0.011$ &$2788	\pm 29 $&$ 1942 \pm 26$ &	$     789	\pm 74  $ \\
Mrk590	   &1 &  48848-49048&	0.02639&  112.1&   $  14.0^{+8.5}_{-8.8}$&$   43.666 \pm  0.011$ &$  43.075 \pm  0.073$  &$  41.522\pm    0.011$ &$3729	\pm 426$&$ 2168	\pm 30$ &	$	   1935	\pm 52  $ \\
Mrk590	   &1 &  49183-49338&	0.02639&  112.1&   $  29.2^{+4.9}_{-5.0}$&$   43.743 \pm  0.010$ &$  43.320 \pm  0.043$  &$  41.690\pm    0.010$ &$2743	\pm 79 $&$1967 \pm 19$ &	$     1251	\pm 72  $ \\
Mrk590	   &1 &  49958-50122&	0.02639&  112.1&   $  28.8^{+3.6}_{-4.2}$&$   43.865 \pm  0.010$ &$  43.589 \pm  0.026$  &$  41.857\pm    0.010$ &$2500	\pm 43 $&$ 1880	\pm 19$ &	$     1201	\pm 130 $ \\
3C120	   &1 &  47837-50388&	0.03301&  140.9&   $  38.1^{+21.3}_{-15.3}$&$ 44.078 \pm  0.012$ &$  44.010 \pm  0.014$  &$  42.306\pm    0.012$ &$2327	\pm 48 $&$ 1249\pm21$ &	$     1166	\pm 50  $ \\
3C120	   &5 &  54726-54920&	0.03301&  140.9&   $  27.9^{+7.1}_{-5.9}$&$   44.116 \pm  0.013$ &$  44.094 \pm  0.013$  &$  42.453\pm    0.012$ &$2386	\pm 52 $& $\ldots$ &    $     1689	\pm 68  $ \\
3C120	   &2 &  55430-55569&	0.03301&  140.9&   $  25.9^{+2.3}_{-2.3}$&$   43.993 \pm  0.012$ &$  43.903 \pm  0.052$  &$  42.298\pm    0.015$ &$1430	\pm 16 $&$1687 \pm 4$ &	$     1514	\pm 65  $ \\
Akn120	   &1 &  48148-48344&	0.03271&  139.6&   $  47.1^{+8.3}_{-12.4}$&$  44.254 \pm  0.010$ &$  43.921 \pm  0.032$  &$  42.553\pm    0.010$ &$6042	\pm 35 $&$1753 \pm 6$ &	$     1959	\pm 109 $ \\
Akn120	   &1 &  49980-50175&	0.03271&  139.6&   $  37.1^{+4.8}_{-5.4}$&$   44.131 \pm  0.010$ &$  43.569 \pm  0.067$  &$  42.390\pm    0.010$ &$6246	\pm 78 $&$1862 \pm 13$ &	$     1884	\pm 48  $ \\
MCG+08-11-011&6& 56639-56797&	0.02048&  86.6 &   $  15.72^{+0.50}_{-0.52}$&$ 43.574 \pm  0.009$ &$  43.282 \pm 0.045$  &$  41.706\pm     0.006$&$1159\pm 8$&$1681 \pm 2$ &$     1466	\pm 143 $ \\
Mrk6	   &7 &  49250-49872&	0.01881&  80.6 &   $  21.2^{+4.}_{-3.2}$&$   43.576 \pm   0.009$ &$  43.351 \pm  0.033$  &$  41.591\pm     0.011$&$   \ldots $&$ 2813 \pm 13$ &$	    2836	\pm 48  $ \\
Mrk6	   &7 &  49980-50777&	0.01881&  80.6 &   $  20.7^{+3.0}_{-2.4}$&$   43.578 \pm   0.009$ &$  43.354 \pm 0.033$  &$  41.632\pm     0.010$&$   \ldots $&$2804 \pm 6$ &$     2626	\pm 37  $ \\
Mrk6	   &7 &  50869-51516&	0.01881&  80.6 &   $  20.5^{+5.6}_{-7.0}$&$   43.523 \pm   0.011$ &$  43.258 \pm 0.042$  &$  41.584\pm     0.013$&$   \ldots $&$2808 \pm 14$ &$	     2626	\pm 37  $ \\
Mrk6	   &7 &  51557-53356&	0.01881&  80.6 &   $  23.9^{+17.0}_{-7.3}$&$   43.431 \pm   0.007$ &$ 43.070 \pm 0.058$  &$  41.449\pm     0.018$&$  \dots  $&$2870 \pm 13$ &$	     3222	\pm 39  $ \\
Mrk6	   &7 &  53611-54804&	0.01881&  80.6 &   $  20.4^{+4.6}_{-4.1 }$&$   43.613 \pm   0.005$ &$ 43.413 \pm 0.027$  &$  41.579\pm     0.012$&$   \ldots $&$2807\pm 8$&$	     2864	\pm 35  $ \\
Mrk6	   &2 &  55340-55569&	0.01881&  80.6 &   $  10.1^{+1.1}_{-1.1}$&$   43.719 \pm   0.008$ &$43.507 \pm   0.029$  &$  41.849\pm     0.012$&$2619 \pm 24 $& $4006 \pm 6$ &$     3714	\pm 68  $ \\
Mrk79	   &1 &  47838-48044&	0.02219&  94.0 &   $  9.0^{+8.3}_{-7.8}$&$   43.668 \pm   0.011$ &$ 43.569 \pm   0.014$  &$  41.818\pm     0.011$&$5056	\pm 85 $&$2314 \pm 23$ &	$     2137	\pm 375 $ \\
Mrk79	   &1 &  48193-48393&	0.02219&  94.0 &   $  16.1^{+6.6}_{-6.6}$&$   43.754 \pm  0.010$ &$  43.675 \pm   0.012$  &$  41.851\pm     0.010$&$4760 \pm 31$ & $ 2281 \pm	26$ & $     1683	\pm 72  $ \\
Mrk79	   &1 &  48905-49135&	0.02219&  94.0 &   $  16.0^{+6.4}_{-5.8}$&$   43.695 \pm  0.010$ &$  43.602 \pm   0.013$  &$  41.820\pm     0.010$&$4766 \pm 71$ & $2312 \pm 21$ &$     1854	\pm 72  $ \\
Mrk374	   &6 &  56663-56795&	0.04263&  183.3&   $  14.84^{+5.76}_{-3.30}$&$ 43.994 \pm 0.009$ &$  43.752 \pm  0.036$  &$  41.764\pm     0.013$&$3250 \pm 19 $ &$1490 \pm	4$ &$     1329	\pm 373 $ \\
PG0804+761 &3 &  48319-51085&	0.10000&  447.5&   $  146.9^{+18.8}_{-18.9}$&$ 44.905 \pm  0.011$ &$  44.849 \pm 0.011$  &$  43.230\pm     0.012$&$3053 \pm 38 $&$1434 \pm	18$ &$     1971	\pm 105 $ \\
NGC2617	   &6 &  56639-56797&	0.01421&  59.8 &   $  4.32^{+1.1}_{-1.35}$&$   43.099 \pm 0.011$ &$ 42.610 \pm   0.096$  &$  41.173\pm     0.012$&$5303 \pm 48$&$2709 \pm 6$ &$     2424	\pm 89  $ \\
Mrk704	   &8 &  55932-55980&	0.02923&  124.5&   $  12.65^{+1.49}_{-2.14}$&$ 43.708 \pm 0.005$ &$ 43.517 \pm   0.025$  &$  41.800\pm     0.007$&$3502 \pm 31$ &$ 2650 \pm	4$ & $     1860	\pm 120 $ \\
Mrk110	   &1 &  48953-49149&	0.03529&  150.9&   $  24.3^{+5.5}_{-8.3}$&$   43.711 \pm   0.011$ &$43.618 \pm   0.014$  &$  42.055\pm     0.011$&$1543 \pm 5$&$962 \pm 15$ &$     1196	\pm 141 $ \\
Mrk110	   &1 &  49751-49874&	0.03529&  150.9&   $  20.4^{+10.5}_{-6.3}$&$   43.771 \pm   0.010$ &$43.691 \pm  0.012$  &$  41.960\pm     0.010$&$1658 \pm 3 $ & $953	\pm 10$  &$     1115	\pm 103 $ \\
Mrk110	   &1 &  50010-50262&	0.03529&  150.9&   $  33.3^{+14.9}_{-10.0}$&$ 43.594 \pm  0.012$ &$  43.468 \pm  0.017$  &$  41.905\pm     0.012$&$1600 \pm 39$&$987 \pm 18$ &$     755	\pm 29  $ \\
Mrk110	   &9 &  51495-51678&	0.03529&  150.9&   $  23.4^{+3.6}_{-3.2}$&$  43.340 \pm  0.007$ &$  43.225 \pm   0.011$  &$  41.769\pm     0.007$&$ \ldots $ & $\ldots$ &$     \ldots$  \\
PG0953+414 &3 &  48319-50997&	0.23410&  1137.2&  $  150.1^{+21.6}_{-22.6}$&$ 45.193 \pm  .010$ &$  45.126 \pm  0.011$  &$  43.390\pm     0.012$&$3071 \pm 27$ &$ 1659 \pm	31$ &$     1306	\pm 144 $ \\
NGC3227	   &10&  54184-54269&	0.00386&  23.7 &   $  3.75^{+0.76}_{-0.82}$&$ 42.629 \pm 0.035$ &$  42.243 \pm   0.068$  &$  40.387\pm     0.035$&$3972 \pm 25$ &$ 1749 \pm	4$ &$     1376	\pm 44  $ \\
NGC3227	   &8 &  55933-56048&	0.00386&  23.7 &   $  1.29^{+1.56}_{-1.27 }$&$42.757 \pm  0.006$ &$  42.424 \pm  0.051$  &$  40.487\pm     0.010$&$1602 \pm 2$ &$1402 \pm 2$ &$     1368	\pm 38  $ \\
Mrk142	   &11&  54506-54618&	0.04494&  193.5&   $  2.74^{+0.73}_{-0.83}$&$   43.709 \pm 0.010$ &$  43.543 \pm 0.015$  &$  41.639\pm     0.010$&$1462 \pm 2 $ &$ 1116 \pm	22$ &$     859	\pm 102 $ \\
Mrk142	   &12&  56237-56413&	0.04494&  193.5&   $  6.4^{+0.8}_{-2.2}$&$   43.610 \pm  0.010$ &$  43.443 \pm   0.016$  &$  41.586\pm     0.010$&$1647	\pm 69 $& $\ldots$ & $     \ldots  $ \\
NGC3516 &14,15&  54181-54300&	0.00884&  37.1 &   $  11.68^{+1.02}_{-1.53}$&$43.299 \pm  0.055$ &$  42.726 \pm  0.133$  &$  40.995\pm     0.057$&$5236 \pm 12$&$ 1584 \pm	1$  &$     1591	\pm 10  $ \\
NGC3516	   &8 &  55932-56072&	0.00884&  37.1 &   $  5.74^{+2.26}_{-2.04}$&$ 43.272 \pm  0.007$ &$  42.529 \pm  0.196$  &$  41.022\pm     0.008$&$3231 \pm 14 $ & $ 2633 \pm 3  $&$     2448	\pm 69  $ \\
SBS1116+583A&11& 54550-54618&	0.02787&  118.5&   $  2.31^{+0.62}_{-0.49 }$&$ 42.995 \pm  0.021$ &$  42.076 \pm 0.224$  &$  40.788\pm     0.015$&$3668 \pm 186$ &$ 1552 \pm 36$ &$	1528	\pm 184 $ \\
Arp151	&11,13&  54506-54618&	0.02109&  89.2 &   $  3.99^{+0.49}_{-0.68}$&$ 42.979 \pm 0.010$ &$  42.497 \pm   0.047$  &$  40.931\pm     0.011$&$3098 \pm 69$&$2006 \pm 24$ &$     1252	\pm 46  $ \\
NGC3783	&14,15&  48607-48833&	0.00973&  25.1 &   $  10.2^{+3.3}_{-2.3}$&$   42.791 \pm 0.025$ &$  42.559 \pm   0.051$  &$  41.009\pm     0.021$&$3770 \pm 68$&$1691 \pm 19  $&$     1753	\pm 141 $ \\
Mrk1310	   &11&  54550-54618&	0.01956&  82.7 &   $  3.66^{+0.59}_{-0.61 }$&$ 42.937 \pm 0.018$ &$  42.231 \pm  0.120$  &$  40.646\pm     0.012$&$2409 \pm 24$&$ 1209 \pm	42 $&$     755	\pm 138 $ \\
NGC4051	   &16&  54180-54311&	0.00234&  15.0 &   $  1.87^{+0.54}_{-0.50}$&$42.290 \pm   0.015$ &$  41.847 \pm  0.080$  &$  40.079\pm     0.018$&$799 \pm 2 $ & $ 1045 \pm	4$ &$     927	\pm 64  $ \\
NGC4051	   &6 &  56645-56864&	0.00234&  15.0 &   $  2.87^{+0.86}_{-1.33}$&$42.265 \pm   0.005$ &$  41.732 \pm  0.106$  &$  39.882\pm     0.012$&$765 \pm 3 $ &$470 \pm 2  $&$     493	\pm 35  $ \\
NGC4151	   &17&  53430-53472&	0.00332&  15.0 &   $  6.59^{+1.12}_{-0.76}$&$42.549 \pm   0.012$ &$  42.004 \pm  0.113$  &$  40.499\pm     0.013$&$5840 \pm 863 $&$ 6158 \pm	47 $&$	   2680	\pm 64  $ \\
NGC4151	   &6 &  55931-56072&	0.00332&  15.0 &   $  6.82^{+0.48}_{-0.57}$&$42.685 \pm   0.007$ &$  42.315 \pm  0.060$  &$  40.956\pm     0.008$&$992 \pm 4 $& $1833 \pm  2  $&$     1894	\pm 9   $ \\
Mrk202	   &11&  54550-54617&	0.02102&  88.9 &   $  3.05^{+1.73}_{-1.12}$&$42.946 \pm   0.016$ &$  42.198 \pm  0.126$  &$  40.477\pm     0.010$&$1471 \pm 18 $&$ 867 \pm	40 $&$     659	\pm 65  $ \\
NGC4253	   &11&  54509-54618&	0.01293&  54.4 &   $  6.16^{+1.63}_{-1.22}$&$42.948 \pm   0.012$ &$  42.509 \pm  0.044$  &$  40.873\pm     0.010$&$1609 \pm 39 $&$ 1088 \pm	37 $&$     \ldots  $ \\
PG1226+023 &3 &  48361-50997&	0.15834&  737.7&   $  306.80^{+68.5}_{-90.9}$&$45.935\pm  0.011$ &$  45.907 \pm  0.011$  &$  44.072\pm     0.014$&$3509 \pm 36$ &$ 1778 \pm 17 $&$     1777	\pm 150 $ \\
3C273      &18&  54795-58194&   0.15834&  737.7&   $146.3^{+8.3}_{-12.1}$& $45.864 \pm 0.011$ &      $45.848 \pm 0.011$  &$44.056 \pm 0.010$&$3256 \pm 36$ & $1701 \pm	15$ &$1090 \pm 121 $ \\
PG1229+204 &3 &  48319-50997&	0.06301&  274.9&   $  37.8^{+27.6}_{-15.3}$&$44.053 \pm   0.010$ &$  43.636 \pm  0.040$  &$  42.275\pm     0.011$&$3828 \pm 54 $& $1608 \pm 24 $&$     1385	\pm 111 $ \\
NGC4593	   &19&  53391-53580&	0.00900&  37.7 &   $  3.73^{+0.75}_{-0.75}$&$43.242 \pm   0.013$ &$  43.005 \pm  0.035$  &$  41.237\pm     0.013$&$5143 \pm 16 $&$1790 \pm	3 $&$     1561	\pm 55  $ \\
NGC4748	   &11&  54550-54618&	0.01463&  61.6 &   $  5.55^{+1.62}_{-2.22}$&$43.072 \pm   0.012$ &$  42.557 \pm  0.060$  &$  41.047\pm     0.010$&$1947 \pm 66$ & $ 1009 \pm 27$ &$     657	\pm 91  $ \\
PG1307+085 &3 &  48319-51042&	0.15500&  718.7&   $  105.6^{+36.0}_{-46.6}$&$44.849 \pm   0.012$ &$  44.790 \pm 0.013$  &$  43.096\pm     0.020$&$5059 \pm 133 $ & $ 1963 \pm 47 $&$	   1820	\pm 122 $ \\
MCG-06-30-15&20& 55988-56079&	0.00775&  25.5 &   $  5.33^{+1.86}_{-1.75}$&$42.393 \pm   0.009$ &$  41.651 \pm  0.197$  &$  39.793\pm     0.011$&$1958 \pm 75 $ &$ 976 \pm	8$ &$     665	\pm 87  $ \\
NGC5273	    &21&  56774-56838&	0.00362&  15.3 &   $  2.21^{+1.19}_{-1.60}$&$42.000 \pm   0.009$ &$  41.465 \pm  0.106$  &$  39.702\pm     0.010$&$5688 \pm 163 $&$ 1821 \pm	53 $&$	   1544	\pm 98  $ \\
Mrk279	    &22&  50095-50289&	0.03045&  129.7&   $  16.7^{+3.9}_{-3.9}$&$43.882 \pm   0.021$ &$  43.643 \pm    0.036$  &$  42.242\pm     0.021$&$5354 \pm 32$&$ 1823 \pm 11 $&$     1420	\pm 96  $ \\
PG1411+442  &3&  48319-51038&	0.08960&  398.2&   $  124.3^{+61.0}_{-61.7}$&$44.603 \pm   0.012$ &$  44.502 \pm 0.014$  &$  42.792\pm     0.014$&$2801 \pm 43 $&$1774 \pm	29$ &$     1607	\pm 169 $ \\
NGC5548&23,24,25&47509-47809&	0.01718&  72.5 &   $  19.7^{+1.5}_{-1.5}$&$ 43.534 \pm   0.021$ &$  43.328 \pm   0.042$  &$  41.728\pm     0.018$&$4674 \pm 63$&$ 1934 \pm 5 $&$     1687	\pm 56  $ \\
NGC5548 &24,25&  47861-48179&	0.01718&  72.5 &   $  18.6^{+2.1}_{-2.3}$&$ 43.390 \pm   0.029$ &$  43.066 \pm   0.068$  &$  41.546\pm     0.029$&$5418 \pm 107$ &$ 2227 \pm 20 $&$	   1882	\pm 83  $ \\
NGC5548	&24,26&  48225-48534&	0.01718&  72.5 &   $  15.9^{+2.9}_{-2.5}$&$ 43.496 \pm   0.017$ &$  43.264 \pm   0.042$  &$  41.645\pm     0.026$&$5236 \pm 87$ &$ 2205 \pm 16 $&$     2075	\pm 81  $ \\
NGC5548	&24,26&  48623-48898&	0.01718&  72.5 &   $  11.0^{+1.9}_{-2.0}$&$ 43.360 \pm   0.020$ &$  42.999 \pm   0.070$  &$  41.457\pm     0.030$&$5986 \pm 95$&$ 3110 \pm 53 $&$     2264	\pm 88  $ \\
NGC5548	&24,27&  48954-49255&	0.01718&  72.5 &   $  13.0^{+1.6}_{-1.4}$&$ 43.497 \pm   0.016$ &$  43.267 \pm   0.040$  &$  41.691\pm     0.016$&$5930 \pm 42$&$ 2486 \pm 13 $&$     1909	\pm 129 $ \\
NGC5548	&24,28&  49309-49636&	0.01718&  72.5 &   $  13.4^{+3.8}_{-4.3}$&$ 43.509 \pm   0.022$ &$  43.287 \pm   0.043$  &$  41.649\pm     0.022$&$7378 \pm 39 $&$ 2877 \pm 17 $&$     2895	\pm 114 $ \\
NGC5548	&24,28&  49679-50008&	0.01718&  72.5 &   $  21.7^{+2.6}_{-2.6}$&$ 43.604 \pm   0.012$ &$  43.436 \pm   0.026$  &$  41.746\pm     0.013$&$6946 \pm 79$&$ 2432 \pm 13 $&$     2247	\pm 134 $ \\
NGC5548	&24,28&  50044-50373&	0.01718&  72.5 &   $  16.4^{+1.2}_{-1.1}$&$ 43.527 \pm   0.020$ &$  43.317 \pm   0.039$  &$  41.656\pm     0.018$&$6623 \pm 93 $&$ 2276 \pm 15 $&$     2026	\pm 68  $ \\
NGC5548	&24,29&  50434-50729&	0.01718&  72.5 &   $  17.5^{+2.0}_{-1.6}$&$ 43.413 \pm   0.018$ &$  43.113 \pm   0.054$  &$  41.622\pm     0.015$&$6298 \pm 65 $&$ 2178 \pm	12$ &$     1923	\pm 62  $ \\
NGC5548	&24,29&  50775-51085&	0.01718&  72.5 &   $  26.5^{+4.3}_{-2.2}$&$ 43.620 \pm   0.020$ &$  43.459 \pm   0.032$  &$  41.762\pm     0.018$&$6177 \pm 36 $&$ 2035 \pm 11 $&$     1732	\pm 76  $ \\
NGC5548	&24,29&  51142-51456&	0.01718&  72.5 &   $  24.8^{+3.2}_{-3.0}$&$ 43.565 \pm   0.017$ &$  43.376 \pm   0.034$  &$  41.719\pm     0.016$&$6247 \pm 57 $&$ 2021 \pm 18 $&$     1980	\pm 30  $ \\
NGC5548	&24,29&  51517-51791&	0.01718&  72.5 &   $  6.5^{+5.7}_{-3.7}$&$  43.327 \pm   0.019$ &$  42.918 \pm   0.081$  &$  41.521\pm     0.017$&$6240	\pm 77 $&$2010 \pm 30 $&	$     1969	\pm 48  $ \\
NGC5548	&24,29&  51878-52174&	0.01718&  72.5 &   $  14.3^{+5.9}_{-7.3}$&$ 43.321 \pm   0.027$ &$  42.903 \pm   0.089$  &$  41.428\pm     0.026$&$6478 \pm 108 $&$ 3111 \pm 131 $&$	   2173	\pm 89  $ \\
NGC5548	&24,30&  53432-53472&	0.01718&  72.5 &   $  6.3^{+2.6}_{-2.3}$&$  43.263 \pm   0.016$ &$  42.526 \pm   0.211$  &$  40.967\pm     0.017$&$6396	\pm 167$&$3210 \pm 642$ & $	   2388	\pm 373 $ \\
NGC5548	&10,24&  54180-54332&	0.01718&  72.5 &   $  12.4^{+2.7}_{-3.9}$&$ 43.287 \pm   0.008$ &$  42.665 \pm   0.140$  &$  40.660\pm     0.070$&$12575\pm 47$&$4736 \pm	23 $&$     1822	\pm 35  $ \\
NGC5548	&11,24&  54508-54618&	0.01718&  72.5 &   $  4.18^{+0.86}_{-1.30}$&$43.214 \pm  0.010$ &$  42.621 \pm   0.129$  &$  41.157\pm     0.017$&$12771 \pm 71$& $ 4266 \pm 65 $&$     4270	\pm 292 $ \\
NGC5548	&8,24 &  55931-56072&	0.01718&  72.5 &   $  2.83^{+0.88}_{-0.90}$&$43.433 \pm  0.005$ &$  43.070 \pm   0.058$  &$  41.543\pm     0.010$&$10587 \pm 82$&$ 3056 \pm	4 $&$     2772	\pm 34  $ \\
NGC5548	  &31 &  56663-56875&	0.01718&  72.5 &   $  4.17^{+0.36}_{-0.36}$&$43.612 \pm  0.003$ &$  43.404 \pm   0.027$  &$  41.666\pm     0.004$&$9496 \pm 418$&$ 3691 \pm	162 $&$	   4278	\pm 671 $ \\
NGC5548	   &32&  57030-57236&	0.01718&  72.5 &   $  7.18^{+1.38}_{-0.70}$&$43.175 \pm  0.005$ &$  42.787 \pm   0.063$  &$  41.630\pm     0.003$&$9912 \pm 362 $&$ 3350 \pm	272 $&$	   3124	\pm 302 $ \\
PG1426+015 &3 &  48334-51042&	0.08657&  383.9&   $  95.0^{+29.9}_{-37.1}$&$44.690 \pm  0.012$ &$  44.568 \pm   0.019$  &$  42.764\pm     0.015$&$7113	 \pm 160 $&$ 2906 \pm 80$&$	   3442	\pm 308 $ \\
Mrk817	   &1 &  49000-49212&	0.03146&  134.2&   $  19.0^{+3.9}_{-3.7}$&$ 43.848 \pm   0.010$ &$  43.726 \pm   0.015$  &$  42.010\pm     0.010$&$4711 \pm 78$&$ 1984\pm 8 $&$     1392	\pm 78  $ \\
Mrk817	   &1 &  49404-49528&	0.03146&  134.2&   $  15.3^{+3.7}_{-3.5}$&$ 43.761 \pm   0.087$ &$  43.608 \pm   0.124$  &$  41.936\pm     0.089$&$5237 \pm 67 $&$ 2098 \pm 13 $&$     1971	\pm 96  $ \\
Mrk817	   &1 &  49752-49924&	0.03146&  134.2&   $  33.6^{+6.5}_{-7.6}$&$ 43.762 \pm   0.009$ &$  43.609 \pm   0.016$  &$  41.860\pm     0.010$&$4767 \pm 72 $&$ 2195 \pm 16 $&$     1729	\pm 158 $ \\
Mrk817	   &10&  54185-54301&	0.03146&  134.2&   $  14.04^{+3.41}_{-3.47}$&$43.901 \pm 0.006$ &$   43.776 \pm  0.010$  &$  41.710\pm     0.016$&$5906 \pm 34$&$2365 \pm 9 $&$     2025	\pm 5   $ \\
Mrk290	   &10&  54180-54321&	0.02958&  126.0&   $  8.72^{+1.21}_{-1.02}$&$ 43.451 \pm   0.028$ &$  43.157 \pm 0.036$  &$  41.747\pm     0.030$&$4521 \pm 24 $&$ 2071 \pm	24 $&$     1609	\pm 47  $ \\
PG1613+658 &3 &  48397-51073&	0.12900&  588.4&   $  40.1^{+15.0}_{-15.2}$&$ 44.948 \pm   0.010$ &$  44.713 \pm 0.019$  &$  42.943\pm     0.014$&$9074 \pm 103 $&$ 3084 \pm	33$&$	   2547	\pm 342 $ \\
PG1617+175 &3 &  48362-51085&	0.11244&  507.4&   $  71.5^{+29.6}_{-33.7}$&$ 44.445 \pm   0.011$ &$  44.330 \pm 0.014$  &$  42.682\pm     0.023$&$6641 \pm 190 $&$ 2313 \pm 69$&$	   2626	\pm 211 $ \\
PG1700+518 &3 &  48378-51084&	0.29200&  1463.3&  $  251.8^{+45.9}_{-38.8}$&$45.600 \pm   0.010$ &$  45.528 \pm 0.011$  &$  43.717\pm     0.020$&$2252\pm 85$&$ 3160 \pm 93 $&$     1700	\pm 123 $ \\
3C382	   &6 &  56679-56864&	0.05787&  251.5&   $  40.49^{+8.02}_{-3.74}$&$44.193 \pm   0.008$ &$  43.792 \pm 0.069$  &$  42.264\pm     0.011$&$3619\pm 203 $&$ 3227 \pm	7$&$	   4552	\pm 190 $ \\
3C390.3	   &33&  49718-50012&	0.05610&  243.5&   $  23.60^{+6.2}_{-6.7}$&$43.902 \pm   0.018$ &$  43.620 \pm   0.039$  &$  42.222\pm     0.015$&$12694 \pm 13 $&$ 3744 \pm	42$ &$     3105	\pm 81  $ \\
3C390.3	   &34&  50100-54300&	0.05610&  243.5&   $  97.0^{+17.0}_{-17.0}$&$44.028 \pm  0.016$ &$  43.913 \pm   0.020$  &$  42.287\pm     0.021$&$11918\pm 325$& $\ldots $&$	   \ldots  $ \\
3C390.3	   &35&  53631-53714&	0.05610&  243.5&   $  46.4^{+3.8}_{-3.2}$&$   44.485 \pm 0.007$ &$  44.434 \pm   0.008$  &$  42.695\pm     0.012$&$13211\pm 278 $&$ 5377 \pm	37 $&$     5455	\pm 278 $ \\
NGC6814	   &11&  54545-54618&	0.00521&  21.6 &   $  6.64^{+0.87}_{-0.90}$&$ 42.500 \pm 0.017$ &$  42.058 \pm   0.057$  &$  40.443\pm     0.010$&$3323 \pm 7 $&$1918 \pm 36 $&$     1610	\pm 108 $ \\
Mrk509	   &1 &  47653-50374&	0.03440&  147.0&   $  79.6^{+6.1}_{-5.4}$&$ 44.240 \pm   0.027$ &$  44.130 \pm   0.028$  &$  42.545\pm     0.027$&$3015 \pm 2 $&$ 1555	\pm 7  $&$     1276	\pm 28  $ \\
PG2130+099 &36&  54352-54450&	0.06298&  274.7&   $  22.9^{+4.7}_{-4.6}$&$ 44.406 \pm   0.012$ &$  44.368 \pm   0.012$  &$  42.667\pm     0.011$&$2853 \pm 39 $&$ 1485 \pm 15 $&$     1246	\pm 222 $ \\
PG2130+099 &2 &  55430-55557&	0.06298&  274.7&   $  9.6^{+1.2}_{-1.2}$&$  44.237 \pm   0.032$ &$  44.150 \pm   0.033$  &$  42.584\pm     0.033$&$1781	\pm 5  $&$ 1769\pm 2	$ &$    1825	\pm 65  $ \\
NGC7469	   &37&  55430-55568&	0.01632&  68.8 &   $  10.8^{+3.4}_{-1.3}$&$ 43.768 \pm   0.009$ &$  43.444 \pm   0.051$  &$  41.557\pm     0.013$&$4369 \pm 6 $&$ 1095	\pm 5 $&$     1274	\pm 126 $
\enddata
\tablecomments{
Columns are
1: AGN name;
2: literature reference for data;
3: Julian Dates of observations;
4: redshift;
5: luminosity distance;
6: \hb\ time lag;
7: log total luminosity at 5100\,\AA;
8: log AGN luminosity at 5100\,\AA;
9: log \hb\ broad-line component luminosity;
10: FWHM of \hb\ broad component in mean spectrum;
11: line dispersion of \hb\ broad component in mean spectrum;
12: line dispersion of \hb\ broad component in rms spectrum.
}
\tablerefs{
1: \cite{Peterson98a};
2: \cite{Grier12};
3: \cite{Kaspi00};
4: \cite{Santos97};
5: \cite{Kollatschny14};
6: \cite{Fausnaugh17};
7: \cite{Doroshenko12};
8: \cite{DeRosa18};
9: \cite{Kollatschny01};
10: \cite{Denney10};
11: \cite{Bentz09b};
12: \cite{Du14};
13: \cite{Bentz08};
14: \cite{Stirpe94};
15: \cite{Onken02};
16: \cite{Denney09b};
17: \cite{Bentz06a};
18: \cite{Zhang19};
19: \cite{Denney06};
20: \cite{Bentz16};
21: \cite{Bentz14};
22: \cite{Santos01};
23: \cite{Peterson91};
24: \cite{Peterson13};
25: \cite{Peterson92}
26: \cite{Peterson94};
27: \cite{Korista95};
28: \cite{Peterson99};
29: \cite{Peterson02};
30: \cite{Bentz07};
31: \cite{Pei17};
32: \cite{Lu16};
33: \cite{Dietrich98};
34: \cite{Shapovalova10};
35: \cite{Dietrich12};
36: \cite{Grier08};
37: \cite{Peterson7469}
}
\label{table:hbrm}
\end{deluxetable}
\end{longrotatetable}

\begin{longrotatetable}
\begin{deluxetable}{lccccccccc}
\tabletypesize{\footnotesize}
\tablewidth{0pt}
\tablecaption{Reverberation-Mapped AGNs (\civ)}
\tablehead{
\colhead{Source} &
\colhead{Ref.} &
\colhead{JD Range} &
\colhead{$z$} &
\colhead{$D_L$} &
\colhead{$\tau(\civ)$} &
\colhead{$\log L(1350)$} &
\colhead{$\fwm(\civ)$} &
\colhead{$\sigm(\civ)$} &
\colhead{$\sigr(\civ)$} \\
\colhead{ } &
\colhead{ } &
\colhead{($-2400000$)}&
\colhead{ } &
\colhead{(Mpc)} &
\colhead{(days)} &
\colhead{(\ergsec)} &
\colhead{(\kms)} &
\colhead{(\kms)} &
\colhead{(\kms)} \\
\colhead{(1)} &
\colhead{(2)} &
\colhead{(3)} &
\colhead{(4)} &
\colhead{(5)} &
\colhead{(6)} &
\colhead{(7)} &
\colhead{(8)} &
\colhead{(9)} &
\colhead{(10)}
}
\startdata
DESJ003-42&1 &56919-57627 & 2.593 & 20723 & $123^{+43}_{-42}$ &$46.510 \pm 0.020$& $4944 \pm 93$ & $3917 \pm 29 $& $6250 \pm 64$ \\
Fairall9 &2,3 & 49473-49713 &0.04702&202.8&$29.6^{+12.9}_{-14.4}$ &$44.530 \pm 0.030$ &$2968 \pm 37 $& $3068 \pm 27$  & $3201 \pm 285 $ \\
DESJ228-04&1&56919-57627 & 1.905 & 1686.4 &$95^{+16}_{-23}$ & $46.430 \pm 0.098$ & $5232 \pm 57$ &$3932 \pm 22$ & $6365 \pm 66$  \\
CT286    &4    &54821-57759& 2.556 & 20,366 &$459^{+71}_{-92}$ & $46.798 \pm 0.009$ & $6256$ & $\ldots$ &$\ldots$ \\
CT406    &4    &54355-57605& 3.183 & 26,533 &$115^{+64}_{-86}$&$46.910 \pm0.040$ & $6236$ &$\ldots$ &$\ldots$ \\
NGC3783	 &5,3   & 48611-48833 &0.00973&25.1   &$3.8^{+1.0}_{-0.9}$ &$43.081 \pm  0.017$ &$2784 \pm 24$& $2476 \pm 18$ &$2948 \pm 160 $ \\
NGC4151	 &6,7   & 47494-47556 &0.00332&15.0  &$3.44^{+1.42}_{-1.24}$ &$42.412 \pm 0.016$ &$2929 \pm 154$&$4922 \pm 51$  &$5426 \pm 196 $ \\
NGC4395	 &8   & 53106  &0.00106&4.0 &$0.033^{+0.017}_{-0.013}$  &$39.494 \pm 0.007$ &$1214 \pm 2$&$1727 \pm 78$ &$3025 \pm 201 $ \\
NGC4395	 &8     & 53190  &0.00106&4.0 & $0.046^{+0.017}_{-0.013}$ &$40.030 \pm 0.012$ &$1532 \pm 6$&$1662\pm34$ &$2859 \pm 376 $ \\
NGC5548	 &9,3   & 47510-47745 &0.01718&72.5  &$9.8^{+1.9}_{-1.5}$  &$43.635 \pm 0.016$ &$5248 \pm 428$&$4351 \pm 37$& $3842 \pm 210 $ \\
NGC5548	 &10,3   & 49060-49135&0.01718&72.5  &$6.7^{+0.9}_{-1.0}$  &$43.552 \pm 0.007$ &$4201 \pm 101$&$3738\pm17$ & $3328 \pm 104 $ \\
NGC5548	 &11    & 56690-56866&0.01718&72.5  &$5.8^{+0.5}_{-0.5}$  &$43.625 \pm 0.007$ &$5236 \pm 87 $&$2205 \pm16$ &$2075 \pm 81  $ \\
3C390.3	 &12,3 & 49718-50147 &0.05610&243.5&$35.7^{+11.4}_{-14.6}$&$44.013 \pm 0.045$ &$6180 \pm 638$&$4578 \pm 65$ &$4400 \pm 186 $ \\
J214355  &4   & 54729-57605& 2.620&20,985& $128^{+91}_{-82}$& $46.962 \pm 0.048$ & $6895$ & $\ldots$ &$\ldots$ \\
J221516  &4   & 54232-57689& 2.706&21821& $165^{+98}_{-13}$ & $47.155 \pm 0.057$& $5888$ & $\ldots$ &$\ldots$ \\
NGC7469	 &13,3 & 50245-50293&0.01632&68.8  &$2.5^{+0.3}_{-0.3}$   &$43.719 \pm 0.016$ &$3112 \pm 54 $&$3650 \pm 27$ &$2619 \pm 118 $
\enddata
\tablecomments{
Columns are
1: AGN name;
2: literature reference for data;
3: Julian Dates of observations;
4: redshift;
5: luminosity distance;
6: \civ\ time lag $\tau(\civ)$;
7: log continuum luminosity at 1350\,\AA;
8: FWHM of \civ\ in the mean spectrum;
9: line dispersion of \civ\ in the mean spectrum;
10: line dispersion of \civ\ in the rms spectrum.
}
\tablerefs{
1: \cite{Hoormann19};
2: \cite{Rod97};
3: \cite{Peterson04};
4: \cite{Lira18};
5: \cite{Reichert94};
6: \cite{Clavel90};
7: \cite{Metzroth06};
8: \cite{Peterson05};
9: \cite{Clavel91};
10: \cite{Korista95};
11: \cite{DeRosa15};
12: \cite{Obrien98};
13: \cite{Wanders97}.
}
\label{table:civrm}
\end{deluxetable}
\end{longrotatetable}

\clearpage

\begin{longrotatetable}
\begin{deluxetable}{lcccccccc}
\tabletypesize{\footnotesize}
\tablewidth{0pt}
\tablecaption{Reverberation-Mapped AGNs (SDSS \hb)}
\tablehead{
\colhead{RMID} &
\colhead{$z$} &
\colhead{$D_L$} &
\colhead{$\tau(\hb)$} &
\colhead{$\log L(5100)$} &
\colhead{$\log L(\hb_{\rm broad})$} &
\colhead{$\fwm(\hb)$} &
\colhead{$\sigm(\hb)$} &
\colhead{$\sigr(\hb)$} \\
\colhead{ } &
\colhead{ } &
\colhead{(Mpc)} &
\colhead{(days)} &
\colhead{(\ergsec)} &
\colhead{(\ergsec)} &
\colhead{(\kms)}&
\colhead{(\kms)}&
\colhead{(\kms)} \\
\colhead{(1)} &
\colhead{(2)} &
\colhead{(3)} &
\colhead{(4)} &
\colhead{(5)} &
\colhead{(6)} &
\colhead{(7)} &
\colhead{(8)} &
\colhead{(9)} }
\startdata
16    & $0.848$& 5240.9 &  $32.0^{+11.6	}_{-15.5}$ & $44.7779 \pm 0.0012$ & $43.0718 \pm 0.0600$ & $7042 \pm 43  $ & $4804 \pm 41 $ & $6477 \pm 54$ \\
17    &	$0.456$& 2466.9 &  $25.5^{+10.9	}_{-5.8 }$ & $44.3552 \pm 0.0005$ & $42.1756 \pm 0.0064$ & $7847 \pm 203 $ & $4295 \pm 47 $ & $6101 \pm 48$ \\
101   &	$0.458$& 2479.8 &  $21.4^{+4.2	}_{-6.4 }$ & $44.3758 \pm 0.0005$ & $42.7316 \pm 0.0449$ & $2207 \pm 7   $ & $1178 \pm 5  $ & $976  \pm 32$ \\
160   &	$0.359$& 1859.7 &  $21.9^{+4.2	}_{-2.4 }$ & $43.7613 \pm 0.0009$ & $42.0456 \pm 0.0047$ & $3988 \pm 23  $ & $2914 \pm 36 $ & $1909 \pm 12$ \\
177   &	$0.482$& 2635.8 &  $10.1^{+12.5	}_{-2.7 }$ & $44.1735 \pm 0.0009$ & $42.2813 \pm 0.0125$ & $4808 \pm 32  $ & $2224 \pm 32 $ & $2036 \pm 39$ \\
191   &	$0.442$& 2377.0 &  $8.5	^{+2.5	}_{-1.4 }$ & $43.9111 \pm 0.0015$ & $41.7344 \pm 0.0131$ & $2023 \pm 32  $ & $1078 \pm 79 $ & $1030 \pm 18$ \\
229   &	$0.47 $& 2557.5 &  $16.2^{+2.9	}_{-4.5 }$ & $43.8259 \pm 0.0017$ & $41.9083 \pm 0.0166$ & $3089 \pm 261 $ & $2178 \pm 156$ & $1781 \pm 38$ \\
265   &	$0.734$& 4388.8 &  $8.5	^{+3.2	}_{-3.9 }$ & $44.3809 \pm 0.0019$ & $42.4400 \pm 0.0273$ & $3655 \pm 323 $ & $2526 \pm 55 $ & $7165 \pm 36$ \\
267   &	$0.587$& 3342.0 &  $20.4^{+2.5	}_{-2.0 }$ & $44.3013 \pm 0.0008$ & $42.5166 \pm 0.0237$ & $2395 \pm 23  $ & $1229 \pm 32 $ & $1202 \pm 33$ \\
272   &	$0.263$& 1298.0 &  $15.1^{+3.2	}_{-4.6 }$ & $43.9119 \pm 0.0009$ & $42.3449 \pm 0.0017$ & $2595 \pm 10  $ & $1590 \pm 5  $ & $1697 \pm 10$ \\
300   &	$0.646$& 3754.6 &  $30.4^{+3.9	}_{-8.3 }$ & $44.6130 \pm 0.0008$ & $42.5889 \pm 0.0379$ & $2376 \pm 33  $ & $1303 \pm 29 $ & $1232 \pm 30$ \\
305   &	$0.527$& 2933.9 &  $53.5^{+4.2	}_{-4.0  }$ & $44.2995 \pm 0.0008$ & $42.5025 \pm 0.0365$ & $2208 \pm 28  $ & $1647 \pm 20 $ & $2126 \pm 35$ \\
316   &	$0.676$& 3968.3 &  $11.9^{+1.3	}_{-1.0  }$ & $44.9958 \pm 0.0004$ & $43.4279 \pm 0.0020$ & $2988 \pm 10  $ & $1884 \pm 5  $ & $7195 \pm 40$ \\
320   &	$0.265$& 1309.4 &  $25.2^{+4.7	}_{-5.7 }$ & $43.6876 \pm 0.0010$ & $41.8663 \pm 0.0096$ & $4061 \pm 26  $ & $3110 \pm 37 $ & $1462 \pm 26$ \\
371   &	$0.472$& 2570.5 &  $13	^{+1.4	}_{-0.8 }$ & $44.0638 \pm 0.0009$ & $42.3726 \pm 0.0086$ & $3506 \pm 26  $ & $1682 \pm 18 $ & $1443 \pm 11$ \\
373   &	$0.884$& 5516.4 &  $20.4^{+5.6	}_{-7.0 }$ & $44.9025 \pm 0.0012$ & $42.7743 \pm 0.0191$ & $5987 \pm 268 $ & $1897 \pm 48 $ & $2491 \pm 26$ \\
377   &	$0.337$& 1727.4 &  $5.9	^{+0.4	}_{-0.6 }$ & $43.7819 \pm 0.0011$ & $41.5130 \pm 0.0156$ & $2746 \pm 118 $ & $1576 \pm 23 $ & $1789 \pm 23$ \\
392   &	$0.843$& 5202.8 &  $14.2^{+3.7	}_{-3.0 }$ & $44.4249 \pm 0.0032$ & $42.4894 \pm 0.0427$ & $2419 \pm 82  $ & $2446 \pm 110$ & $3658 \pm 56$ \\
399   &	$0.608$& 3487.6 &  $35.8^{+1.1	}_{-10.3}$ & $44.3272 \pm 0.0020$ & $42.2823 \pm 0.0281$ & $2689 \pm 88  $ & $1989 \pm 89 $ & $1619 \pm 38$ \\
428   &	$0.976$& 6233.7 &  $15.8^{+6.0	}_{-1.9 }$ & $45.4013 \pm 0.0015$ & $43.2816 \pm 0.0048$ & $2795 \pm 29  $ & $1836 \pm 18 $ & $7568 \pm 70$ \\
551   &	$0.68 $& 3997.0 &  $6.4	^{+1.5	}_{-1.4 }$ & $44.1196 \pm 0.0021$ & $42.4389 \pm 0.0842$ & $2101 \pm 45  $ & $1255 \pm 59 $ & $1298 \pm 36$ \\
589   &	$0.751$& 4513.8 &  $46	^{+9.5	}_{-9.5 }$ & $44.4877 \pm 0.0015$ & $42.6421 \pm 0.0107$ & $3738 \pm 62  $ & $2835 \pm 62 $ & $5013 \pm 49$ \\
622   &	$0.572$& 3238.9 &  $49.1^{+11.1	}_{-2.0 }$ & $44.3737 \pm 0.0006$ & $42.5966 \pm 0.0062$ & $2389 \pm 36  $ & $1147 \pm 11 $ & $1423 \pm 32$ \\
645   &	$0.474$& 2583.6 &  $20.7^{+0.9	}_{-3.0 }$ & $44.1342 \pm 0.0008$ & $42.2965 \pm 0.0047$ & $6428 \pm 163 $ & $2799 \pm 13 $ & $1438 \pm 17$ \\
720   &	$0.467$& 2538.0 &  $41.6^{+14.8	}_{-8.3 }$ & $44.3176 \pm 0.0008$ & $42.4324 \pm 0.0029$ & $2829 \pm 15  $ & $1679 \pm 17 $ & $1232 \pm 16$ \\
772   &	$0.249$& 1219.6 &  $3.9	^{+0.9	}_{-0.9 }$ & $43.7867 \pm 0.0005$ & $41.5251 \pm 0.0081$ & $2381 \pm 33  $ & $1983 \pm 40 $ & $1026 \pm 14$ \\
775   &	$0.172$& 805.9  &  $16.3^{+13.1	}_{-6.6 }$ & $43.7943 \pm 0.0003$ & $41.7848 \pm 0.0021$ & $2744 \pm 36  $ & $2028 \pm 10 $ & $1818 \pm 8 $ \\
776   &	$0.116$& 524.6  &  $10.5^{+1.0	}_{-2.2 }$ & $43.3829 \pm 0.0004$ & $41.4179 \pm 0.0220$ & $3060 \pm 20  $ & $3178 \pm 19 $ & $1409 \pm 11$ \\
781   &	$0.263$& 1298.0 &  $75.2^{+3.2	}_{-3.3 }$ & $43.7604 \pm 0.0034$ & $41.8863 \pm 0.0155$ & $2506 \pm 19  $ & $1290 \pm 17 $ & $1089 \pm 22$ \\
782   &	$0.362$& 1877.9 &  $20	^{+1.1	}_{-3.0 }$ & $44.0941 \pm 0.0006$ & $41.9722 \pm 0.0044$ & $3027 \pm 35  $ & $1527 \pm 16 $ & $1353 \pm 23$ \\
790   &	$0.237$& 1153.2 &  $5.5	^{+5.7	}_{-2.1 }$ & $43.8222 \pm 0.0014$ & $41.8443 \pm 0.0272$ & $8365 \pm 44  $ & $5069 \pm 47 $ & $6318 \pm 38$ \\
840   &	$0.244$& 1191.8 &  $5	^{+1.5	}_{-1.4 }$ & $43.6987 \pm 0.0005$ & $41.5724 \pm 0.0074$ & $6116 \pm 267 $ & $3286 \pm 254$ & $4457 \pm 60$
\enddata
\tablecomments{
Columns are
1: Reverberation mapping identifier (RMID) --- see \cite{Shen15};
2: redshift;
3: luminosity distance;
4: \hb\ time lag;
5: log AGN continuum luminosity at 5100\,\AA;
6: log broad \hb\ luminosity;
7: FWHM of \hb\ in the mean spectrum;
8: line dispersion of \hb\ in the mean spectrum;
9: line dispersion of \hb\ in the rms spectrum.}
\label{table:sdssrmhb}
\end{deluxetable}
\end{longrotatetable}

\begin{longrotatetable}
\begin{deluxetable}{lccccccc}
\tabletypesize{\footnotesize}
\tablewidth{0pt}
\tablecaption{Reverberation-Mapped AGNs (SDSS \civ)}
\tablehead{
\colhead{RMID} &
\colhead{$z$} &
\colhead{$D_L$} &
\colhead{$\tau(\civ)$} &
\colhead{$\log L(1350\,{\rm \AA})$} &
\colhead{\fwm(\civ)} &
\colhead{\sigm(\civ)} &
\colhead{\sigr(\civ)} \\
\colhead{ } &
\colhead{ } &
\colhead{(Mpc)} &
\colhead{(days)} &
\colhead{(\ergsec)} &
\colhead{(\kms)} &
\colhead{(\kms)} &
\colhead{(\kms)} \\
\colhead{(1)} &
\colhead{(2)} &
\colhead{(3)} &
\colhead{(4)} &
\colhead{(5)} &
\colhead{(6)} &
\colhead{(7)} &
\colhead{(8)} }
\startdata
0   &	1.463	&10283	&$131.1	_{-36.6	}^{+42.9}$&	$44.847	\pm 0.004$&	$3967	\pm 107$&	$1968	\pm 160$&	$2144	\pm 46 $\\
32  &    1.72	&12554	&$22.8	_{-3.6	}^{+3.5 }$&    $44.492	\pm 0.021$&	$2999	\pm 34 $&     $1770	\pm 24 $&     $2017	\pm 10 $\\
36  &    2.213	&17094	&$188.4	_{-29	}^{+15.6}$&	$45.909	\pm 0.001$&	$4830	\pm 24 $&     $2890	\pm 24 $&     $3900	\pm 34 $\\
52  &    2.311	&18020	&$56.5	_{-5.9	}^{+3.1 }$&    $45.499	\pm 0.002$&	$2258	\pm 14 $&     $1809	\pm 15 $&     $1322	\pm 22 $\\
57  &    1.93	&14461	&$208.3	_{-5.6	}^{+10.6}$&	$45.393	\pm 0.003$&	$2692	\pm 11 $&     $1626	\pm 8  $&     $1682	\pm 12 $\\
58  &    2.299	&17906	&$186.1	_{-7.4	}^{+5.9 }$&    $45.353	\pm 0.002$&	$3627	\pm 45 $&     $2611	\pm 31 $&     $3412	\pm 30 $\\
130 &    1.96	&14737	&$224.3	_{-37.9	}^{+12.4}$&	$45.534	\pm 0.001$&	$5619	\pm 30 $&     $4078	\pm 55 $&     $4324	\pm 36 $\\
144 &    2.295	&17868	&$179.4	_{-42.3	}^{+31.2}$&	$45.516	\pm 0.001$&	$6153	\pm 53 $&     $2762	\pm 19 $&     $2792	\pm 19 $\\
145 &    2.138	&16390	&$180.9	_{-4.7	}^{+4.7 }$&    $45.113	\pm 0.004$&	$4472	\pm 74 $&     $3287	\pm 40 $&     $3408	\pm 16 $\\
158 &    1.477	&10405	&$36.7	_{-26.1	}^{+18.6}$&	$44.999	\pm 0.004$&	$3603	\pm 101$&	$2099	\pm 60 $&     $2136	\pm 31 $\\
161 &    2.071	&15764	&$180.1	_{-6.4	}^{+5.6 }$&    $45.491	\pm 0.001$&	$3163	\pm 28 $&     $2323	\pm 25 $&     $2524	\pm 20 $\\
181 &    1.678	&12177	&$102.6	_{-10.1	}^{+5   }$&    $44.545	\pm 0.015$&	$2998	\pm 35 $&     $2127	\pm 44 $&     $2721	\pm 34 $\\
201 &    1.797	&13248	&$41.3	_{-19.5	}^{+32  }$&    $46.240	\pm 0.001$&	$5438	\pm 56 $&     $1833	\pm 9  $&     $2408	\pm 117$\\
231 &    1.646	&11892	&$80.4	_{-7.5	}^{+6.3 }$&    $45.736	\pm 0.001$&	$5975	\pm 98 $&     $3267	\pm 102$&	$3803	\pm 18 $\\
237 &    2.394	&18810	&$49.9	_{-4.4	}^{+6.6 }$&    $45.866	\pm 0.001$&	$5455	\pm 39 $&     $2734	\pm 18 $&     $2779	\pm 23 $\\
245 &    1.677	&12168	&$107.1	_{-28.6	}^{+22.9}$&	$45.351	\pm 0.004$&	$9496	\pm 107$&	$4174	\pm 54 $&     $3953	\pm 86 $\\
249 &    1.721	&12562	&$24.9	_{-3.1	}^{+9.7 }$&    $44.984	\pm 0.010$&	$1871	\pm 15 $&     $1432	\pm 12 $&     $1640	\pm 15 $\\
256 &    2.247	&17414	&$43	_{-11.9	}^{+16.3}$&	$45.089	\pm 0.003$&	$2544	\pm 54 $&     $1742	\pm 29 $&     $1802	\pm 24 $\\
269 &    2.4	&18868	&$197.2	_{-12.6	}^{+2.4 }$&    $45.193	\pm 0.003$&	$3930	\pm 312$&	$3280	\pm 50 $&     $3547	\pm 30 $\\
275 &    1.58	&11307	&$81	_{-24.4	}^{+8.2 }$&    $45.611	\pm 0.001$&	$3213	\pm 20 $&     $2108	\pm 9  $&     $2406	\pm 5  $\\
295 &    2.351	&18400	&$163.8	_{-5.3	}^{+8.2 }$&    $45.605	\pm 0.001$&	$4311	\pm 41 $&     $2501	\pm 23 $&     $2446	\pm 19 $\\
298 &    1.633	&11777	&$106.1	_{-31.7	}^{+18.7}$&	$45.596	\pm 0.001$&	$3160	\pm 30 $&     $2066	\pm 26 $&     $2549	\pm 35 $\\
312 &    1.929	&14452	&$56.9	_{-6.7	}^{+11.4}$&	$45.077	\pm 0.004$&	$7663	\pm 166$&	$4273	\pm 74 $&     $4291	\pm 30 $\\
332 &    2.58	&20598	&$81.6	_{-11.4	}^{+5.6 }$&    $45.551	\pm 0.002$&	$3799	\pm 14 $&     $3009	\pm 63 $&     $4277	\pm 33 $\\
346 &    1.592	&11413	&$71.9	_{-11.3	}^{+23.8}$&	$44.905	\pm 0.003$&	$3389	\pm 168$&	$2220	\pm 131$&	$3055	\pm 29 $\\
386 &    1.862	&13838	&$38.2	_{-19.3	}^{+13.2}$&	$45.279	\pm 0.002$&	$2972	\pm 40 $&     $1782	\pm 38 $&     $2187	\pm 41 $\\
387 &    2.427	&19126	&$30.3	_{-3.4	}^{+19.6}$&	$45.687	\pm 0.001$&	$3676	\pm 24 $&     $2123	\pm 14 $&     $2451	\pm 23 $\\
389 &    1.851	&13738	&$224.3	_{-18	}^{+7.1 }$&    $45.564	\pm 0.002$&	$5222	\pm 111$&	$3839	\pm 16 $&     $4064	\pm 15 $\\
401 &    1.823	&13484	&$47.4	_{-8.9	}^{+15.2}$&	$45.564	\pm 0.002$&	$3273	\pm 21 $&     $2457	\pm 12 $&     $3321	\pm 12 $\\
411 &    1.734	&12679	&$248.3	_{-39	}^{+21.1}$&	$44.887	\pm 0.007$&	$4256	\pm 67 $&     $2511	\pm 61 $&     $2490	\pm 39 $\\
418 &    1.419	&9903	&$82.5	_{-16.9	}^{+27.6}$&	$45.040	\pm 0.003$&	$3143	\pm 44 $&     $2662	\pm 94 $&     $3110	\pm 23 $\\
470 &    1.883	&14030	&$19.9	_{-4	}^{+43.2}$&	$44.821	\pm 0.006$&	$4022	\pm 52 $&     $2441	\pm 34 $&     $2317	\pm 60 $\\
485 &    2.557	&20376	&$133.4	_{-5.2	}^{+22.6}$&	$46.119	\pm 0.001$&	$5342	\pm 48 $&     $2924	\pm 32 $&     $3961	\pm 41 $\\
496 &    2.079	&15839	&$197.9	_{-6.6	}^{+9.7 }$&    $45.560	\pm 0.001$&	$2364	\pm 27 $&     $2137	\pm 34 $&     $2409	\pm 45 $\\
499 &    2.327	&18172	&$168.5	_{-35.9	}^{+20.4}$&	$45.058	\pm 0.003$&	$3261	\pm 41 $&     $2968	\pm 41 $&     $3085	\pm 26 $\\
506 &    1.753	&12850	&$231.6	_{-11.1	}^{+13.3}$&	$45.075	\pm 0.003$&	$5046	\pm 52 $&     $3507	\pm 27 $&     $3510	\pm 24 $\\
527 &    1.651	&11937	&$52.3	_{-12.2	}^{+15.1}$&	$44.788	\pm 0.003$&	$5154	\pm 110$&	$3384	\pm 62 $&     $3587	\pm 34 $\\
549 &    2.277	&17698	&$69.8	_{-7.2	}^{+5.3 }$&    $45.369	\pm 0.002$&	$3907	\pm 59 $&     $1818	\pm 47 $&     $2176	\pm 21 $\\
554 &    1.707	&12437	&$194	_{-12.2	}^{+20.4}$&	$45.573	\pm 0.002$&	$3690	\pm 65 $&     $2253	\pm 47 $&     $2229	\pm 35 $\\
562 &    2.773	&22476	&$158.5	_{-34.2	}^{+18.2}$&	$46.302	\pm 0.001$&	$4379	\pm 113$&	$2036	\pm 29 $&     $2078	\pm 27 $\\
686 &    2.13	&16315	&$64.7	_{-6.3	}^{+12.6}$&	$45.444	\pm 0.002$&	$3827	\pm 34 $&     $2135	\pm 25 $&     $2203	\pm 27 $\\
689 &    2.007	&15170	&$157.6	_{-42.2	}^{+22.9}$&	$45.223	\pm 0.003$&	$2258	\pm 23 $&     $1292	\pm 8  $&     $1407	\pm 5  $\\
734 &    2.324	&18144	&$87.2	_{-11	}^{+13.9}$&	$45.530	\pm 0.001$&	$5701	\pm 121$&	$2982	\pm 65 $&     $3405	\pm 40 $\\
809 &    1.67	&12106	&$108.6	_{-50.7	}^{+27.7}$&	$45.204	\pm 0.005$&	$4811	\pm 38 $&     $5210	\pm 60 $&     $4749	\pm 96 $\\
827 &    1.966	&14792	&$137.7	_{-19.4	}^{+18.3}$&	$44.999	\pm 0.006$&	$2542	\pm 35 $&     $971	\pm 13 $&     $1443	\pm 13 $
\enddata
\tablecomments{
Columns are
1: Reverberation mapping identifier (RMID) --- see \cite{Shen15};
2: redshift;
3: luminosity distance;
4: \civ\ time lag;
5: log continuum luminosity at 1350\,\AA;
6: FWHM of \civ\ in the mean spectrum;
7: line dispersion of \civ\ in the mean spectrum;
8: line dispersion of \civ\ in the rms spectrum.}
\label{table:sdssrmciv}
\end{deluxetable}
\end{longrotatetable}

\startlongtable
\begin{deluxetable}{lccccc}
\tabletypesize{\footnotesize}
\tablewidth{0pt}
\tablecaption{\civ\ Single-Epoch Masses (VP06) }
\tablehead{
\colhead{Source} &
\colhead{\fwm(\civ)} &
\colhead{\sigm(\civ)} &
\colhead{$\log L(1350)$} &
\colhead{$\mu_{\rm SE}$(VP06)} &
\colhead{$\mu_{\rm SE}$(SDSS-RM)}\\
\colhead{ } &
\colhead{(\kms)} &
\colhead{(\kms)} &
\colhead{(\ergsec)} &
\colhead{($\msun$)} &
\colhead{($\msun$)} \\
\colhead{(1)} &
\colhead{(2)} &
\colhead{(3)} &
\colhead{(4)} &
\colhead{(5)} &
\colhead{(6)}
}
\startdata
Mrk335	   &$ 2291 \pm 27   $ & $ 2116 \pm  160	 $&$ 44.173 \pm 0.020$   & $ 6.663 \pm   0.337 $   &$7.079 \pm   0.145$\\
Mrk335	   &$ 1741 \pm 99   $ & $ 1806 \pm  360	 $&$ 44.291 \pm 0.078$   & $ 6.588 \pm   0.375 $   &$7.080 \pm   0.187$\\
Mrk335	   &$ 2023 \pm 17   $ & $ 2140 \pm  93	 $&$ 44.262 \pm 0.013$   & $ 6.720 \pm   0.332 $   &$7.153 \pm   0.140$\\
PG0026+129 &$ 1837 \pm 136  $ & $ 3364 \pm  70	 $&$ 45.165 \pm 0.025$   & $ 7.591 \pm   0.331 $   &$8.092 \pm   0.140$\\
PG0052+251 &$ 3983 \pm 370  $ & $ 5118 \pm  486	 $&$ 45.265 \pm 0.037$   & $ 8.009 \pm   0.341 $   &$8.402 \pm   0.150$\\
PG0052+251 &$ 5192 \pm 251  $ & $ 5083 \pm  437	 $&$ 45.176 \pm 0.041$   & $ 7.956 \pm   0.339 $   &$8.331 \pm   0.149$\\
Fairall9   &$ 2593 \pm 65   $ & $ 2981 \pm  197	 $&$ 44.470 \pm 0.028$   & $ 7.118 \pm   0.335 $   &$7.496 \pm   0.144$\\
Fairall9   &$ 2831 \pm 40   $ & $ 3532 \pm  92	 $&$ 44.582 \pm 0.011$   & $ 7.325 \pm   0.331 $   &$7.676 \pm   0.139$\\
Fairall9   &$ 2370 \pm 151  $ & $ 2978 \pm  508	 $&$ 44.759 \pm 0.126$   & $ 7.270 \pm   0.368 $   &$7.715 \pm   0.193$\\
Mrk590	   &$ 4839 \pm 59   $ & $ 3574 \pm  141	 $&$ 44.119 \pm 0.029$   & $ 7.089 \pm   0.332 $   &$7.330 \pm   0.141$\\
3C120	   &$ 3302 \pm 75   $ & $ 3199 \pm  169	 $&$ 44.943 \pm 0.039$   & $ 7.430 \pm   0.334 $   &$7.895 \pm   0.144$\\
3C120	   &$ 3278 \pm 105  $ & $ 3409 \pm  286	 $&$ 44.617 \pm 0.056$   & $ 7.312 \pm   0.339 $   &$7.682 \pm   0.152$\\
Ark120	   &$ 3989 \pm 451  $ & $ 3795 \pm  165	 $&$ 44.634 \pm 0.021$   & $ 7.414 \pm   0.332 $   &$7.755 \pm   0.141$\\
Ark120	   &$ 3945 \pm 42   $ & $ 3240 \pm  149	 $&$ 44.482 \pm 0.022$   & $ 7.197 \pm   0.333 $   &$7.551 \pm   0.141$\\
Mrk79	   &$ 3182 \pm 521  $ & $ 3344 \pm  222	 $&$ 43.879 \pm 0.039$   & $ 6.904 \pm   0.336 $   &$7.110 \pm   0.146$\\
Mrk79	   &$ 3049 \pm 128  $ & $ 2971 \pm  248	 $&$ 43.495 \pm 0.058$   & $ 6.598 \pm   0.339 $   &$6.752 \pm   0.152$\\
Mrk79	   &$ 3113 \pm 122  $ & $ 3803 \pm  388	 $&$ 43.726 \pm 0.065$   & $ 6.935 \pm   0.343 $   &$7.065 \pm   0.157$\\
Mrk110	   &$ 2990 \pm 64   $ & $ 2601 \pm  272	 $&$ 43.770 \pm 0.050$   & $ 6.628 \pm   0.343 $   &$6.887 \pm   0.155$\\
Mrk110	   &$ 1638 \pm 59   $ & $ 2576 \pm  231	 $&$ 43.876 \pm 0.081$   & $ 6.676 \pm   0.342 $   &$6.962 \pm   0.159$\\
PG0953+414 &$ 2873 \pm 57   $ & $ 3512 \pm  361	 $&$ 45.588 \pm 0.031$   & $ 7.853 \pm   0.342 $   &$8.438 \pm   0.151$\\
NGC3516	   &$ 4675 \pm 538  $ & $ 3311 \pm  372	 $&$ 42.830 \pm 0.093$   & $ 6.340 \pm   0.348 $   &$6.306 \pm   0.167$\\
NGC3516	   &$ 4875 \pm 17   $ & $ 3132 \pm  64	 $&$ 42.823 \pm 0.017$   & $ 6.288 \pm   0.331 $   &$6.270 \pm   0.139$\\
NGC3516	   &$ 5147 \pm 103  $ & $ 3245 \pm  84	 $&$ 43.192 \pm 0.013$   & $ 6.514 \pm   0.331 $   &$6.570 \pm   0.139$\\
NGC3516	   &$ 4729 \pm 28   $ & $ 3430 \pm  92	 $&$ 43.143 \pm 0.013$   & $ 6.536 \pm   0.331 $   &$6.564 \pm   0.139$\\
NGC3516	   &$ 4525 \pm 97   $ & $ 3137 \pm  79	 $&$ 43.030 \pm 0.012$   & $ 6.399 \pm   0.331 $   &$6.428 \pm   0.139$\\
NGC3516	   &$ 3940 \pm 18   $ & $ 2834 \pm  95	 $&$ 42.485 \pm 0.034$   & $ 6.022 \pm   0.332 $   &$5.957 \pm   0.142$\\
NGC3516	   &$ 4912 \pm 23   $ & $ 3973 \pm  36	 $&$ 42.793 \pm 0.012$   & $ 6.479 \pm   0.330 $   &$6.380 \pm   0.138$\\
NGC3783	   &$ 2831 \pm 22   $ & $ 3273 \pm  100	 $&$ 43.601 \pm 0.014$   & $ 6.738 \pm   0.331 $   &$6.886 \pm   0.139$\\
NGC3783	   &$ 2308 \pm 17   $ & $ 3179 \pm  185	 $&$ 43.744 \pm 0.022$   & $ 6.789 \pm   0.334 $   &$6.979 \pm   0.143$\\
NGC4051    &$ 1319 \pm 13   $ & $ 1713 \pm  227	 $&$ 41.373 \pm 0.058$   & $ 4.995 \pm   0.351 $   &$4.830 \pm   0.163$\\
NGC4151    &$ 6929 \pm 76   $ & $ 5220 \pm  123	 $&$ 43.224 \pm 0.010$   & $ 6.944 \pm   0.331 $   &$6.860 \pm   0.139$\\
NGC4151    &$ 5418 \pm 150  $ & $ 4604 \pm  249	 $&$ 43.340 \pm 0.019$   & $ 6.896 \pm   0.333 $   &$6.878 \pm   0.142$\\
NGC4151    &$ 5062 \pm 51   $ & $ 4651 \pm  371	 $&$ 43.396 \pm 0.029$   & $ 6.935 \pm   0.338 $   &$6.926 \pm   0.147$\\
NGC4151    &$ 5246 \pm 44   $ & $ 4675 \pm  397	 $&$ 43.396 \pm 0.031$   & $ 6.939 \pm   0.339 $   &$6.929 \pm   0.148$\\
NGC4151    &$ 5752 \pm 144  $ & $ 4585 \pm  321	 $&$ 43.418 \pm 0.023$   & $ 6.934 \pm   0.336 $   &$6.935 \pm   0.144$\\
NGC4151    &$ 5173 \pm 593  $ & $ 4664 \pm  475	 $&$ 43.354 \pm 0.044$   & $ 6.915 \pm   0.342 $   &$6.896 \pm   0.153$\\
NGC4151    &$ 3509 \pm 10   $ & $ 4384 \pm  66	 $&$ 43.038 \pm 0.006$   & $ 6.694 \pm   0.330 $   &$6.621 \pm   0.138$\\
PG1229+204 &$ 3391 \pm 205  $ & $ 3241 \pm  457	 $&$ 44.654 \pm 0.028$   & $ 7.288 \pm   0.352 $   &$7.682 \pm   0.160$\\
PG1307+085 &$ 3465 \pm 168  $ & $ 3687 \pm  290	 $&$ 45.012 \pm 0.039$   & $ 7.590 \pm   0.338 $   &$8.027 \pm   0.148$\\
Mrk279	   &$ 4126 \pm 487  $ & $ 3118 \pm  414	 $&$ 43.795 \pm 0.118$   & $ 6.799 \pm   0.355 $   &$7.007 \pm   0.181$\\
Mrk279	   &$ 3876 \pm 99   $ & $ 3286 \pm  511	 $&$ 43.754 \pm 0.127$   & $ 6.823 \pm   0.363 $   &$7.005 \pm   0.189$\\
NGC5548    &$ 4790 \pm 67   $ & $ 4815 \pm  257	 $&$ 43.654 \pm 0.022$   & $ 7.102 \pm   0.333 $   &$7.142 \pm   0.142$\\
NGC5548    &$ 4096 \pm 14   $ & $ 3973 \pm  34	 $&$ 43.568 \pm 0.006$   & $ 6.889 \pm   0.330 $   &$6.969 \pm   0.138$\\
NGC5548    &$ 3280 \pm 27   $ & $ 5050 \pm  787	 $&$ 43.773 \pm 0.069$   & $ 7.206 \pm   0.359 $   &$7.259 \pm   0.171$\\
PG1426+015 &$ 3778 \pm 448  $ & $ 4101 \pm  391	 $&$ 45.295 \pm 0.023$   & $ 7.832 \pm   0.340 $   &$8.301 \pm   0.149$\\
Mrk817	   &$ 4027 \pm 71   $ & $ 4062 \pm  289	 $&$ 44.123 \pm 0.022$   & $ 7.203 \pm   0.336 $   &$7.404 \pm   0.145$\\
PG1613+658 &$ 5902 \pm 136  $ & $ 3965 \pm  215	 $&$ 45.221 \pm 0.023$   & $ 7.764 \pm   0.334 $   &$8.226 \pm   0.142$\\
PG1617+175 &$ 4558 \pm 1763 $ & $ 3383 \pm  1036 $&$ 44.784 \pm 0.108$   & $ 7.394 \pm   0.428 $   &$7.805 \pm   0.234$\\
Mrk509	   &$ 5035 \pm 298  $ & $ 3558 \pm  205	 $&$ 44.641 \pm 0.029$   & $ 7.362 \pm   0.334 $   &$7.725 \pm   0.143$\\
Mrk509	   &$ 4345 \pm 49   $ & $ 3426 \pm  115	 $&$ 44.532 \pm 0.015$   & $ 7.272 \pm   0.331 $   &$7.621 \pm   0.140$\\
Mrk509	   &$ 4973 \pm 233  $ & $ 3647 \pm  172	 $&$ 44.803 \pm 0.020$   & $ 7.469 \pm   0.333 $   &$7.862 \pm   0.141$\\
Mrk509	   &$ 4961 \pm 218  $ & $ 3127 \pm  226	 $&$ 44.552 \pm 0.033$   & $ 7.203 \pm   0.336 $   &$7.585 \pm   0.146$\\
Mrk509	   &$ 3716 \pm 228  $ & $ 3174 \pm  448	 $&$ 44.706 \pm 0.071$   & $ 7.297 \pm   0.354 $   &$7.710 \pm   0.168$\\
PG2130+099 &$ 2113 \pm 119  $ & $ 2390 \pm  184	 $&$ 44.692 \pm 0.025$   & $ 7.044 \pm   0.337 $   &$7.541 \pm   0.146$\\
NGC7469    &$ 3094 \pm 53   $ & $ 3379 \pm  182	 $&$ 43.774 \pm 0.016$   & $ 6.858 \pm   0.333 $   &$7.036 \pm   0.142$\\
NGC7469    &$ 2860 \pm 12   $ & $ 3266 \pm  110	 $&$ 43.679 \pm 0.015$   & $ 6.778 \pm   0.331 $   &$6.945 \pm   0.140$
\enddata
\tablecomments{
Data sources are listed in Table 2 of VP06.
Columns are
1: AGN name;
2: FWHM of \civ;
3: line dispersion of \civ;
4: AGN continuum luminosity at 1350\,\AA;
5: single-epoch virial product from VP06;
6: single-epoch virial product based on the data
in this table and equation ({\ref{eq:predictciv}}).
}
\label{table:civVP06}
\end{deluxetable}


\clearpage

\begin{thebibliography}

\bibitem[Allen et al.(2011)]{Allen11} 
Allen, J.~T., Hewett, P.~C., Maddox, N., et al.\ 2011, \mnras, 410:860

\bibitem[Antonucci \& Cohen(1983)]{Antonucci83}
Antonucci, R.R.J., \& Cohen, R.D. 1983, \apj, 271:564

\bibitem[Assef et al.(2011)]{Assef11}
Assef, R.~J., Denney, K.~D., Kochanek, C.~S., et al.\ 2011, \apj, 742:93

\bibitem[Bahk, Woo, \& Park(2019)]{Bahk19}
Bahk, H., Woo, J.-H., \& Park, D. 2019, \apj, 875:50

\bibitem[Baldwin(1977)]{Baldwin77} 
Baldwin, J.~A.\ 1977, \apj, 214:679

\bibitem[Barth et al.(2015)]{Barth15} 
Barth, A.~J., Bennert, V.~N., Canalizo, G., et al.\ 2015, \apjs, 217:26

\bibitem[Baskin \& Laor(2004)]{Baskin04} 
Baskin, A., \& Laor, A.\ 2004, \mnras, 350:L31

\bibitem[Baskin \& Laor(2005)]{Baskin05}
Baskin, A., \& Laor, A. 2005, \mnras, 356:1029

\bibitem[Batiste et al.(2017)]{Batiste17}
Batiste, M., Bentz, M.~C., Raimundo, S.~I., Vestergaard, M., \& Onken, C.~A.\ 2017, \apjl, 838:L10

\bibitem[Bentz et al.(2016)]{Bentz16}
Bentz, M.~C., Cackett, E.~M., Crenshaw, D.~M., et al.\ 2016, \apj, 830:136

\bibitem[Bentz et al.(2006a)]{Bentz06a}
Bentz, M.~C., Denney, K.~D., Cackett, E.~M., et al.\ 2006a, \apj, 651:775

\bibitem[Bentz et al.(2007)]{Bentz07}
Bentz, M.~C., Denney, K.~D., Cackett, E.~M., et al.\ 2007, \apj, 662:205

\bibitem[Bentz et al.(2013)]{Bentz13}
Bentz, M.~C., Denney, K.~D., Grier, C.~J., et al.\ 2013, \apj, 767:149

\bibitem[Bentz et al.(2019)]{Bentz19}
Bentz, M.~C., Ferrarese, L., Onken, C.~A.,
Peterson, B.~M., \& Valluri, M. 2019, \apj, 885:161

\bibitem[Bentz et al.(2014)]{Bentz14}
Bentz, M.~C., Horenstein, D., Bazhaw, C., et al.\ 2014, \apj, 796:8

\bibitem[Bentz \& Katz(2015)]{Bentz15}
Bentz, M.~C., \& Katz, S.\ 2015, \pasp, 127:67

\bibitem[Bentz et al.(2009a)]{Bentz09a}
Bentz, M.~C., Peterson, B.~M., Netzer, H., Pogge, R.~W., \& Vestergaard, M. 2009a, \apj, 697:160

\bibitem[Bentz(2006b)]{Bentz06b}
Bentz, M.~C., Peterson, B.~M., Pogge, R.~W., Vestergaard, M., \& Onken, C.~A. 2006b, \apj, 644:133

\bibitem[Bentz et al.(2008)]{Bentz08}
Bentz, M.~C., Walsh, J.~L., Barth, A.~J., et al.\ 2008, \apjl, 689:L21

\bibitem[Bentz et al.(2009b)]{Bentz09b}
Bentz, M.~C., Walsh, J.~L., Barth, A.~J., et al.\ 2009b, \apj, 705:199

\bibitem[Bentz et al.(2010)]{Bentz10}
Bentz, M.~C., Walsh, J.~L., Barth, A.~J., et al.\ 2010, \apj, 726:993

\bibitem[Bian et al.(2012)]{Bian12}
Bian, W.-H., Fang, L.-L., Huang, K.-L., \& Wang, J.-M. 
2012, \mnras, 427:2881 

\bibitem[Bisogni et al.(2017)]{Bisogni17} 
Bisogni, S., di Serego Alighieri, S., Goldoni, P., et al.\ 2017, \aap, 603:A1

\bibitem[Blandford \& McKee(1982)]{Blandford82}
Blandford, R.D., \& McKee, C.F. 1982, \apj, 255:419

\bibitem[Boller, Brandt, \& Fink(1996)]{Boller96} 
Boller, Th., Brandt, W.~N., \& Fink, H.\ 1996, \aap, 305, 53

\bibitem[Boroson(2002)]{Boroson02} 
Boroson, T.~A.\ 2002, \apj, 565:78

\bibitem[Boroson \& Green(1992)]{Boroson92} 
Boroson, T.~A., \& Green, R.~F.\ 1992, \apjs, 80:109

\bibitem[Brotherton et al.(2015a)]{Brotherton15a}
Brotherton, M. S., Runnoe, J. C., Shang, Z., \& DiPompeo, M. A. 
2015a, \mnras, 451:1290 

\bibitem[Brotherton, Singh \& Runnoe(2015b)]{Brotherton15b} 
Brotherton, M. S., Singh, V., \& Runnoe, J. 2015b, \mnras, 454:3864 

\bibitem[Burbidge \& Burbidge(1967)]{Burbidge67}
Burbidge, G., \& Burbidge, M. 1967, {\it Quasi-Stellar Objects},
(San Francisco: Freeman)

\bibitem[Cackett et al.(2015)]{Cackett15}
Cackett, E.~M., G\"{u}ltekin,~K., Bentz, M.~C., Fausnaugh, M.~M., 
Peterson, B.~M., Troyer, J., \& Vestergaard, M. 2015, \apj, 810:86 

\bibitem[Cackett \& Horne(2006)]{Cackett06} 
Cackett, E.~M., \& Horne, K.\ 2006, \mnras, 365:1180

\bibitem[Cappellari et al.(2013)]{Cappellari13}
Cappellari, M., Scott, N., Alatalo, K., et al.\ 2013, \mnras, 432:1709

\bibitem[Clavel et al.(1990)]{Clavel90} 
Clavel, J., Boksenberg, A., Bromage, G.~E., Elvius, A., Penston, M.~V., 
Perola, G.~C., Santos-Lle\'{o}, M., Snijders, M.~A.~J., \& Ulrich, M.-H. 1990, 
\mnras, 246:668 

\bibitem[Clavel et al.(1991)]{Clavel91}
Clavel, J., Reichert, G.~A., Alloin, D., et al.\ 1991, \apj, 366:64

\bibitem[Clavel, Wamsteker, \& Glass(1989)]{CWG89}
Clavel, J., Wamsteker, W., \& Glass, I.~S. 1989, \apj, 337:236

\bibitem[Coatman et al.(2016)]{Coatman16} 
Coatman, L., Hewett, P.~C., Banerji, M., et al.\ 2016, \mnras, 461:647

\bibitem[Coatman et al.(2017)]{Coatman17}
Coatman, L., Hewett, P.~C., Banerji, M., et al.\ 2017, \mnras, 465:2120

\bibitem[Collin et al.(2006)]{Collin06}
Collin, S., Kawaguchi, T., Peterson, B.M., \& Vestergaard, M. 2006, \aap, 456:75

\bibitem[Corbett et al.(2003)]{Corbett03}
Corbett, E.~A., Croom, S.~M., Boyle, B.~J., et al.\ 2003, \mnras, 343:705

\bibitem[Corbin(1990)]{Corbin90} 
Corbin, M.~R.\ 1990, \apj, 357:346

\bibitem[Czerny et al.(2019)]{Czerny19} 
Czerny, B., Olejak, A., Rałowski, M., et al. 2019, \apj, 880:46 

\bibitem[Dalla Bont{\`a} et al.(2018)]{DallaBonta18}
Dalla Bont{\`a}, E., Davies, R.~L., Houghton, R.~C.~W., et al.\ 2018, \mnras, 474:339

\bibitem[Davidson(1972)]{Davidson72}
Davidson, K.\ 1972, \apj, 171:213

\bibitem[Davidson \& Netzer(1979)]{Davidson79} 
Davidson, K., \& Netzer, H.\ 1979, Reviews of Modern Physics, 51, 715

\bibitem[De Rosa et al.(2018)]{DeRosa18}
De Rosa, G., Fausnaugh, M.~M., Grier, C.~J., et al.\ 2018, \apj, 866:133

\bibitem[De Rosa et al.(2015)]{DeRosa15}
De Rosa, G., Peterson, B.~M., Ely, J., et al.\ 2015, \apj, 806:128



\bibitem[Denney(2012)]{Denney12} 
Denney, K.~D.\ 2012, \apj, 759:44

\bibitem[Denney et al.(2006)]{Denney06}
Denney, K.~D., Bentz, M.~C., Peterson, B.~M., et al.\ 2006, \apj, 653:152

\bibitem[Denney et al.(2009a)]{Denney09a}
Denney, K.D., Peterson, B.M., Dietrich, M., Vestergaard, M., \& Bentz, M.C. 2009a, \apj, 692:246

\bibitem[Denney et al.(2010)]{Denney10}
Denney, K.~D., Peterson, B.~M., Pogge, R.~W., et al.\ 2010, \apj, 721:715

\bibitem[Denney et al.(2013)]{Denney13} Denney, K.~D., Pogge, R.~W., Assef, R.~J., et al.\ 2013,
\apj, 775:60

\bibitem[Denney et al.(2009b)]{Denney09b}
Denney, K.~D., Watson, L.~C., Peterson, B.~M., et al.\ 2009b, \apj, 702:1353

\bibitem[Dibai(1980)]{Dibai80}
Dibai, E.A. 1980, Soviet Astronomy, 24:389

\bibitem[Dietrich et al.(1998)]{Dietrich98}
Dietrich, M., Peterson, B.~M., Albrecht, P., et al.\ 1998, \apjs, 115:185

\bibitem[Dietrich et al.(2012)]{Dietrich12}
Dietrich, M., Peterson, B.~M., Grier, C.~J., et al.\ 2012, \apj, 757:53



\bibitem[Doroshenko et al.(2012)]{Doroshenko12}
Doroshenko, V.~T., Sergeev, S.~G., Klimanov, S.~A., Pronik, V.~I., \& Efimov, Y.~S.\ 2012, \mnras, 426:416

\bibitem[Du et al.(2014)]{Du14}
Du, P., Hu, C., Lu, K.-X., et al.\ 2014, \apj, 782:45

\bibitem[Du et al.(2016)]{Du16}
Du, P., Lu, K.-X., Zhang, Z.-X., et al.\ 2016, \apj, 825:126

\bibitem[Du \& Wang(2019)]{Du19}
Du, P., \& Wang, J.-M. 2019, \apj, 886:42

\bibitem[Du et al.(2018)]{Du18}
Du, P., Zhang, Z.-X., Wang, K., et al.\ 2018, \apj, 856:6

\bibitem[Edelson et al.(2017)]{Edelson17} 
Edelson, R., Gelbord, J., Cackett, E., et al.\ 2017, \apj, 840:41

\bibitem[Edelson et al.(2019)]{Edelson19} 
Edelson, R., Gelbord, J., Cackett, E., et al.\ 2019, \apj, 870:123

\bibitem[Edelson et al.(2015)]{Edelson15} 
Edelson, R., Gelbord, J.~M., Horne, K., et al.\ 2015, \apj, 806:129

\bibitem[Espey et al.(1989)]{Espey89} 
Espey, B.~R., Carswell, R.~F., Bailey, J.~A., et al.\ 1989, \apj, 342:666

\bibitem[Fausnaugh et al.(2016)]{Fausnaugh16} 
Fausnaugh, M.~M., Denney, K.~D., Barth, A.~J., et al.\ 2016, \apj, 821:56

\bibitem[Fausnaugh et al.(2017)]{Fausnaugh17}
Fausnaugh, M.~M., Grier, C.~J., Bentz, M.~C., et al.\ 2017, \apj, 840:97

\bibitem[Ferland \& Shields(1985)]{Ferland85}
Ferland, G.~J., \& Shields, G.~A. 1985, in 
{\em Astrophysics of Active Galaxies and Quasi-Stellar Objects}
(Mill Valley: University Science Books), pp.\ 157--184

\bibitem[Ferrarese \& Merritt(2000)]{Ferrarese00}
Ferrarese, L., \& Merritt, D.\ 2000, \apjl, 539:L9

\bibitem[Fine et al.(2008)]{Fine08} 
Fine, S., Croom, S.~M., Hopkins, P.~F., et al.\ 2008, \mnras, 390:1413

\bibitem[Fonseca Alvarez et al.(2020)]{Alvarez20}
Fonseca Alvarez, G., Trump, J.~R., Homayouni, Y. et al. 2020,
\apj, 899:1

\bibitem[Fromerth \& Melia(2000)]{Fromerth00}
Fromerth, M.J., \& Melia, F. 2000, \apj, 533:172

\bibitem[Gaskell(1982)]{Gaskell82} 
Gaskell, C.~M.\ 1982, \apj, 263:79

\bibitem[Gaskell \& Pe\-ter\-son(1987)]{Gaskell87}
Gaskell, C.~M., \& Peterson, B.~M.\ 1987, \apjs, 65:1

\bibitem[Gebhardt et al.(2000)]{Gebhardt00}
Gebhardt, K., Bender, R., Bower, G., et al.\ 2000, \apjl, 539:L13

\bibitem[Gilbert \& Peterson(2003)]{Gilbert03} 
Gilbert, K.~M., \& Peterson, B.~M.\ 2003, \apj, 587:123

\bibitem[Goad, Korista, \& Knigge(2004)]{Goad04} 
Goad, M.~R., Korista, K.~T., \& Knigge, C.\ 2004, \mnras, 352:277

\bibitem[Greene \& Ho(2005)]{Greene05}
Greene, J.~E., \& Ho, L.~C.\ 2005, \apj, 630:122

\bibitem[Greene \& Ho(2007)]{Greene07}
Greene, J.~E., \& Ho, L.~C.\ 2007, \apj, 667:131

\bibitem[Greene et al.(2010)]{Greene10a}
Greene, J.~E., Hood, C.~E., Barth, A.~J., et al.\ 2010, \apj, 723:409

\bibitem[Greene, Peng, \& Ludwig(2010)]{Greene10b}
Greene, J.~E., Peng, C.~Y., \& Ludwig, R.~R.\ 2010, \apj, 709:937

\bibitem[Grier et al.(2013)]{Grier13}
Grier, C.~J., Martini, P., Watson, L.~C., et al.\ 2013, \apj, 773:90

\bibitem[Grier et al.(2017a)]{Grier17a}
Grier, C.~J., Pancoast, A., Barth, A.~J., et al.\ 2017a, \apj, 849:146

\bibitem[Grier et al.(2017b)]{Grier17b}
Grier, C.~J., Trump, J.~R., Shen, Y.,  et al.\ 2017b, \apj, 851:21.
Erratum: 2018, \apj, 868:76

\bibitem[Grier et al.(2008)]{Grier08}
Grier, C.~J., Peterson, B.~M., Bentz, M.~C., et al.\ 2008, \apj, 688:837


\bibitem[Grier et al.(2012)]{Grier12}
Grier, C.~J., Peterson, B.~M., Pogge, R.~W., et al.\ 2012, \apj, 755:60

\bibitem[Grier et al.(2019)]{Grier19}
Grier, C.~J., Shen, Y., Horne, K., et al.\ 2019, \apj, 887:38

\bibitem[G{\"u}ltekin et al.(2009)]{Gultekin09}
G{\"u}ltekin, K., Richstone, D.~O., Gebhardt, K., et al.\ 2009, \apj, 698:198

\bibitem[Hall et al.(2002)]{Hall02} 
Hall, P.~B., Anderson, S.~F., Strauss, M.~A., et al.\ 2002, \apjs, 141:267

\bibitem[Hemler et al.(2019)]{Hemler19} 
Hemler, Z.~S., Grier, C.~J., Brandt, W.~N., et al.\ 2019, \apj, 872:21

\bibitem[Hewett \& Foltz(2003)]{Hewett03} 
Hewett, P.~C., \& Foltz, C.~B.\ 2003, \aj, 125:1784

\bibitem[Homayouni et al.(2020)]{Homayouni20}
Homayouni, Y., Trump, J.R., Grier, C.J., et al.\ 2020, submitted to \apj\
(arXiv:2005.03663)

\bibitem[Hoormann et al.(2019)]{Hoormann19}
Hoormann, J.K., Martini, P., Davis, T.M., et al.\ 2019, \mnras, 487:3650

\bibitem[Ili{\'c} et al.(2017)]{Ilic17}
Ili{\'c}, D., Shapovalova, A.~I., Popovi{\'c}, L. {\v{C}}., et al.\ 2017,
Frontiers in Astronomy and Space Sciences, 4:12

\bibitem[Joly et al.(1985)]{Joly85}
Joly, M., Collin-Souffrin, S., Masnou, J.~L., \& Nottale, L.\ 1985, 
\aap, 152:282

\bibitem[Kaspi et al.(2007)]{Kaspi07} 
Kaspi, S., Brandt, W.~N., Maoz, D., Netzer, J., Schneider, D.P.,
\& Shemmer, O.\ 2007, \apj, 659:997

\bibitem[Kaspi et al.(2005)]{Kaspi05}
Kaspi, S., Maoz, D., Netzer, H., Peterson, B.M., Vestergaard, M., \& Jannuzi, B.T. 2005,
\apj, 629:61

\bibitem[Kaspi et al.(2000)]{Kaspi00} 
Kaspi, S., Smith, P. S., Netzer, H., Maoz, D., Jannuzi, B. T., 
\& Giveon, U. 2000, \apj, 533:631 

\bibitem[Kollatschny(2003)]{Kollatschny03}
Kollatschny, W. 2003, \aap, 407:461

\bibitem[Kollatschny et al.(2001)]{Kollatschny01}
Kollatschny, W., Bischoff, K., Robinson, E.~L., Welsh, W.~F., \& Hill, G.~J.\ 2001, \aap, 379:125

\bibitem[Kollatschny et al.(2014)]{Kollatschny14}
Kollatschny, W., Ulbrich, K., Zetzl, M., Kaspi, S., \& Haas, M.\ 2014, \aap, 566:A106

\bibitem[Kollatschny \& Zetzl(2013)]{Kollatschny13}
Kollatschny, W., \& Zetzl, M. 2013, \aap, 549:A100

\bibitem[Kollmeier et al.(2006)]{Kollmeier06}
Kollmeier, J.~A., Onken, C.~A., Kochanek, C.~S., et al.\ 2006, \apj, 648:128

\bibitem[Koratkar \& Gaskell(1991)]{Koratkar91}
Koratkar, A.~P., \& Gaskell, C.~M.\ 1991, \apjl, 370:L61

\bibitem[Korista et al.(1995)]{Korista95}
Korista, K.~T., Alloin, D., Barr, P., et al.\ 1995, \apjs, 97:285

\bibitem[Korista, Baldwin, \& Ferland(1998)]{Korista98} 
Korista, K., Baldwin, J., \& Ferland, G.\ 1998, \apj, 507:24

\bibitem[Krolik et al.(1991)]{Krolik91}
Krolik, J.~H., Horne, K., Kallman, T.~R., et al.\ 1991, \apj, 371:541

\bibitem[Laor(1998)]{Laor98}
Laor, A. 1998, \apjl, 505:L83

\bibitem[Li et al.(2019)]{Li19}
Li, J.~I., Shen, Y., Brandt, W.~N., et al.\ 2019, \apj, 884:119

\bibitem[Lira et al.(2018)]{Lira18} 
Lira, P., Kaspi, S., Netzer, H., Botti, I., Morrell, N., 
Mej\'{\i}a-Restrepo, J., S\'{a}nchez-S\'{a}ez, P., \&
Mart\'{\i}nez-Palomera, J., L\'{o}pez, P. 2018, \apj, 865:56 

\bibitem[Lu et al.(2016)]{Lu16}
Lu, K.-X., Du, P., Hu, C., et al.\ 2016, \apj, 827:118

\bibitem[Marconi et al.(2008)]{Marconi08}
Marconi, A., Axon, D.~J., Maiolino, R., et al.\ 2008, \apj, 678:693

\bibitem[Marconi et al.(2009)]{Marconi09} 
Marconi, A., Axon, D.~J., Maiolino, R., et al.\ 2009, \apjl, 698:L103

\bibitem[Mart{\'\i}nez-Aldama et al.(2019)]{Martinez19} 
Mart{\'\i}nez-Aldama, M.~L., Czerny, B., Kawka, D., et al.\ 2019, 
\apj, 883, 170

\bibitem[Mart\'{\i}nez-Aldama et al.(2020)]{Martinez20} 
Mart\'{\i}nez-Aldama, M.~L., Zaja\v{c}ek, M., Czerny, B., \& 
Panda, S. 2020, arXiv:2007.09955 

\bibitem[Marziani et al.(2019)]{Marziani19} 
Marziani, P., del Olmo, A., Mart{\'\i}nez-Carballo, M.~A., et al.\ 2019, \aap, 627:A88

\bibitem[Marziani et al.(2018)]{Marziani18} 
Marziani, P., Dultzin, D., Sulentic, J.~W., et al.\ 2018, 
Frontiers in Astronomy and Space Sciences, 5:6

\bibitem[McHardy et al.(2014)]{McHardy14} 
McHardy, I.~M., Cameron, D.~T., Dwelly, T., et al.\ 2014, \mnras, 444:1469

\bibitem[McHardy et al.(2018)]{McHardy18} 
McHardy, I.~M., Connolly, S.~D., Horne, K., et al.\ 2018, \mnras, 480:2881

\bibitem[McLure \& Jarvis(2002)]{McLure02}
McLure, R.J., \& Jarvis, M.J. 2002, \mnras, 337:109

\bibitem[Mej{\'\i}a-Restrepo et al.(2018)]{Mejia18}
Mej{\'\i}a-Restrepo, J.~E., Trakhtenbrot, B., Lira, P., et al.\ 2018, \mnras, 478:1929

\bibitem[Metzroth, Onken, \& Peterson(2006)]{Metzroth06}
Metzroth, K.~G., Onken, C.~A., \& Peterson, B.~M.\ 2006, \apj, 647:901

\bibitem[Netzer(2019)]{Netzer19}
Netzer, H. 2019, \mnras, 488:5185

\bibitem[Netzer et al.(2007)]{Netzer07}
Netzer, H., Lira, P., Trakhtenbrot, B., et al.\ 2007, \apj, 671:1256

\bibitem[Netzer \& Marziani(2010)]{Netzer10}
Netzer, H., \& Marziani, P.\ 2010, \apj, 724:318

\bibitem[O'Brien et al.(1998)]{Obrien98}
O'Brien, P.~T., Dietrich, M., Leighly, K., et al.\ 1998, \apj, 509:163

\bibitem[Onken \& Peterson(2002)]{Onken02}
Onken, C.~A., \& Peterson, B.~M.\ 2002, \apj, 572:746

\bibitem[Osterbrock(1985)]{Osterbrock85}
Osterbrock, D.~E. 1985, in {\em Astrophysics of
Active Galaxies and Quasi-Stellar Objects}
(Mill Valley: University Science Books), pp.\ 111-155

\bibitem[Osterbrock \& Pogge(1985)]{OsterbrockPogge85} 
Osterbrock, D.~E., \& Pogge, R.~W.\ 1985, \apj, 297:166

\bibitem[Padovani \& Rafanelli(1988)]{Padovani88}
Padovani, P., \& Rafanelli, P. 1988, \aap, 205:53

\bibitem[Pancoast et al.(2014)]{Pancoast14}
Pancoast, A., Brewer, B.~J., Treu, T., et al.\ 2014, \mnras, 445:3073

\bibitem[Park et al.(2017)]{Park17}
 Park, D., Barth, A.~J., Woo, J.-H., Malkan, M.~A., Treu, T., Bennert, V.~N., Assef, R.~J., 
\& Pancoast, A. 2017, \apj, 839:93 

\bibitem[Pei et al.(2017)]{Pei17}
Pei, L., Fausnaugh, M.~M., Barth, A.~J., et al.\ 2017, \apj, 837:131

\bibitem[Peterson(1993)]{Peterson93}
Peterson, B.M. 1993, \pasp, 105:247

\bibitem[Peterson(2014)]{Peterson14}
Peterson, B.M. 2014,  Space Sci.\ Rev., 183:253

\bibitem[Peterson et al.(1992)]{Peterson92}
Peterson, B.~M., Alloin, D., Axon, D., et al.\ 1992, \apj, 392:470

\bibitem[Peterson et al.(1991)]{Peterson91}
Peterson, B.~M., Balonek, T.~J., Barker, E.~S., et al.\ 1991, \apj, 368:119

\bibitem[Peterson et al.(1999)]{Peterson99}
Peterson, B.~M., Barth, A.~J., Berlind, P., et al.\ 1999, \apj, 510:659

\bibitem[Peterson et al.(2005)]{Peterson05}
Peterson, B.~M., Bentz, M.~C., Desroches, L.-B., et al.\ 2005, \apj, 632:799

\bibitem[Peterson et al.(1994)]{Peterson94}
Peterson, B.~M., Berlind, P., Bertram, R., et al.\ 1994, \apj, 425:622

\bibitem[Peterson et al.(2002)]{Peterson02}
Peterson, B.~M., Berlind, P., Bertram, R., et al.\ 2002, \apj, 581:197

\bibitem[Peterson et al.(2013)]{Peterson13}
Peterson, B.~M., Denney, K.~D., De Rosa, G., et al.\ 2013, \apj, 779:109

\bibitem[Peterson et al.(2004)]{Peterson04}
Peterson, B.~M., Ferrarese, L., Gilbert, K.~M., et al.\ 2004, \apj, 613:682

\bibitem[Peterson et al.(2014)]{Peterson7469}
Peterson, B.~M., Grier, C.~J., Horne, K., et al.\ 2014, \apj, 795:149

\bibitem[Peterson et al.(1985)]{Peterson85}
Peterson, B.~M., Meyers, K.~A., Capriotti, E.~R., et al.\ 1985, \apj, 292:164

\bibitem[Peterson \& Wandel(1999)]{PetersonWandel99}
Peterson, B.~M., \& Wandel, A.\ 1999, \apjl, 521:L95

\bibitem[Peterson \& Wandel(2000)]{PetersonWandel00}
Peterson, B.~M., \& Wandel, A.\ 2000, \apjl, 540:L13

\bibitem[Peterson et al.(1998a)]{Peterson98a}
Peterson, B.~M., Wanders, I., Bertram, R., Hunley, J.~F., Pogge, R.~W., \& Wagner, R.~M.
1998a, \apj, 501:82

\bibitem[Peterson et al.(1998b)]{Peterson98b}
Peterson, B.~M., Wanders, I., Horne, K., et al.\ 1998b, \pasp, 110:660

\bibitem[Phillips(1978)]{Phillips78} 
Phillips, M.~M.\ 1978, \apjs, 38:187

\bibitem[Pogge \& Peterson(1992)]{Pogge92} 
Pogge, R.~W., \& Peterson, B.~M.\ 1992, \aj, 103:1084

\bibitem[Rafiee \& Hall(2011)]{Rafiee11}
Rafiee, A., \& Hall, P.~B.\ 2011, \mnras, 415:2932

\bibitem[Reichert et al.(1994)]{Reichert94}
Reichert, G.~A., Rodriguez-Pascual, P.~M., Alloin, D., et al.\ 1994, \apj, 425:582

\bibitem[Richards et al.(2011)]{Richards11} 
Richards, G.~T., Kruczek, N.~E., Gallagher, S.~C., et al.\ 2011, \aj, 141, 167

\bibitem[Richards et al.(2002)]{Richards02} 
Richards, G.~T., Vanden Berk, D.~E., Reichard, T.~A., Hall, P.~B., Schneider, D.~P., 
SubbaRao, M., Thakar, A.~R., \& York, D.~G. 2002, \aj, 124:1 

\bibitem[Rodr{\'\i}guez-Pascual et al.(1997)]{Rod97}
Rodr{\'\i}guez-Pascual, P.~M., Alloin, D., Clavel, J., et al.\ 1997, \apjs, 110:9

\bibitem[Rousseeuw \& van Driessen(2006)]{Rousseeuw06}
Rousseeuw, P., \& van Driessen, K. 2006, Data Min.\ Knowl.\ Discovery, 12:29

\bibitem[Runnoe et al.(2013a)]{Runnoe13a} 
 Runnoe, J. C., Brotherton, M. S., Shang, Z., \& DiPompeo, M.~A. 2013a, \mnras, 434:848

\bibitem[Runnoe et al.(2013b)]{Runnoe13b}
Runnoe, J.~C., Brotherton, M.~S., Shang, Z., Wills, B.~J., \&  DiPompeo, M.~A. 
2013b, \mnras, 429:135 

\bibitem[Santos-Lle{\'o} et al.(1997)]{Santos97}
Santos-Lle{\'o}, M., Chatzichristou, E., de Oliveira, C.~M., et al.\ 1997, \apjs, 112:271

\bibitem[Santos-Lle{\'o} et al.(2001)]{Santos01}
Santos-Lle{\'o}, M., Clavel, J., Schulz, B., et al.\ 2001, \aap, 369:57

\bibitem[Schlafly \& Finkbeiner(2011)]{Schlafly11}
Schlafly, E.~F., \& Finkbeiner, D.~P.\ 2011, \apj, 737:103

\bibitem[Schlegel, Finkbeiner, \& Davis(1998)]{Schlegel98}
Schlegel, D.~J., Finkbeiner, D.~P., \& Davis, M.\ 1998, \apj, 500:525


\bibitem[Shapovalova et al.(2010)]{Shapovalova10}
Shapovalova, A.~I., Popovi{\'c}, L.~{\v C}., Burenkov, A.~N., et al.\ 2010, \aap, 517:A42

\bibitem[Shappee et al.(2014)]{Shappee14} 
Shappee, B.~J., Prieto, J.~L., Grupe, D., et al.\ 2014, \apj, 788:48

\bibitem[Shen et al.(2008a)]{Shen08a}
Shen, J., Vanden Berk, D.~E., Schneider, D.~P., et al.\ 2008a, \aj, 135:928

\bibitem[Shen(2013)]{Shen13}
Shen, Y.\ 2013, Bulletin of the Astronomical Society of India, 41:61

\bibitem[Shen(2016)]{Shen16a} 
Shen, Y.\ 2016, \apj, 817:55

\bibitem[Shen et al.(2015)]{Shen15}
Shen, Y., Brandt, W.~N., Dawson, K.~S., et al.\ 2015, \apjs, 216:4

\bibitem[Shen et al.(2008b)]{Shen08b}
Shen, Y., Greene, J.~E., Strauss, M.~A., et al.\ 2008b, \apj, 680:169

\bibitem[Shen et al.(2019)]{Shen19}
Shen, Y., Hall, P.~B., Horne, K., et al.\ 2019, \apjs, 241:34

\bibitem[Shen \& Ho(2014)]{Shen14} 
Shen, Y., \& Ho, L.~C.\ 2014, \nat, 513:210

\bibitem[Shen et al.(2016)]{Shen16b}
Shen, Y., Horne, K., Grier, C.~J., et al.\ 2016, \apj, 818:30

\bibitem[Shen \& Kelly(2012)]{ShenKelly12}
Shen, Y., \& Kelly, B.~C.\ 2012, \apj, 746:169

\bibitem[Shen \& Liu(2012)]{Shen12}
Shen, Y., \& Liu, X.\ 2012, \apj, 753:125

\bibitem[Steinhardt \& Elvis(2010)]{Steinhardt10}
Steinhardt, C.~L., \& Elvis, M.\ 2010, \mnras, 402:2637

\bibitem[Stirpe et al.(1994)]{Stirpe94}
Stirpe, G.~M., Winge, C., Altieri, B., et al.\ 1994, \apj, 425:609

\bibitem[Sulentic et al.(2007)]{Sulentic07}
Sulentic, J.~W., Bachev, R., Marziani, P., et al.\ 2007, \apj, 666:757

\bibitem[Sulentic et al.(2017)]{Sulentic17} 
Sulentic, J.~W., del Olmo, A., Marziani, P., et al.\ 2017, \aap, 608:A122

\bibitem[Sulentic et al.(2000)]{Sulentic00} 
Sulentic, J.~W., Zwitter, T., Marziani, P., \&
Dultzin-Hacyan, D. 2000, \apjl, 536:L5

\bibitem[Sun \& Shen(2015)]{Sun15} 
Sun, J., \& Shen, Y.\ 2015, \apjl, 804:L15

\bibitem[Tarter \& McKee(1973)]{Tarter73}
Tarter, C.~B., \& McKee, C.~F.\ 1973, \apjl, 186:L63

\bibitem[Trakhtenbrot \& Netzer(2012)]{Trak12}
Trakhtenbrot, B., \& Netzer, H.\ 2012, \mnras, 427:3081

\bibitem[Trevese et al.(2014)]{Trevese14} 
Trevese, D., Perna, M., Vagnetti, F., Saturni, F.G., \&
Dadina, M.\ 2014, \apj, 795:164

\bibitem[Vanden Berk et al.(2004)]{Vandenberk04}
Vanden Berk, D.E., et al.\ 2004, in AGN Physics with the
Sloan Digital Sky Survey, ed. G.~T.\ Richards and P.~B.\ Hall
(San Francisco: Astronomical Society of the Pacific), p.\ 21

\bibitem[Vestergaard(2002)]{Vestergaard02}
Vestergaard, M. 2002, \apj, 571:733

\bibitem[Vestergaard(2004)]{Vestergaard04}
Vestergaard, M. 2004, \apj, 601:676

\bibitem[Vestergaard et al.(2008)]{Vestergaard08}
Vestergaard, M., Fan, X., Tremonti, C.~A., et al.\ 2008, \apjl, 674:L1

\bibitem[Vestergaard \& Peterson(2006)]{Vestergaard06}
Vestergaard, M., \& Peterson, B.M. 2006, \apj, 641:689 (VP06)

\bibitem[Vietri et al.(2018)]{Vietri18} 
Vietri, G., Piconcelli, E., Bischetti, M., et al.\ 2018, \aap, 617:A81

\bibitem[Wandel, Peterson, \& Malkan(1999)]{Wandel99}
Wandel, A., Peterson, B.M., \& Malkan, M.A. 1999, \apj, 526:579

\bibitem[Wandel \& Yahil(1985)]{Wandel85}
Wandel, A., \& Yahil, A.\ 1985, \apjl, 295:L1

\bibitem[Wanders et al.(1997)]{Wanders97}
Wanders, I., Peterson, B.~M., Alloin, D., et al.\ 1997, \apjs, 113:69

\bibitem[Wang et al.(2009)]{Wang09}
Wang, J.-G., Dong, X.-B., Wang, T.-G., et al.\ 2009, \apj, 707:1334

\bibitem[Wang et al.(2019)]{Wang19}
Wang, S., Shen, Y., Jiang, L., et al.\ 2019, \apj, 882:4

\bibitem[Wang et al.(2020)]{Wang20}
Wang, S., Shen, Y., Jiang, L., et al.\ 2020, submitted to \apj\ (arXiv:2006.06178)

\bibitem[Weedman(1976)]{Weedman76}
Weedman, D.W. 1976, QJRAS, 17:227

\bibitem[Weymann et al.(1991)]{Weymann91} 
Weymann, R.~J., Morris, S.~L., Foltz, C.~B., et al.\ 1991, \apj, 373:23

\bibitem[White \& Peterson(1994)]{White94}
White, R.~J., \& Peterson, B.~M.\ 1994, \pasp, 106:879

\bibitem[Wilkes(1984)]{Wilkes84} 
Wilkes, B.~J.\ 1984, \mnras, 207:73

\bibitem[Wilkes(1986)]{Wilkes86} 
Wilkes, B.~J.\ 1986, \mnras, 218:331

\bibitem[Wills et al.(1993)]{Wills93} 
Wills, B.~J., Brotherton, M.~S., Fang, D., et al.\ 1993, \apj, 415:563

\bibitem[Wills \& Browne(1986)]{Wills86}
Wills, B.~J., Browne, I.~W.~A. 1986, \apj, 302:56

\bibitem[Woltjer(1959)]{Woltjer59}
Woltjer, L.\ 1959, \apj, 130:38

\bibitem[Woo et al.(2013)]{Woo13}
Woo, J.-H., Schulze, A., Park, D., et al.\ 2013, \apj, 772:49

\bibitem[Yee(1980)]{Yee80}
Yee, H.K.C. 1980, \apj, 241:894

\bibitem[Yu et al.(2020)]{Yu20}
Yu, Z., Kochanek, C.~S., Peterson, B.~M., et al.\ 2020, \mnras, 491:6045

\bibitem[Zaja\v{c}ek et al.(2020)]{Zajacek20} 
Zaja\v{c}ek, M., Czerny, B., 
Mart\'{\i}nez-Aldama, M. L., et al. 2020, \apj, 896:146 

\bibitem[Zhang et al.(2019)]{Zhang19}
Zhang, Z.-X., Du, P., Smith, P.~S., et al.\ 2019, \apj, 876:49

\bibitem[Zu, Kochanek, \& Peterson(2011)]{Zu11}
Zu, Y., Kochanek, C.~S., \& Peterson, B.~M.\ 2011, \apj, 735:80

\end{thebibliography}
\end{document}